\documentclass[a4paper,12pt]{article}
\usepackage{enumerate}
\usepackage{boxedminipage}
\usepackage{dcolumn}
\newcolumntype{d}[1]{D{.}{.}{#1}}

\usepackage[fleqn]{amsmath}      
\numberwithin{equation}{section}
\usepackage[titletoc,title]{appendix}
\usepackage[english]{babel}
\usepackage[font=small,labelfont=bf]{caption}
\usepackage[margin=2cm]{geometry}
\usepackage[authoryear,round]{natbib}

\usepackage[flushleft]{threeparttable}
\usepackage[all]{xy}
\usepackage{dcolumn}
\usepackage{amssymb,amsthm,array,arydshln,authblk,bbm,bigints,blkarray,bm,booktabs,csquotes,datetime,econometrics,enumitem,float,graphicx,rotating,mathrsfs,mathtools,mdframed,microtype,multicol,multirow,pdflscape,relsize,setspace,soul,subcaption,subfiles,tablefootnote,thmtools,url,verbatim,xfrac,xr}
\usepackage[hypertexnames=false,linktocpage=true]{hyperref}
\hypersetup{colorlinks=true,anchorcolor=blue,citecolor=blue,filecolor=blue,urlcolor=blue,bookmarksnumbered=true,pdfview=FitB}
\usepackage[normalem]{ulem}
\usepackage{titlefoot}

\usepackage{adjustbox}

\definecolor{royalazure}{rgb}{0.0, 0.2, 0.65}
\definecolor{amaranth}{rgb}{0.9, 0.17, 0.31}
\definecolor{darkgreen}{RGB}{0.0, 128, 0.0}
\definecolor{darkgoldenrod}{rgb}{0.72, 0.53, 0.04}

\DeclareMathOperator*{\argmin}{argmin}

\theoremstyle{plain}
\newtheorem{theorem}{Theorem}

\newtheorem{proposition}{Proposition}
\newtheorem{lemma}{Lemma}

\theoremstyle{definition}
\newtheorem{definition}{Definition}
\newtheorem{example}{Example}
\newtheorem{remark}{Remark}

\usepackage{multibib}
\newcites{apndx}{References in Appendix}

\newtheoremstyle{assumptionstyle}%
{0.2cm}
{0.2cm}
{\itshape}
{}
{\bfseries}
{}
{ }
{\thmname{#1}}%
\theoremstyle{assumptionstyle}
\newtheorem{assumptions}{Assumptions:}

\makeatletter

\newcommand{\Rmnum}[1]{\expandafter\@slowromancap\romannumeral #1@}
\makeatother

\newcommand{\ind}{\dto}
\newcommand{\indic}[1]{\mathbbm{1}{\left\{#1\right\}}}

\newcommand{\inp}{\pto}

\def\mytitle{Identification, estimation, and robust inference of structural autoregressions for multidimensional data
}

\newdateformat{monthyeardate}{%
	\monthname[\THEMONTH], \THEYEAR}

\newdateformat{daymonthyeardate}{%
	\monthname[\THEMONTH] \THEDAY, \THEYEAR}

\title{\mytitle}
\author[1]{Bernhard van der Sluis\thanks{ Email: \href{mailto:vandersluis@ese.eur.nl}{vandersluis@ese.eur.nl}. I thank Jonas Arias, Xu Cheng, Frank Schorfheide, Eva Janssens, Hugo Freeman, Zhongjun Qu, Annika Camehl, Erik Kole, Wendun Wang, Richard Paap, Yicong Lin, and the participants of the 2026 SBIES NSF/CEME Conference and the Tinbergen Institute PhD Job Market Jamboree,
for their valuable feedback. }\\ Econometric Institute, Erasmus University Rotterdam}

\date{\today}

\begin{document}

\thispagestyle{empty}
\setcounter{page}{0}
\begin{spacing}{1.2}
\maketitle
\begin{abstract}
\noindent
Many modern datasets such as cross-country panels have a multidimensional structure. Conventional approaches on structural vector autoregressions for identifying structural shocks are not directly suited to these settings where data are naturally organized as matrices or tensors, since the number of parameters grows rapidly and identifying restrictions become hard to impose. We introduce a class of structural autoregressions for multidimensional data that achieves tractability and global identification by assuming a Kronecker structure on the model coefficients. We develop a minimum-distance estimator and establish its asymptotic distribution. To accommodate approximate Kronecker separability, we develop confidence intervals for the structural impulse response functions that are robust to local misspecification, and we show their asymptotic validity. Monte Carlo simulations demonstrate the validity of the asymptotic approximations and good finite-sample performance of the estimator and their (robust) confidence intervals. We illustrate the methods on a country panel dataset to analyze the dynamic and spillover effects of US monetary policy shocks.


\end{abstract}
\end{spacing}

\clearpage
\newpage
\newpage
\section{Introduction} \label{sec:intro}
A large and growing body of empirical work confronts data that are inherently multidimensional: macroeconomic variables are not observed in isolation, but they are recorded at the intersection of multiple classifications. For instance, analyzing transmissions of monetary shocks across countries concerns data of two dimensions \citep[see, e.g., ][]{canova1998sources, canova2009estimating, georgiadis2016determinants}; evaluation of sectoral policies on international trade data across sectors is a four-dimensional problem [import country, export country, sector, time] \citep[see, e.g., ][]{proudman2000evolving, mancusi2008international, cai2022knowledge};  and monitoring financial banking networks across countries for different money channels requires observing data of four dimensions [country, bank, bank, channel] \citep[see, e.g., ][]{cerutti2015financial}. Conventional approaches on structural vector autoregressions (SVARs), pioneered by \cite{sims1980macroeconomics} and \cite{blanchard1989}, to identify structural shocks are not directly suited to such settings, as the numbers of parameters grows rapidly and identifying restrictions become hard to impose. 

To perform a structural analysis on multidimensional data, there at least two main challenges. First, interpretation of identifying restrictions is an important feature of structural analyses, but not straightforward for multidimensional data due to its large and complex nature. Ideally we may adopt existing economic theory as input of the setup, while respecting the multidimensional nature of the data. Second, limiting the number of parameters while preserving heterogeneous dynamics is crucial for multidimensional data. Suppose we have data for five countries, three sectors, and four economic indicators. A standard single lag SVAR model would include $2(5\times3\times4)^2 = 7200$ coefficients, let alone if more dimensions and more lags are included. At the same time, the unrestricted SVAR captures heterogeneous dynamics, an important feature of multidimensional data to capture \citep{canova2007price, canova2009estimating}. Working with multidimensional data, the propagation of shocks may operate through distinct channels corresponding to the different dimensions in the data, and thus, some degree of heterogeneity is necessary for a multidimensional SVAR.

This paper presents methods to conduct structural inference in a multidimensional VAR, while respecting interpretation, the size of the system, and heterogeneous dynamics. We restrict the large coefficient matrices to depend on dimension-specific matrices. Each of these coefficient matrices capture, for example, monetary relationships common across countries, or, country spillovers common across monetary variables. Their Kronecker product determines the heterogeneous dynamics. The structure implies that relationships across the system are separated per dimension, and hence, economic theory and existing identifying restrictions can be readily applied per dimension. Adopting the identification framework of \cite{rubio2010structural} per dimension yields our identification results. We propose a minimum-distance (MD) estimator to estimate the structural parameters. The identifying zero restrictions imply moments and thus can be used for estimation of the structural parameters. This idea has recently been used for structural factor models by \cite{HAN2025106057}. Our MD estimator and corresponding structural impulse response functions are consistent and asymptotically normal as the number of time periods diverges to infinity.  

Importantly, the Kronecker structure that we impose on the coefficients implies a hierarchical factor structure. To see this, consider a three-dimensional example of economic indicators per sector per country. The output factor corresponds to the linear combination of output in all sectors and countries. The loadings correspond to the country-specific and sector-specific relationships within the economic indicator. Hence, the expected value of the output in some sector in some country is modeled as a particular linear combination of the output across all sectors and country, and then reweighted by the economic relationships within that sector and country.  The hierarchical or block structure introduced for large panel data, see \cite{moench2013dynamic} and \cite{diebold2008global}, bears resemblance to our structure. \cite{chen2021autoregressive} and \cite{chan2025large} impose the same Kronecker structure in a reduced-form autoregression for matrix time series (which corresponds to two-dimensional data). However, we consider data with potentially more than two dimensions and are interested in structural inference. Alternative approaches have been used to conduct structural inference in large SVAR systems including structural factor models \citep{HAN2025106057}, shrinkage methods \citep{arias2026inference}, clustering \citep{janssens2025restricted}, large Bayesian VARs \citep{chan2019large}, among others. Closest to us are structural factor models, that estimate a weighted combination of the variables to capture the most variability, and where the loadings correspond to the heterogeneous effects of these factors on the variables. Our approach differentiates in at least two ways. First, factor estimation requires the number of variables to grow, whereas we keep the sizes of each dimension and the number of dimensions fixed. Second, existing SVAR identifying restrictions from economic theory can be employed directly on our model due to the dimension-separation, whereas the structural factor model requires restrictions on a weighted combination of the data.

The related literature on existing approaches on tensor and large SVAR shares several similarities with our approach. First, \cite{billio2023bayesian} and \cite{qi2025largebayesiantensorautoregressions} consider a tensor autoregression and employ a low-rank representation of the coefficients through so-called ``marginals'' to reduce the number of parameters. Their structure is more flexible than ours, but is less suitable for structural approaches since the marginals have no interpretation. The structure we employ (i) reduces the number of parameters, (ii) preserves interpretation for structural inference, and (iii) implies a tractable factor structure that can be used for policy purposes (impulse responses, forecasting, coincident indicators, ...). Furthermore, in the large SVAR literature, panel VAR models \citep{canova2013panel} consider each cross-sectional unit (say country) separately and let all units enter as regressors. Factorization is a common approach in panel VAR to reduce the size of the system, losing interpretability and separation of propagation channels (see Section \ref{sec:compare}). Alternatively for two-dimensional data, we can consider the global VAR model proposed by \cite{pesaran2004modeling}, in which foreign weighted averages enter as regressors in a country-specific VAR model. The weights are exogenously determined and the number of parameters remains large, making structural identification and interpretation less straightforward than the present model. Finally, the network VAR presented by \cite{zhu2017nar} is a VAR model tailored for time series with network interpretation where each node’s response is assumed to be a linear combination of its previous value, the average of its connected nodes, covariates and noise. We instead depart from a network interpretation and allow for more than two dimensions in the data. 

One may view the Kronecker structure as an approximation to the true structure in finite sample, and thus, we provide inference methods in case there is local misspecification. Here, the specification error and the sampling error are of the same order of magnitude, such that it may not be clear in finite-sample whether the Kronecker structure is correctly specified. Such local misspecification has been used by others in different settings, including but not limited to, \cite{newey1985generalized, newey1985maximum}, \cite{guggenberger2012asymptotic}, \cite{berkowitz2012validity}, and \cite{bugni2019inference}. Notably, \cite{kitamura2013robustness} and \cite{bonhomme2022minimizing} consider an alternative approach based on misspecification defined by notions of statistical distance. Our conventional approach to construct confidence intervals for structural impulse response functions is biased in the case of local misspecification. We therefore propose robust confidence intervals, based on the approach of \cite{armstrong2021sensitivity} for locally misspecified moment conditions. In this approach, one adds and subtracts a critical value that accounts for misspecification bias, and sharpens the length by using a misspecification-optimal weighting matrix, to compute robust confidence intervals. \cite{andrews2017measuring, andrews2020informativeness} consider a similar approach on inference under local misspecification, in which influence function weights are used for sensitivity analysis. \cite{armstrong2021sensitivity} sharpen their approach of sensitivity analysis using the misspecification-optimal weighting matrix. We adopt the latter in Section \ref{sec:robust}.

We establish the asymptotic distribution of the MD estimator under correct specification of Kronecker separability and under local misspecification. We construct an asymptotically valid specification test for overidentifying restrictions in the structural autoregression model. Our simulation study shows that asymptotic approximations perform well for inference in finite sample. Furthermore, we illustrate our methods on a two-dimensional dataset of four monetary variables across five countries. We adopt recursive ordering in both dimensions to identify the dynamic effects of a US monetary policy shock on domestic and foreign economies. We find contemporaneous effects on foreign interest rates and lagged effects on foreign output. Our robust confidence intervals show that these findings still hold under certain degrees of approximate of Kronecker separability.

The paper proceeds as follows. Section \ref{sec:methods} introduces a class of structural autoregressions for multidimensional data and its identification, and compares the framework to related existing approaches. Section \ref{sec:estim and props} establishes global identification of the structural impulse response functions, and outlines the MD estimator and its asymptotic properties. Section \ref{sec:robust} presents our local misspecification robust confidence intervals. Section \ref{sec:simulation} and \ref{sec:empirics} present the simulation study and empirical application, respectively. Section \ref{sec:conclusion} concludes.

\textbf{Notation.} Finally, we present some notation. A vector is written in lower-case and bold ($\vv$), a matrix in upper-case and bold ($\mV$), and a tensor in calligraphic and bold ($\mcalV$). An $N$-dimensional real-valued tensor is defined as a $N$-dimensional array $\mcalX \in \SR^{n_1\times \ldots \times n_N}$ with entries $\mcalX_{i_1,\ldots,i_N}$ with $i_l = 1,\ldots,n_l$ and $l=1,\ldots,N$. For $N=1$ and $N=2$, the tensor reduces to a vector and matrix, respectively. We refer to the dimensions of a tensor as \emph{modes}. The vectorization operator $\vec(\cdot)$ stacks the columns within each dimension of a tensor into a single vector in $\SR^{n_1\cdot n_2 \cdots n_d}$. For a matrix $\mX$. let $[\mX]_i$ denote the $i$-th column of $\mX$. The Kronecker product between matrices $\mA \in \SR^{n_1\times n_2}$ and $\mB^{m_1\times m_2}$ is denoted as $\mA \otimes \mB \in \SR^{n_1 m_1 \times n_2 m_2}$. Moreover, the mode-$k$ product of a tensor $\mcalX \in \SR^{n_1 \times \cdots \times n_N}$ with a matrix $\mA \in \SR^{m_k \times n_k}$ is denoted $\mcalX \times_k \mA \in \SR^{n_1 \times \cdots \times n_{k-1} \times m_k \times n_{k+1} \times \cdots \times n_N}$, which multiplies $\mA$ along the $k$-th mode of $\mcalX$. For successive mode-$k$ products across all modes, we use the shorthand $\mcalY=\mcalX \times_{k=1}^{N} \mA_k \coloneqq \mcalX \times_1 \mA_1 \times_2 \mA_2 \cdots \times_N \mA_N$, or equivalently $\vec(\mcalY) = \left(\mA_N\otimes \cdots \otimes \mA_1\right)\vec(\mcalX)$. The determinant of a matrix $\mV$ is denoted as $|\mV|$. We write $||\mV||_F = \sqrt{\sum^n_{i=1}\sum^m_{j=1}|v_{ij}|^2}$ for some $\mV$ to denote its Frobenius norm. Furthermore, for a $D\times D$ square positive semi-definite matrix $\mA$ with eigenvalues $\lambda_1\geq\ldots\geq\lambda_D$, we define the spectral radius of $\mA$ as $\rho(\mA) = \lambda_1$. For a set $\sS$, we denote its cardinality as $|\sS|$. Throughout the paper, $C$ denotes a universal constant that may take different values.

\section{Methodology} \label{sec:methods}
Let $\mcalY_t = (Y_{_{i_1i_2\cdots i_N,t}}) \in \SR^{n_1\times n_2\times\cdots\times n_N}$ be a $N$-dimensional tensor of observable endogenous variables at time $t$, with sizes $n_1, n_2, \ldots, n_N$. We propose the following structural autoregressive model of order $p$ as
\begin{align}\label{eq:sar}
    \mcalY_t \times^N_{k=1} \mA_{0,k} = \widetilde{\mcalC} + \sum^p_{l=1}\mcalY_{t-l}\times^N_{k=1}\mA_{l,k} + \mcalE_t, \quad \text{for }t=1,\ldots,T,
\end{align}
where $T$ denotes the sample size. The term $\mcalE_t$ denotes an $N$-dimensional tensor of exogenous structural shocks. Conditional on past information and initial conditions $\mcalY_0, \ldots,\mcalY_{1-p}$, the vector $\Vec\left(\mcalE_t\right)$ is assumed to be Gaussian with mean zero and covariance matrix $\mI_{n_1 n_2\cdots n_N}$, the $n_1n_2\cdots n_N$ identity matrix. The matrices $\mA_{l,k} \in \SR^{n_k\times n_k}$ correspond to the structural coefficients for lag $l$ at mode $k$. Each coefficient matrix captures the structural relations within each dimension. For example, for an analysis on monetary policy across countries over time ($N=2$), $\mA_{0,1}$ captures the contemporaneous effects across monetary variables, whereas $\mA_{0,2}$ captures the contemporaneous monetary spillover across countries. The tensor $\widetilde{\mcalC} \in \SR^{n_1\times n_2\times\cdots \times n_N}$ holds the intercept terms. As desired, our model reduces to a standard SVAR when there is a single dimension ($N=1$). 

We assume that for each mode $k=1,\ldots,N$, the contemporaneous effect matrix $\mA_{0,k}$ is invertible, to obtain the reduced-form representation of \eqref{eq:sar}:
\begin{align} \label{eq:rf-sar}
    \mcalY_t = \mcalC + \sum^p_{l=1}\mcalY_{t-l}\times^N_{k=1}\mB_{l,k} + \mcalU_t, \quad \text{for }t=1,\ldots,T,
\end{align}
where $\mcalC = (\widetilde{\mcalC} \times^N_{k=1} \mA^{-1}_{0,k})$, $\mB_{l,k} = \mA_{0,k}^{-1}\mA_{l,k}$, and $\mcalU_t = \mcalE_t \times^N_{k=1} \mA^{-1}_{0,k}$. The covariance matrix of $\vec(\mcalU_t)$ is given by $\E(\vec(\mcalU_t)\vec(\mcalU_t)^\prime) = \mSigma = (\mSigma_N \otimes \cdots \otimes \mSigma_1)$, where $ \mSigma_k= (\mA_{0,k}^\prime\mA_{0,k})^{-1}$ for all $k=1,\ldots,N$. 
The reduced-form presentation of the data is equivalent to the standard VAR model for $N=1$ and to the matrix-autoregressive (MAR) model studied by \cite{chen2021autoregressive} for $N=2$. Our reduced-form extends the MAR interpretation for data with potentially more than two dimensions. In particular, the reduced-form representation can be viewed as a special case of the standard VAR$(p)$ model, in which the autoregressive coefficient matrices admit a Kronecker structure.

To see the Kronecker structure and its benefits, we vectorize the data $\mcalY_t$ and specify a standard SVAR model,
\begin{align} \label{eq:var}
    \vec(\mcalY_t)^\prime\widetilde{\mPhi}_0 = \sum^p_{l=1}\vec(\mcalY_{t-l})^\prime\widetilde{\mPhi}_l + \vec(\mcalE_t)^\prime, \quad \text{for }t=1,\ldots,T,
\end{align}
where the structural coefficient matrix $\widetilde{\mPhi}_l$ is an $n\times n$ matrix with $n=\prod^N_{k=1}n_k$, for all lags $l=1,\ldots p$. This SVAR model counts $(p+1)n^2$ parameters. If we impose a Kronecker structure, $\widetilde{\mPhi}_{l} = (\mA^\prime_{l,N}\otimes \ldots \otimes \mA^\prime_{l,1})$ for each lag $l$, we obtain representation \eqref{eq:sar}. The number of structural parameters in \eqref{eq:sar} (without intercept) is $(p+1)\sum^N_{k=1}n^2_k$ instead of $(p+1)n^2$. Even when $N$ and the $n_k$'s are moderate the number of parameters in our model is much smaller.\footnote{Appendix \ref{app:methods} documents the number of parameters in our model relative to the SVAR model for several data sizes. In short, the number of parameters in SVAR explodes relative to \eqref{eq:sar}, even for moderate sizes of $N$ and $n_k$.} For estimation and inference, this result is particularly important. Moreover, we can see in \eqref{eq:var} that a linear SVAR representation for matrix data neglects the matrix structure, as the roles of modes are no longer separated. Model \eqref{eq:sar} allows for straightforward interpretation, which is crucial for policymakers as impulse response functions can be deduced directly from the mode-specific coefficients, similar to impulse response functions in SVARs in a single data dimension. 

\subsection{Examples}
To illustrate the interpretation of our approach, we consider three examples.
\begin{example}[A multicountry model]
    Suppose we observe three monetary variables - Output (GDP), Inflation (Inf), Interest Rate (Int) - for the U.S., the Euro area (EUR), and a country representing the Rest of the world (RoW) such that $n_1=n_2=3$. Model \eqref{eq:sar} without intercept amd a single lag corresponds to
    \begin{align*}
    &
       \mA_{0,1}
        \begin{pmatrix}
           Y_{GDP,US} & Y_{GDP,EUR} & Y_{GDP,RoW}\\
            Y_{Inf,US} & Y_{Inf,EUR} & Y_{Inf,RoW}\\
Y_{Int,US} & Y_{Int,EUR} & Y_{Int,RoW}
       \end{pmatrix}_t
        \mA_{0,2}^\prime \\
       &= 
        \mA_{1,1}
        \begin{pmatrix}
           Y_{GDP,US} & Y_{GDP,EUR} & Y_{GDP,RoW}\\
            Y_{Inf,US} & Y_{Inf,EUR} & Y_{Inf,RoW}\\
Y_{Int,US} & Y_{Int,EUR} & Y_{Int,RoW}
       \end{pmatrix}_{t-1}
        \mA_{1,2}^\prime +
        \begin{pmatrix}
           \epsilon_{GDP,US} & \epsilon_{GDP,EUR} & \epsilon_{GDP,RoW}\\
            \epsilon_{Inf,US} & \epsilon_{Inf,EUR} & \epsilon_{Inf,RoW}\\
\epsilon_{Int,US} & \epsilon_{Int,EUR} & \epsilon_{Int,RoW}
       \end{pmatrix}, 
    \end{align*}
    where, for each lag $l=0,1$, $\mA_{l,1}$ and $\mA_{l,2}$ govern the variable and country relationships, respectively. For example, consider $\mA_{0,1}$: it captures the contemporaneous monetary interactions, irrespective of the country. Heterogeneity across countries of the contemporaneous effect appears due to $\mA_{0,2}$. To see this, use that $\vec(\mA_{0,1} \mcalY_t \mA_{0,2}^\prime) = \left(\mA_{0,2}\otimes \mA_{0,1}\right) \vec(\mcalY_t)$, and thus, the large SVAR coefficient matrix $\widetilde{\mPhi}_0$ in \eqref{eq:var} corresponds
    \begin{align*}
        &\widetilde{\mPhi}_0 = \mA_{0,2}^\prime\otimes \mA^\prime_{0,1}\\
        &=\begin{pmatrix}
            a_{0,2,US,US} & a_{0,2,EUR,US} & a_{0,2,RoW,US}\\
            a_{0,2,US,EUR} & a_{0,2,EUR,EUR} & a_{0,2,RoW,EUR}\\
            a_{0,2,US,RoW} & a_{0,2,EUR,RoW} & a_{0,2,RoW,RoW}            
        \end{pmatrix} \otimes
        \begin{pmatrix}
            a_{0,1,GDP,GDP} & a_{0,1,Inf,GDP} & a_{0,1,Int,GDP}\\
            a_{0,1,GDP,Inf} & a_{0,1,Inf,Inf} & a_{0,1,Int,Inf}\\
            a_{0,1,GDP,Int} & a_{0,1,Inf,Int} & a_{0,1,Int,Int}            
        \end{pmatrix}\\
        &= \begin{pmatrix}
            a_{0,2,US,US}\times a_{0,1,GDP,GDP} & a_{0,2,US,US}\times a_{0,1,Inf,GDP} &\cdots & a_{0,2,EUR,US}\times a_{0,1,Int,GDP}\\
            a_{0,2,US,US}\times a_{0,1,GDP,Inf} & a_{0,2,US,US}\times a_{0,1,Inf,Inf} & \cdots & \vdots\\
            \vdots & \vdots&\ddots & \vdots\\
            a_{0,2,US,RoW}\times a_{0,1,GDP,Int} &\cdots & \cdots & a_{0,2,RoW,RoW}\times a_{0,1,Int, Int}
        \end{pmatrix}
    \end{align*}
    Thus, the contemporaneous effect of US output on Euro inflation is the effect of output on inflation ($a_{0,1,Inf,GDP}$) proportional to the monetary effect the U.S. has on the Euro area ($a_{0,2,EUR,US}$). The heterogeneity of monetary effect across countries is thus proportional to an ``average" monetary effect. The proportional heterogeneity-argument carries over to any $N$-dimensional case. For example, in international trade of types of goods, the heterogeneity in the effect of an export restriction for good 1 in country A is proportional to importing magnitude with country A and to trading good 1. 

    The Kronecker structure separates the monetary effect and country spillovers. As a primer, this separation allows us to identify the system by identifying each mode (variable, country) separately. Since the economic literature has proposed several identification strategies for monetary effects within a country, they can be readily applied in our framework. The same holds for structural spillovers of countries, for example by recursive ordering of counties.

\end{example}

\begin{example}[DSGE model]
    We consider a standard three-equation New Keynesian model for $n_2$ countries, augmented with spillovers. For country $i$, we specify the equations for output ($x_{it}$), inflation ($\pi_{it}$), and interest rate ($r_{it}$) as 
    \begin{align*}
        \text{(IS)} \quad &(x_{it} + \sigma^{-1}r_{it}) + \sum_{j\neq i}v_{ij}(x_{jt} + \sigma^{-1}r_{jt}) \\
        &\qquad= \E_{t}\left(x_{it+1} + \sigma^{-1}\pi_{it+1}\right) + \sum_{j\neq i}w_{ij}\E_{t}\left(x_{jt+1}+ \sigma^{-1}\pi_{jt+1}\right) + r^n_{it}\\
        \text{(PC)} \quad &(\pi_{it} - \kappa x_{it}) + \sum_{j\neq i}v_{ij}(\pi_{jt} - \kappa x_{jt}) = \beta\E_t \pi_{it+1} + \sum_{j\neq i}w_{ij}\beta\E_t \pi_{jt+1} + u^s_{it}\\
        \text{(TR)} \quad & \left(r_{it} - \phi_{\pi}\pi_{it} - \phi_{x}x_{it}\right) + \sum_{j\neq i}v_{ij}\left(r_{jt} - \phi_{\pi}\pi_{jt} - \phi_{x}x_{jt}\right) = \epsilon^{mp}_{it}
    \end{align*}
    Let $[\mA_{0,2}]_{ij} = v_{ij}$ and $[\mA_{1,2}]_{ij} = w_{ij}$ with $v_{ii} = w_{ii}=1$. Stacking all $n_2$ countries, the system takes the form $\left(\mA_{0,2}\otimes \mA_{0,1}\right)\vec(\mcalY_t) = \left(\mA_{1,2}\otimes \mA_{1,1}\right) \E_t\vec(\mcalY_{t+1}) + \vec(\mcalE_t)$, where
    \begin{align*}
        \mA_{0,1} = \begin{pmatrix}
            1 & 0 & \sigma^{-1}\\
            -\kappa & 1 & 0\\
            -\phi_x & -\phi_{\pi} & 1
        \end{pmatrix}, \quad \mA_{1,1} = \begin{pmatrix}
            1&-\sigma^{-1} & 0\\
            0 & \beta & 0\\
            0 & 0 & 0
        \end{pmatrix}.
    \end{align*}
    Matrices $\left(\mA_{0,2},\mA_{1,2}\right)$ capture the spillover intensities in the equations at each lag. Identification is separated into two components. First, $\mA_{0,1}$ is identified from within-country timing restrictions, see e.g. \cite{christiano1999monetary}. Second, $\mA_{0,2}$ is identified from the network structure. For instance, a dominant country ordering renders a lower triangular structure $\mA_{0,2}$.
    
    The global VAR approach of \citet{pesaran2004modeling, dees2007exploring, chudik2016theory, smith2013gvar} captures spillovers through exogeneous, fixed (trade) weights, while our weights are endogenous, structurally interpretable, and directly restrict the SVAR coefficient matrix for identification.
\end{example}

\begin{example}[Factor model]
The Kronecker structure implies a hierarchical factor interpretation. Consider $N=3$ with dimensions corresponding to economic indicators ($n_1$), sectors ($n_2$), and countries ($n_3$). The reduced-form \eqref{eq:rf-sar} without intercept is

\begin{equation}
    \mcalY_t = \mcalY_{t-1} \times_1 \mB_{1,1} \times_2 \mB_{1,2} \times_3 \mB_{1,3} + \mcalU_t.
\end{equation}
The conditional mean $\mcalY_t - \mathcal{U}_t$ can be build hierarchically through three successive mode products. First, define the country-aggregated data
\begin{equation}
    \mcalX^{(3)}_{t-1} := \mcalY_{t-1} \times_3 \mB_{1,3} \in \mathbb{R}^{n_1 \times n_2 \times n_3},
\end{equation}
with entries $X^{(3)}_{j_1 j_2 i_3, t-1} = \sum_{j_3=1}^{n_3} [\mB_{1,3}]_{i_3 j_3}Y_{j_1 j_2 j_3, t-1}$. Each slice $\mcalX^{(3)}_{\cdot, \cdot, i_3, t-1}$ is an $n_1 \times n_2$ matrix that aggregates all countries into a single panel for country $i_3$, with weights given by the $i_3$-th row of $\mB_{1,3}$. The factor $F^{(3)}_{\text{GDP}, \text{Manuf}, \text{US}, t-1}$, for instance, represents a weighted average of manufacturing GDP across all countries, where the weights reflect how strongly each country's manufacturing output feeds into the U.S. dynamics.

Second, apply the sector coefficients to obtain
\begin{equation}
    \mcalX^{(2)}_{t-1} := \mcalX^{(3)}_{t-1} \times_2 \mB_{1,2} = \mcalY_{t-1} \times_2 \mB_{1,2} \times_3 \mB_{1,3},
\end{equation}
with entries $X^{(2)}_{j_1 i_2 i_3, t-1} = \sum_{j_2=1}^{n_2} [\mB_{1,2}]_{i_2 j_2}\, X^{(3)}_{j_1 j_2 i_3, t-1}$. At this stage, $X^{(2)}_{j_1 i_2 i_3, t-1}$ is a factor for indicator $j_1$ at position $(i_2, i_3)$: it aggregates indicator $j_1$ across all sectors and countries, localized to sector $i_2$ and country $i_3$. For example, $X^{(2)}_{\text{GDP}, \text{Manuf}, \text{US}, t-1}$ summarizes how GDPs from all sectors and countries contribute to U.S. manufacturing, after both country and sector reweighting.

Finally, the conditional mean is completed by loading through $\mB_{1,1}$:
\begin{equation}
    \mcalY_t - \mcalU_t = \mcalX^{(2)}_{t-1} \times_1 \mB_{1,1},
\end{equation}
where $\mB_{1,1}$ acts as a loading matrix along the indicator dimension. The entry for indicator $i_1$ in sector $i_2$ of country $i_3$ is
\begin{equation}\label{eq:factor_entry}
    Y_{i_1 i_2 i_3, t} - U_{i_1 i_2 i_3, t} = \sum_{j_1=1}^{n_1} [\mB_{1,1}]_{i_1 j_1}\, X^{(2)}_{j_1 i_2 i_3, t-1}.
\end{equation}
Equation \eqref{eq:factor_entry} says that, for example, the expected manufacturing output in the U.S. is a linear combination of $n_1$ indicator-specific factors - an output factor, an inflation factor, an interest rate factor - each of which has already been aggregated across countries by $\mB_{1,3}$ and across sectors by $\mB_{1,2}$. The row of $\mB_{1,1}$ corresponding to output determines how these indicator factors combine.

The structure resembles the hierarchical factor models of \cite{diebold2008global} and \cite{moench2013dynamic}, where block-level and global factors capture variation at different levels of aggregation. The key distinction from standard factor models is that the factors here are not latent: they are known linear combinations of the observables, with coefficients that are dimension-specific and separately interpretable. In the structural model \eqref{eq:sar}, the contemporaneous matrices $\mA_{0,k}$ play an analogous role, and existing identifying restrictions from Section \ref{sec:identification} can be imposed on each level of the hierarchy separately.
\end{example}

\subsection{Structural impulse response functions}
To keep notation simple, we assume that there is no intercept term $\widetilde{\mC}$ and set $p=1$ throughout the paper. In what follows, we derive the structural impulse response function for $p=1$ and suppress subscript notation for lagged reduced-form coefficients (e.g., $\mB_k \equiv \mB_{1,k}$). Using the single-lag form, we can show that the reduced-form model admits a causal and stationary solution if the spectral radius of the autoregressive coefficient matrix in the reduced-form representation of the data is strictly less than one.

\begin{proposition} \label{prop:causality}
    If $\prod^N_{k=1}\rho(\mB_{k}) < 1$ for a reduced-form process with lag order one, then this process is stationary and stable.
\end{proposition}
Following Proposition \ref{prop:causality} we obtain the following causal representation,
\begin{align*}
    \vec(\mcalY_t) = \sum^{\infty}_{s=0}\left(\mB_N^s\otimes \cdots \otimes \mB_1^s\right)\vec(\mcalU_{t-s}),
\end{align*}
with the convention that $\mB_N^s \otimes \cdots \otimes \mB_1^s = \mI_{n}$ for $s=0$. We can further rewrite this in terms of structural $\mA_{0,k}$,
\begin{align*}
    \vec(\mcalY_t) = \sum^{\infty}_{s=0}\left(\mB_N^s \otimes \cdots\otimes\mB_1^s\right)(\mA^{-1}_{0,N}\otimes \cdots \otimes\mA^{-1}_{0,1})\vec(\mcalE_{t-s}).
\end{align*}
Let $\mcalE_{t-s}$ be a matrix with nonzero element $[\mcalE_{t-s}]_{i_1,i_2,\ldots, i_N}$ and zero elsewhere. Then the structural impulse response function of $\mcalY_t$ to shock $[\mcalE_{t-s}]_{i_1,i_2,\ldots,i_N}$ can be computed as
 \begin{align}
\frac{\partial \vec(\mcalY_t)}{\partial [\mcalE_{t-s}]_{i_1,i_2,\ldots,i_N}}
&=
\left(\mB_N^s\otimes\cdots\otimes\mB_1^s\right)\left(\Big[\mA_{0,N}^{-1}\Big]_{i_N}\otimes \cdots \otimes \Big[\mA_{0,1}^{-1}\Big]_{i_1}\right)
\\
&= \left(\mB_N^s\Big[\mA_{0,N}^{-1}\Big]_{i_N}\right) \otimes \cdots \otimes \left(\mB_1^s\Big[\mA_{0,1}^{-1}\Big]_{i_1}\right)
\label{eq:sr_irf_vec}
 \end{align}
 In tensor form, we can write
\begin{align}
     \mcalF_{i_1,i_2,\ldots,i_N}(s) \equiv \mcalE^{(i_1,\ldots,i_N)}\times^N_{k=1}\left(\mB_{k}^s\mA_{0,k}^{-1}\right),
\label{eq:sr_irf_tensor}
\end{align}
where $\mcalE^{(i_1,\ldots,i_N)}$ denotes a $n_1\times\ldots\times n_N$ tensor with a one at $(i_1,\ldots,i_N)$ and zeros elsewhere. In the case that $N=2$, the structural impulse response functions $\mcalF_{ij}(s)$ to a unit shock in row $i$ and column $j$ simplify to a matrix
\begin{align} \label{eq:sr_irf_matrix}
    \mcalF_{ij}(s) = \left[\mB_1^s \mA^{-1}_{0,1}\right]_i \left[\mB_2^s \mA^{-1}_{0,2}\right]^\prime_j.
\end{align}
In general, the impulse response functions $\mcalF_{i_1,i_2,\ldots,i_N}(s)$ are $N$-dimensional tensors, with each element reflecting the response of an individual unit in the data. We discuss estimation of the structural parameters and the IRFs in Section \ref{sec:estim and props}. To enhance intuition of the impulse response functions, we provide a discussion for the $2$-dimensional case in the supplement. 

\subsection{Identification} \label{sec:identification}
\subsubsection{Normalization of coefficients}
Suppose that the Kronecker product $\mB_N\otimes \cdots \otimes \mB_1$ in \eqref{eq:rf-sar} has been identified from the data. However, for any nonzero scalars $c_1,\ldots,c_N$ with $\prod^N_{k=1}c_k=1$, one can replace $\mB_k$ with $c_k\mB_k$ for all $k$, and the Kronecker product would remain unchanged. The same issue occurs in $\mSigma = \mSigma_N\otimes\cdots \otimes \mSigma_1$. Therefore, if the coefficients are of interest, $N-1$ normalization constraints are necessary to fix the scale of the coefficients and covariance matrices.

\begin{definition}
A \textbf{normalization scheme} $\SZ^B$ ($\SZ^\Sigma$)is a set of restrictions on $\{\mB_k\}^N_{k=1}$ ($\{\mSigma_k\}^N_{k=1}$) such that they can be uniquely determined from the Kronecker product $\mB_N\otimes \cdots \otimes \mB_1$ ($\mSigma_N\otimes \cdots\otimes\mSigma_1$), up to a sign flip.
\end{definition}
A plausible normalization is to impose that the Frobenius norm $||\mB_k||_F$ is equal to one for $k=1,\ldots,N-1$, and similarly for $\mSigma_k$. We follow this convention.

 \subsubsection{Global identification of structural parameters}
 Let $\mC_0 \equiv \mA^{-1}_{0,N}\otimes \cdots \otimes \mA^{-1}_{0,1}$.
 Following \cite{rothenberg1971identification}, two structural parameters $\mC_0$ and $\Bar{\mC}_0$ are observationally equivalent if they imply the same distribution of the process $\{\mcalY_t\}^T_{t=1}$. In the linear Gaussian model that we consider, this means that, for given normalization rule $\SZ$, $\mC_0$ and $\Bar{\mC}_0$ are observationally equivalent if and only if they imply the same reduced-form representation. Due to the structure of the reduced-form parameters $\mB_k = \mA^{-1}_{0,k}\mA_k$, the parameters $\mC_0$ and $\Bar{\mC}_0$ are observationally equivalent if and only if $\mA_{0,k} = \mP_{k}\Bar{\mA}_{0,k}$ and $\mA_{k} = \mP_{k}\Bar{\mA}_{k}$ for some orthogonal matrices $\mP_k \in \calO(n_k)$, for all $k=1,\ldots,N$. The set $\calO(\ell)$ denotes the set of all $\ell \times \ell$ orthogonal matrices. Equivalently, we can say that $\mC_0$ and $\Bar{\mC}_0$ are observationally equivalent if and only if $\left(\mA_{0,N} \otimes \cdots \otimes \mA_{0,1}\right) = \mP \left(\Bar{\mA}_{0,N} \otimes \cdots \otimes \Bar{\mA}_{0,1}\right)$ and $\left(\mA_{N} \otimes \cdots \otimes \mA_{1}\right) = \mP \left(\Bar{\mA}_N \otimes \cdots \otimes \Bar{\mA}_1\right)$, for some matrix $\mP \in \calP \subset \calO(n)$ where $\calP$ is defined by $\calP = \{\mP \in \calO(n)|\mP = \mP_N \otimes \cdots \otimes \mP_1, \mP_k \in \calO(n_k)\}$.

 \begin{definition} \label{def:global_ident}
     Given a normalization rule $\SZ$, a parameter point $\mC_0$ is globally identified if and only if there is no other parameter point that is observationally equivalent.
 \end{definition}
 
Since model \eqref{eq:sar} can be written as a large SVAR, it is easy to see that an unrestricted model \eqref{eq:sar} is not globally identified. We need to impose restrictions on the structural parameters. The first class of restrictions we consider is the class of zero restrictions on the structural parameters or functions thereof. For example, we can adopt triangular identification scheme in model \eqref{eq:sar}. Let us consider a two-dimensional example: we consider monetary policy for two economies, one small open economy and one large closed economy. Commonly, block exogeneity restrictions following \cite{cushman1997identifying} imply that the monetary policies of the closed economy influence the policy in the small open economy contemporaneously, but that there is no feedback from the small open economy to the large closed economy. In our model this would imply a recursive ordering in the, say, first dimension,\begin{align*}
    \mA^\prime_{0,1} = \begin{pmatrix}
    a_{0,1,11} & a_{0,1,21}\\
    0 & a_{0,1,22}
\end{pmatrix}.
\end{align*}

Secondly, we consider nonlinear restrictions on the structural parameters. This set of restrictions includes restrictions on long-run effects à-la \cite{blanchard1989}. Let $\mcalX^{(2)}_{t-l} = \mcalY_{t-l}\times^N_{k=2} \mA_{l,k}$ such that \eqref{eq:sar} can be written as
\begin{align*}
    \mA_{0,1} \mcalX^{(2)}_t = \sum^p_{l=1}\mA_{l,1} \mcalX^{(2)}_{t-l} + \widetilde{\mcalC} + \mcalE_t.
\end{align*}
Following the derivation of \cite{blanchard1989} for SVAR in this reparameterized model, if the $(i_1,\ldots,i_N)$-th variable is in first-difference, its long-run impulse response to a shock indexed by $(j_1,\ldots,j_N)$ is obtained from the corresponding $n_1\times n_1$ block of
\begin{align*}
    \mathbf{IR}^1_{\infty}(\mA_{0,1}, \{\mA_{l,1}\}^p_{l=1}) = \left(\mI_{n_2 \cdots n_N} \otimes \mA_{0,1} - \sum^p_{l=1}\mI_{n_2 \cdots n_N}\otimes \mA_{l,1}\right)^{-1} = \mI_{n_2 \cdots n_N} \otimes \left(\mA_{0,1} - \sum^p_{l=1}\mA_{l,1}\right)^{-1}.
\end{align*}
We interpret matrix $\mathbf{IR}^1_{\infty}(\mA_{0,1}, \{\mA_{l,1}\}^p_{l=1})$ as the long-run propagation matrix along the first dimension, which applies identically for every fixed configuration of the remaining indices $(i_2,\ldots,i_N)$.

Since $(\mA_{0,k}, \mA_k) = (\mP_k \Bar{\mA}_{0,k}, \mP_k \Bar{\mA}_k)$ implies $(\mA^\prime_{0,k}, \mA^\prime_k) = (\Bar{\mA}^\prime_{0,k}\mP_k, \Bar{\mA}_k^\prime \mP_k)$ and $\mP_k^\prime \in O(n_k)$, we relabel $\mP_k^\prime$ as $\mP_k$ and consider zero restrictions on any mapping $f_K(\mA_{0,k}^\prime, \mA_{k}^\prime)$ from the $k$-mode specific structural parameters $\left(\mA^\prime_{0,k}, \mA^\prime_{k}\right)$ to the space of $z_k\times n_k$ matrices, for all $k=1,\ldots,N$. The functions $\{f_k\}^N_{k=1}$ need to satisfy some regularity conditions, see \cite{rubio2010structural} and \cite{arias2018inference}, and more importantly the functions need to be \textit{admissible}, 
\begin{align*}
    f_k(\mA_{0,k}^\prime, \mA_{k}^\prime) \mP_k = f_k(\mA^\prime_{0,k}\mP_k, \mA^\prime_k\mP_k),
\end{align*}
for every $\mP \in \calP$. For example, for direct zero restrictions on the structural row parameters, we can define the identity mapping
\begin{align*}
    f_k(\mA^\prime_{0,k}, \mA^\prime_k) = \begin{pmatrix}
        \mA^\prime_{0,k}\\ \mA^\prime_k
    \end{pmatrix},
\end{align*}
similarly as in \cite{rubio2010structural}, which trivially satisfies the admissibility condition. Widely used restrictions in the SVAR literature such as recursive ordering can thus be applied to model \eqref{eq:sar}. Furthermore, we can impose linear restrictions on the short-run IRFs by decomposing the IRF into $N$ parts:
\begin{align*}
    f_k(\mA^\prime_{0,k}, \mA^\prime_k) = \mB_k \mA^{-1}_{0,k} = \mA^{-1}_{0,k}\mA_k \mA^{-1}_{0,k}, \quad \text{for }k=1,\ldots,N.
\end{align*}
At horizon $s=1$, as above, we can clearly see that the admissible condition is satisfied. The argument extends for further horizons by writing each mode of the IRF in a compact form, see the supplement and Example 2 in \cite{rubio2010structural}.

Let $\mQ_{k,j}$ be an $z_{k} \times z_k$ full rank matrix, with $0\leq z_{k} \leq n_k-j$ for $1\leq j\leq n_k$. Given the function $f_k(\cdot)$, the zero restrictions on column $j$ of $f_k(\cdot)$ can be represented by the matrix $\mQ_{j,k}$ such that
\begin{align}\label{eq:lin_restict}
    \mQ_{j,k}f_k(\mA^\prime_{0,k}, \mA^\prime_k) \ve_j = \vzeros,
\end{align}
for $1\leq j\leq n_k$ where $\ve_j$ is the $j$-th column of the $n_k\times n_k$ identity matrix. Hence, the matrices $\mQ_{j,k}$ linearly transform the $j$-th column of $f_k(\mA^\prime_{0,k}, \mA^\prime_k)$ into zero. To illustrate, consider the mappings $f_k(\mA^\prime_{0,k}, \mA^\prime_{k}) = \mA^\prime_{0,k}$ and $\mQ_{j,k} = \diag(1,0,\ldots,0)$, then \eqref{eq:lin_restict} implies that the first entry in the $j$-th column of $\mA^\prime_{0,k}$ is equal to zero. Let $q_{j,k}$ denote the number of restrictions on the $j$-th column of $f_k(\cdot)$. Thus, the rank of $\mQ_{j,k}$ is $q_{j,k}$. Without loss of generality, we assume that
\begin{align*}
    q_{1,k} \geq q_{2,k} \geq \cdots \geq q_{n_k,k}, \quad \text{for all }k=1,\ldots,N.
\end{align*}
The total number of restrictions is equal to $\sum^N_{k=1}\sum^{n_k}_{i=1} q_{i,k}$,. Following \cite{rothenberg1971identification}, global identification requires $\sum^{n_k}_{i=1} q_{i,k} \geq n_k(n_k-1)/2$ for each mode $k$ separately.

Without restrictions on the sign of each equation in \eqref{eq:lin_restict}, the system \eqref{eq:lin_restict} will not be globally identified. Therefore, we must choose an additional normalization rule $\mathbb{Z}^D$ that determines the sign of each equation. We define the sign normalization rule in Definition \ref{def:signnorm} below, which follows Definition 3 in \cite{rubio2010structural} for each mode $k=1,\ldots,N$. In total, our framework requires the parameters to satisfy normalizations $\mathbb{Z} \equiv \mathbb{Z}^{A}\times \mathbb{Z}^{\Sigma} \times \mathbb{Z}^D$.
\begin{definition}\label{def:signnorm}
    A sign normalization rule can be characterized by a set $\mathbb{Z}^D \equiv \mathbb{Z}^{D_1} \times \cdots \times \mathbb{Z}^{D_N}$ such that, for each mode $k=1,\ldots,N$, for any equation-form structural parameter point $(\mA^\prime_{0,k}, \mA^\prime_{k})$, there exists a unique $n_k\times n_k$ diagonal matrix $\mD_k$ with plus or minus ones along the the diagonal such that $(\mA^\prime_{0,k}\mD_k, \mA^\prime_k \mD_k) \in \mathbb{Z}^{D_k}$.
\end{definition}

\subsection{A comparison to alternative models} \label{sec:compare}
Though our framework is, to our knowledge, the first to provide a structural modelling framework for multidimensional data with an arbitrary number of dimensions $N$, several alternative approaches to structural analysis exist for the special case of panel VAR data where $N=2$. We briefly highlight the most prominent ones below.

\subsubsection{Large SVAR}
As seen in equation \eqref{eq:var}, our representation of the data corresponds to a large SVAR where the coefficients follow a Kronecker product structure. Imposing this structure has several implications. First, identification in a large SVAR system becomes quickly infeasible as dimensions grow. The Kronecker structure reduces the dimensionality of the system by using natural separations (countries, variables, sectors, ...), allowing for identification. In particular, identifying restrictions can be layered on each data dimension separately, while restrictions in a large SVAR must be applied to the complete system. Second, our representation trades off flexibility by structuring the coefficient matrix. Although the separation is natural, the bilinear propagation may not be how variables influence each other. For this reason, \cite{chen2021autoregressive} provide a test to verify the Kronecker structure in MAR models. Third, if one accepts the network interpretation of MAR models, the structural parameters $\mA_{0,k}$ reflect structural networks across modes. In SVAR such networks across dimensions are hard to separate from a single coefficient matrix. 

\subsubsection{Panel VAR}
In case $N=2$, the reduced-form model \eqref{eq:rf-sar} relates closely to a panel VAR model. The panel VAR model treats each unit at a time and captures interdependencies by letting all units enter the model for each unit and by allowing cross-sectional dependence in the reduced-form shocks. That is, each row or column in the matrix data is analyzed separately. 

Let us consider the classic panel VAR with one-lag and no intercept \citep{canova2013panel},
\begin{align}\label{eq:pvar}
    \vy_{i,t} = \mPhi_{i}\vec(\mcalY_{t-1}) + \ve_{i,t},
\end{align}
where $\vy_{i,t}$ denotes row $i$ of $\mcalY_t$ with dimension $n_2$, $\ve_{i,t}$ row $i$ of $\mcalE_t$, and $\mPhi_i$ a $n_2\times n_1n_2$ matrix of slope coefficients.\footnote{We use the row $i$ for ease of exhibition. The arguments also hold for column $i$ of $\mcalY_t$.} For now we assume that the error terms $\ve_{i,t}$ follow a standard multivariate normal distribution, and are cross-sectionally independent. The number of reduced-form parameters of this simplified panel VAR model is $n_1^2n_2^2$, which is substantially more than the $n_1^2 + n_2^2$ parameters for the MAR$(1)$. 

To reduce complexity in the panel VAR, a common approach is to factorize the coefficients. The coefficients are stacked in a vector and factorized as a linear combination of a lower-dimensional vector. We illustrate factorization for the example above. Let $\vtheta_1$ and $\vtheta_2$ denote $n_1\times 1$ and $n_2\times 1$ vectors of coefficients, respectively. Then, factorizing $\vphi = [\vec(\mPhi_1)' \cdots \vec(\mPhi_{n_1})']'$ yields
\begin{align*}
    \vphi = \mXi_1 \vtheta_1 + \mXi_2 \vtheta_2 + \vu_t,
\end{align*}
where $\mXi_1$ and $\mXi_2$ are matrices of dimension $n_1^2n_2^2 \times n_1$ and $n_1^2 n_2^2\times n_1$, respectively. Intuitively, factor $\vtheta_1$ captures row-specific effects and $\vtheta_2$ column-specific effects, with idiosyncratic error term $\vu_t$. The advantage of factorization is a reduction of the number of parameters, here only $n_1 + n_2$. It allows one to obtain more accurate estimates of the parameters and lower the standard errors, and thus, factorization is an appealing approach for forecasting and impulse response function estimation. However, if one is interested in the structural parameters, a factorization of the parameters is too parsimonious for consistent recovery of the parameters. 

Factorization or not, the panel VAR neglects the matrix structure in the data by treating each row (column) equation as conditionally independent given regressors. Hence, the panel VAR cannot distinguish propagation across rows from propagation across columns. Each unit is related to the others and thus, we would like to model them jointly. The reduced-form MAR (Appendix \ref{app:mar}) does so by imposing a strong relationship between rows and between columns. In particular, the panel VAR model \eqref{eq:pvar} corresponds to a restricted MAR model with $\mB_{1}=\mI_{n_1}$ and $\mB_2 = \mPhi_i$. Hence, the heterogeneity of $\mPhi_i$ in a panel VAR allows for row-wise interaction, at a price of a factor $n_1$ more coefficients, while the MAR establishes the row-wise interaction with only $n_1^2$ additional parameters irrespective of the column size $n_2$. 

Furthermore, identification of contemporaneous effects is not straightforward in a panel VAR model. Identification in panel VAR requires restrictions on an $n_1n_2\times n_1n_2$ matrix of contemporaneous coefficients. As the dimensions grow, it becomes difficult to justify the identifying restrictions economically, as well as inference becomes increasingly challenging. Identification in \eqref{eq:sar}, however, is straightforward. The Kronecker structure decomposes the contemporaneous effects into two lower-dimensional objects. As a result, identifying restrictions can be imposed on $\mA_{0,k}$'s separately, such that fewer restrictions per dimensions are necessary and can be easily interpreted.

In short, the Kronecker structure in the MAR allows for a rich model that can be interpreted in a structural manner without any factorization necessary, while the rich panel VAR is overparameterized and needs a less-interpretable factorization for practical use.

\subsubsection{Global VAR}
Methodologically, the global VAR (GVAR) is closest to model \eqref{eq:sar} with two dimensions. \cite{pesaran2004modeling} proposed the GVAR as a global model for the world economy. Each country $i$ follows a country-specific VAR model with foreign averages augmented as regressors. Each country's VAR model can be estimated separately, while cross-sectional dependence enters through the weighted averages of other units' variables. In this sense, both the GVAR and our framework aim to model large systems with many cross-sectional units while avoiding the curse of dimensionality.

In particular, the GVAR regressions can be written as
\begin{align} \label{eq:gvar}
    \vy_{i,t} = \sum^p_{l=1} \mPhi_{il} \vy_{i,t-l} + \sum^{\tilde{p}}_{\tilde{l}=0}\mPhi^*_{i,l} \vy^*_{i,t-\tilde{l}} + \ve_{i,t},
\end{align}
where $\vy^*_{i,t} = \mW_i\vec(\mcalY_t)$ are the foreign averages weighted by $n_1\times n_1n_2$ country-specific weight matrix $\mW_i$ with elements $w_{ij}$. Typically, $\mW_i$ is determined exogenously from alternative data such as trade shares. For the comparison we abstract from contemporaneous foreign effects by setting $\mPhi^*_{i,0}=\vzeros$ and set $p^*=p$. Then we can write the GVAR in stacked form as \begin{align*}
    \vec(\mcalY_t) = \sum^p_{l=1}\mPhi_{l}\vec(\mcalY_{t-l}) + \mXi^{-1}_{0}\vec(\mcalE_t),
\end{align*}
where 
\begin{align*}
    &\mPhi_{l} = \mXi^{-1}_0\mXi_{l}, \\
    &\mXi_{l} = \begin{pmatrix}
        \mOmega_{1,l}\mW_1\\
        \mOmega_{2,l}\mW_2\\
        \vdots\\
        \mOmega_{n_2,l}\mW_{n_2}
    \end{pmatrix},\\
    &\mOmega_{i0} = \left(\mI_{n_1}, -\mPhi^*_{i0}\right), \quad \mOmega_{il} = \left(\mPhi_{il},\mPhi^*_{il}\right), \quad \text{for }l=1,\ldots,p; \quad i=1,\ldots,n_2.
\end{align*}
We can see that the GVAR allows the coefficient matrices to be individual specific, varying freely. In that way coefficient matrix $\mPhi_l$ has little to no structure compared to the Kronecker product in \eqref{eq:sar}. Our model forces each $n_1\times n_1$ block in $\mPhi_l = \mB_{l,2}\otimes \mB_{l,1}$ to share the same shape $\mB_{l,1}$, with cross-unit heterogeneity entering only through scalar weights given by the entries of $\mB_{l,2}$. If, for each lag $l$,  $\mPhi_{il} = \alpha_{il}\mB_{l,1}$ and $\mPhi^*_{il} = \beta_{il}\mB_{l,1}$ for some scalars $\alpha_{il}$ and $\beta_{il}$, then the GVAR reduces to a reduced-form model with row-coefficients
\begin{align}\label{eq:gvar_relation}
    [\mB_{l,2}]_{ij}=\indic{i=j}\alpha_{il} + w_{ij}\beta_{il}
\end{align}
To see this, we can plug in $\mPhi_{i,l} = \alpha_{il}\mB_{l,1}$ and $\mPhi^*_{i,l} = \beta_{il}\mB_{l,1}$ into \eqref{eq:gvar} to obtain 
\begin{align*}
\vy_{i,t}&=\sum^p_{l=1}\alpha_{il}\mB_{l,1}\vy_{i,t-l}+\sum^{n_2}_{j=1}w_{ij}\beta_{il}\mB_{l,1}\vy_{j,t-l}+\vu_{i,t}\\
&\sum^p_{l=1}\sum^{n_2}_{j=1}\left(\indic{i=j}\alpha_{il} + w_{ij}\beta_{il}\right)\mB_{l,1}\vy_{j,t-l} + \vu_{i,t}.
\end{align*}
Vectorizing this GVAR with coefficients \eqref{eq:gvar_relation} system yields
\begin{align*}
    \vec(\mcalY_t) &= \sum^p_{l=1}\begin{pmatrix}
        [\mB_{l,2}]_{11}\mB_{l,1} & [\mB_{l,2}]_{12}\mB_{l,1}  & \cdots & [\mB_{l,2}]_{1n_2}\mB_{l,1} \\
        [\mB_{l,2}]_{21}\mB_{l,1} & [\mB_{l,2}]_{22}\mB_{l,1}  & \cdots & [\mB_{l,2}]_{2n_2}\mB_{l,1} \\
        \vdots & \vdots & \ddots & \vdots\\
        [\mB_{l,2}]_{n_1 1}\mB_{l,1} & [\mB_{l,2}]_{n_1 2}\mB_{l,1}  & \cdots & [\mB_{l,2}]_{n_1 n_2}\mB_{l,1} \\
    \end{pmatrix}
    \vec(\mcalY_{t-l}) + \vec(\mcalU_t)\\
    &= \sum^p_{l=1}(\mB_{l,2} \otimes \mB_{l,1})\vec(\mcalY_{t-l}) + \vec(\mcalU_t)
\end{align*}
In words, the exogenous weights $w_{ij}$ appear as off-diagonal elements of the column-matrix $\mB_{l,2}$, scaled by foreign-sensitivity parameters $\beta_{il}$, while the domestic dynamics $\alpha_{il}$ determine the diagonal elements of $\mB_{l,2}$. All the within-variable influences are captured by $\mB_{l,1}$. 
Hence, the reduced-form \eqref{eq:rf-sar} is a particular (restricted) form of the GVAR model. 

There are several important difference between GVAR and our framework. In GVAR, cross-sectional dependence enters through exogenous weights, while in our case, cross-sectional dependence enter endogenously through the model parameters. This implies that identifying restrictions in model \eqref{eq:sar} can be directly imposed on the transmission network through the mode-specific objects $\mA_{l,k}$ and this can be done separately. Importantly, the ``network" of countries is estimable and testable instead of fixed. 

The structural identification in GVAR is done a large system of parameters, where identifying restrictions implicitly depend on the choice of weights. Interpretation of structural shocks may become less straightforward than in our model. For example, structural IRFs are decomposed in row- and column-channels in the model \eqref{eq:sar}, whereas IRFs in the global VAR are partially governed by the weights.

\subsubsection{Block exogeneity}
A common approach for identification in multi-unit SVAR systems is to consider block exogeneity, where a block of variables does not enter the equations for another block of variables. Suppose we partition the stacked vector $\vec(\mcalY_t)$ into two blocks with sizes $r$
and $n_1 n_2-r$, respectively: $\vec(\mcalY_t) = [\vy_{1,t}^\prime, \vy_{2,t}^\prime]^\prime$. If the first block $\vy_{1,t}$ has no contemporaneous effect on the second block $\vy_{2,t}$, then model \eqref{eq:sar} without lags can be written as
\begin{align*}
    \begin{bmatrix}
        \widetilde{\mPhi}_{0,11}^\prime & \widetilde{\mPhi}^\prime_{0,12}\\
        \vzeros & \widetilde{\mPhi}^\prime_{0,22}
    \end{bmatrix}^\prime\begin{bmatrix}
        \vy_{1,t}\\ \vy_{2,t}
    \end{bmatrix} = \begin{bmatrix}
        \ve_{1,t}\\ \ve_{2,t}
    \end{bmatrix}.
\end{align*}
Such a block restriction can be achieved in model \eqref{eq:sar} with $N=2$ by a zero restriction on the row-matrix $\mA_{0,1}$. To see this, we partition $\mA_{0,1}$ into four blocks, using a $r_1/(n_1 -r_1)$ split,
\begin{align*}
    \mA_{0,1} = \begin{bmatrix}
        \mA_{0,1,11} & \mA_{0,1,12}\\
        \mA_{0,1,21} & \mA_{0,1,22}
    \end{bmatrix},
\end{align*}
where $\mA_{11}$ is $r_1\times r_1$. The Kronecker product $\widetilde{\mPhi}_0^\prime = \mA_{0,2} \otimes \mA_{0,1}$ is then equal to
\begin{align*}
\mA_{0,2} \otimes \mA_{0,1}^\prime = 
    \begin{bmatrix}
        \mA_{0,2} \otimes \mA_{0,1,11} & \mA_{0,2} \otimes \mA_{0,1,12}\\
        \mA_{0,2} \otimes \mA_{0,1,21} & \mA_{0,2} \otimes \mA_{0,1,22}.
    \end{bmatrix}
\end{align*}
Therefore, if $\mA_{0,1,21}=\vzeros$, we obtain the original block-exogeneity restriction in stacked SVARs. In other words, if a group of variables is excluded from another group, then this is the case for every country.

\subsection{Examples} \label{sec:examples}
We provide an example to illustrate identification restrictions in our framework.
\begin{example}
    Commonly used short-run restrictions in the SVAR literature can be directly applied in the present context. For example, consider the lower triangular structure or \emph{causal ordering}, as described by \citep{christiano1996effects}. If a causal ordering has been established between $n_1$ variables, we can directly impose the same triangular structure on $\mA_{0,1}$. Such adaption of known identification schemes is challenging for panel VAR models and GVAR models.

    At the same time, the bilinear structure of \eqref{eq:sar} allows us to implement and establish causal ordering between entities. Consider an example where we observe three monetary variables for the U.S., the Euro area, and a country representing the Rest of the word. We impose the causal ordering Output, Inflation, and Interest Rate, reflected in $\mA_{0,1}$. Suppose we believe that the entities can also be ordered, where monetary shocks originating in the United States can contemporaneously affect the Euro area and the Rest of the world, while Euro area shocks can affect only the Rest of the world contemporaneously, and the Rest of the world does not exert contemporaneous effects on the other two regions. This ordering is easily imposed trough $\mA_{0,2}$. Therefore, the presented framework allows us to capture directional spillovers of monetary policy across economic regions through $\mA_{0,2}$, while preserving the structural interpretation of $\mA_{0,1}$ at the variable level.

    For such contemporaneous restrictions, the transformation functions are
    \begin{align*}
        f_k(\mA^\prime_{0,k}, \mA^\prime_k) = \mA^\prime_{0,k},\quad \text{for }k=1,2. 
    \end{align*}
    The two lower triangular structures on $\mA_{0,1}$ and $\mA_{0,2}$ respectively, imply an block upper triangular matrix $\mA^\prime_{0,2} \otimes \mA^\prime_{0,1}$ where each block is upper triangular. For the case $n_1=n_2=3$, as above, the Kronecker product looks as follows,
    \begin{align*}
        &\mA_{0,2}^\prime \otimes \mA^\prime_{0,1} =
\begin{pmatrix}
a_{0,2,11}\mA^\prime_{0,1} & a_{0,2, 21}\mA^\prime_{0,1} & a_{0,2,31}\mA^\prime_{0,1} \\[6pt]
0 & a_{0,2,22}\mA^\prime_{0,1} & a_{0,2,32}\mA^\prime_{0,1} \\[6pt]
0 & 0 & a_{0,2,33}\mA^\prime_{0,1}
\end{pmatrix} 
    \end{align*}
Since we separate the identifying restrictions per dimension, we can specify the linear transform matrices $\mQ_1$ and $\mQ_2$ separately. In this example, we have $N=2$ and $n_1=n_2=3$ and we impose causal ordering on both dimensions. Hence, $\mQ_{j,1} = \mQ_{j,2}$ for $j=1,2,3$. Let $\mJ_{i_1,\ldots, i_k} = [\ve_{i_1} \ve_{i_2}\cdots \ve_{i_k}]$ denote a matrix of indicator columns. Then, the zero restrictions in the above example can be expressed through matrices $\mQ_{j,1}$ for $j=1,2,3$ as
\begin{align*}
    \mQ_{1,k} = \mJ_{2,3}, \quad \mQ_{2,k} = \mJ_{3}, \quad \mQ_{3,k} = \vzeros_{3\times 3},
\end{align*}
for modes $k=1,2$ The total number of restrictions, in this case, is $3+3$, which is sufficient for identification according to \cite{rothenberg1971identification}.

\begin{remark}
One may argue to impose the identifying restrictions on the Kronecker product directly instead. However, due to the Kronecker structure, some of these restrictions become linearly dependent, whereas \cite{rothenberg1971identification} argues we need to count the number of linearly independent restrictions. It is not trivial to count the number of linearly independent restrictions in a Kronecker product. We therefore argue to separate the identifying restrictions per dimension.
\end{remark}

\end{example}
\section{Properties of IRF} \label{sec:estim and props}
We first formally define global identification of structural IRFs in Proposition \ref{prop:global_irfs}. Due to the Kronecker structure in the coefficients, sign indeterminacy across modes does not affect the uniqueness of the IRF. Moreover, the choice of the normalization rule does not affect the uniqueness of the IRFs, as long as the structural parameters are globally identified under the normalization rule. Proposition \ref{prop:global_irfs} is similar to Proposition 1 in \cite{HAN2025106057} that defines the global identification of structural IRFs in structural factor models, where normalization is necessary to fix the rotation of the factors (and loadings).
\begin{proposition} \label{prop:global_irfs}
    If $\mC_0$ is globally identified under normalization rule $\SZ$, then the structural IRFs $\mcalF_{i_1,i_2,\ldots,i_N}(s)$ are globally identified in the sense that the value of $\mcalF_{i_1,i_2,\ldots,i_N}(s)$ is unique.
\end{proposition}
We note that Proposition \ref{prop:global_irfs} assumes that the structural parameters are globally identified under a normalization rule. However, as discussed in Section \ref{sec:identification}, we need identifying restrictions to ensure this. The identifying restrictions take the form in \eqref{eq:lin_restict}. Let $U_{k,\SZ}$ denote the domain of $f_k(\cdot)$ under normalization rule $\SZ$, and let
\begin{align*}
    \SD_{\SZ} = \{\vtheta \in U_{k,\SZ}|\mQ_{j,k} f_k\left(\mA^\prime_{0,k}, \mA^\prime_k\right) \ve_j = \vzeros \quad \text{for }j=1,\ldots,n_k \text{ and } k=1,\ldots,N\}
\end{align*}
be the set of structural parameters under normalization scheme $\SZ$ that satisfy the identifying restrictions. Furthermore, for any $z_k \times n_k$ matrix $\mPsi$, we define $\mM_{j,k}(\mPsi)$ by
\begin{align*}
    \mM_{j,k}(\mPsi) = \left[
    \begin{array}{cc}
       \multicolumn{2}{c}{\mQ_{j,k} \mPsi}  \\
        \mI_j & \vzeros_{j\times (n_k-j)}
    \end{array}
    \right]_{(z_k+j)\times n_k}
\end{align*}
for all $1\leq j\leq m$, for each mode $k$. Theorem \ref{thm:global_param} characterizes the global identification of the structural IRFs using identifying restrictions.

\begin{theorem} \label{thm:global_param}
    Consider model \eqref{eq:sar} with parameters $\vtheta \in \SD_{\SZ}$ under normalization $\SZ$, and 
    \begin{align*}
        \mathrm{rank}\Bigg[\mM_{j,k}\left(f_k\left(\mA^\prime_{0,k}, \mA^\prime_k\right)\right)\Bigg] = n_k, \quad \text{ for }j=1,\ldots,n_k;\quad \text{for all }k=1,\ldots,N.       
    \end{align*}
    If $f_k(\cdot)$ is admissible for any mode $k$, then the structural IRFs are structurally identified.
\end{theorem}

\subsection{Estimation of structural IRFs} \label{sec:gmm}
To estimate the structural IRFs under identification schemes satisfying admissibility and bilinearity, we first need to estimate the reduced-form parameters $\vpi$, defined as $\vpi \equiv (\vb^\prime, \vsigma^\prime)^\prime$ with $\vb \equiv (\vec(\mB_1)^\prime, \ldots, \vec(\mB_N)^\prime)^\prime$ and $\vsigma \equiv (\vec(\mSigma_1)^\prime, \ldots, \vec(\mSigma_N)^\prime)^\prime$. Following \cite{chen2021autoregressive} for the MAR ($N=2$), one could consider three methods to estimate $\vpi$ consistently, namely the projection method, iterated least squares, or the maximum likelihood estimator (MLE). We consider the Gaussian QMLE since it is more tractable for $N>2$, and most efficient if the covariance matrix of the reduced-form errors follows the Kronecker structure $\mSigma_{N} \otimes \cdots \otimes \mSigma_1$ \citep{chen2021autoregressive} as we have assumed.

Let $\hat{\vpi}$ denote the QMLE for the reduced-form parameters. Given the reduced-form parameters, under some normalization rule $\mathbb{Z}$, the following identifying conditions must hold:
\begin{align}\label{eq:id_cond_estim}
    \bar{\mQ}_{j,k} f_k(\mA^\prime_{0,k}, \mA^\prime_{k}|\vpi)\ve_j = \vzeros^{\prime}_{c_{j,k}\times 1}, \quad \text{for }j=1,\ldots,n_k,
\end{align}
where $n_k\times c_{j,k}$ matrix $\bar{\mQ}_{j,k}$ denote the nonzero rows in $\mQ_{j,k}$ for each mode $k$. We condition the function $f_k(\cdot)$ on $\vpi$, since the reduced-form parameters are implicitly embedded in the structural parameters and we want to exploit this later. We can rewrite equation \eqref{eq:id_cond_estim} in matrix form as
\begin{align} \label{eq:constraints_stacked}
    \begin{bmatrix}
        \bar{\mQ}_{1,k} & & & \\
        & \bar{\mQ}_{2,k} & &\\
        & & \ddots &\\
        & & & \bar{\mQ}_{n_k,k}
    \end{bmatrix} [\mI_{m}\otimes f_k(\mA^\prime_{0,k}, \mA^\prime_k|\vpi)]  \vec(\mI_{n_k})= \vzeros_{c_k\times 1},
\end{align}
with $c_k=\sum^{n_k}_{i=1}c_{i,k}$. Let $\vg_k$ denote the function on the left-hand side, such that  \eqref{eq:constraints_stacked} can be written as
\begin{align}
    \vg_k(\mA_{0,k}, \mA_k|\vpi) = \vzeros_{c_k\times 1},
\end{align}
where we denote the total number of moments as $c\equiv\sum^N_{k=1}c_k$. We can then estimate the structural parameters using a GMM-type objective function. We first stack moments $\vg_k$, $k=1,\ldots,N$, as follows,
\begin{align}\label{eq:stacked_gmm_moments}
    \vg(\{(\mA_{0,k}, \mA_k)\}^N_{k=1}|\vpi) = \begin{bmatrix}
        \vg_1\left(\mA_{0,1}, \mA_1|\vpi\right)\\
        \vg_2\left(\mA_{0,2}, \mA_2|\vpi\right)\\
        \vdots\\
        \vg_N\left(\mA_{0,N}, \mA_N|\vpi\right)
    \end{bmatrix}.
\end{align}
Replacing $\vpi$ with a consistent estimator $\hat{\vpi}$, we minimize the following objective function for $\{(\mA_{0,k}, \mA_k)\}^N_{k=1}$,
\begin{align} \label{eq:GMM_constr}
    &Q_{T}(\{(\mA_{0,k}, \mA_k)\}^N_{k=1}|\hat{\vpi}) = \vg(\{(\mA_{0,k},\mA_k)\}^N_{k=1}|\hat{\vpi})'\mW_{T}\vg(\{(\mA_{0,k}, \mA_k)\}^N_{k=1}|\hat{\vpi})\\
    &\quad \text{s.t. }\\
    &\qquad \mA_{0,k}^{\prime} \mA_{0,k} = \hat{\mSigma}^{-1}_k, \quad \mA^{-1}_{0,k} \mA_{k} = \hat{\mB}_k, \quad \text{ for all }k=1,\ldots,N,
\end{align}
where $\mW_{T}$ is some $c\times c$ positive definite matrix. By Definition \ref{def:global_ident}, there must exist unique orthogonal matrices $\mP_k$ such that $\mA^{-1}_{0,k} =\mL_k \mP_k$ for each $k$ where $\mL_k\mL_k^\prime = \mSigma_k$. 
Each matrix $\hat{\mL}_k$ can be obtained using a Cholesky decomposition. Hence, the constrained GMM objective function can be reformulated as an unconstrained problem. We can rewrite \eqref{eq:GMM_constr} as
{\small
\begin{align}
    \hspace{-1cm}Q_{T}(\{\mP_k\}^N_{k=1}|\hat{\vpi}) = \vg\left(\{\hat{\mL}_k \mP_k\}^N_{k=1}, \{(\hat{\mL}_k \mP_k)^{-1}\hat{\mB}_k\}^N_{k=1}\right)^\prime\mW_{T}\vg\left(\{\hat{\mL}_k \mP_k\}^N_{k=1}, \{(\hat{\mL}_k \mP_k)^{-1}\hat{\mB}_k\}^N_{k=1}\right),
\end{align}
}%
where $\{\mP_r\}^N_{k=1}$ are the only parameters. As suggested by \cite{HAN2025106057} for structural factor models, we consider the Cayley transform. If a $\mP_k$ does not have an eigenvalue equal to $-1$, then 
\begin{align*}
    \mP_k = (\mI_{n_k} - \mC_k)(\mI_{n_k} + \mC_k)^{-1}
    \end{align*}
and $\mC_k$ is a skew-symmetric matrix ($\mC_k^\prime = -\mC_k$). Therefore, we need to find a solution for the skew-symmetric matrices $\mC_k$, which have $n_k(n_k-1)/2$ free entries. Hence, $\vtheta= (\vtheta_1^\prime, \ldots, \vtheta_N^\prime)^\prime$ with $\vtheta_k = \mathrm{veck}(\mC_k)$. We define our minimum-distance (MD) estimator as
\begin{align}\label{eq:gmm_uncon}
    \hat{\vtheta} = \argmin_{\vtheta \in \mTheta} Q_{T}(\vtheta|\hat{\vpi}), \quad \text{where }Q_{T}(\vtheta|\hat{\vpi}) = \vg(\vtheta| \hat{\vpi})^\prime\mW_{T}\vg(\vtheta| \hat{\vpi}),
\end{align}
where $\mTheta$ is a compact subset of $\mathbb{R}^{\sum^N_{k=1}n_k(n_k-1)/2}$. We can recover the structural parameters and impulse response functions from $\hat{\vtheta}$. First, recover $\mC_k$ for each $k$ from $\vtheta$ and set the structural parameters as
\begin{align}\label{eq:transforms_estims}
    &\widehat{\mA}^{-1}_{0,k} = \hat{\mL}_k (\mI_{n_k} - \hat{\mC}_k)(\mI_{n_k} + \hat{\mC}_k)^{-1},\\
    &\widehat{\mA}_k = \widehat{\mA}_{0,k} \hat{\mB}_k.
\end{align}
Note that if weight matrix $\mW_{T}$ has a block-diagonal structure, then the GMM reduces to $N$ separate GMM objective functions that can be solved in parallel. 

Second, we can recover the impulse response functions using the estimated reduced-form parameters. To illustrate, if $N=2$ and $p=1$, the estimator of the structural impulse response function to a shock in $(i,j)$ is given by
\begin{align}
    \hat{\mcalF}_{ij}(s) = \Bigg[\hat{\mB}_1^s \widehat{\mA}^{-1}_{0,1}\Bigg]_i \Bigg[\hat{\mB}^s_2 \widehat{\mA}^{-1}_{0,2}\Bigg]^\prime_j.
\end{align}

\subsection{Large-sample properties}
We estimate the reduced-form parameters $\vpi$ by maximum likelihood estimation, since it is shown to be more efficient than two similar alternatives for $N=2$ by \cite{chen2021autoregressive}. The Gaussian log-likelihood function $\ell(\vpi)$ can be written as
\begin{align}\label{eq:ll}
    \ell(\vpi) = -\frac{T-1}{2}\sum^N_{k=1}\frac{n}{n_k}\log|\mSigma_k| - \sum^T_{t=2}\vec(\mcalU_t)'\mSigma^{-1}\vec(\mcalU_t). 
\end{align}
We maximize the log-likelihood $\ell(\vpi)$ iteratively using the gradient conditions, and finally yields the maximum likelihood estimator $\hat{\vpi}$. The computational details for finding $\hat{\vpi}$ can be found in the supplement.

We make the following assumptions.
\begin{assumptions}
    \begin{enumerate}\setcounter{enumi}{0}
        \item\label{a1:iid}The sequence $\{\vec(\mcalE)_t\}^T_{t=1}$ is i.i.d with mean zero, $\E(\vec(\mcalE_t)\vec(\mcalE_t)^\prime)=\mI_n$, is absolutely continuous with respect to the Lebesgue measure and satisfies $\E||\vec(\mcalE_t)||^{4+\delta}<\infty$ for some $\delta>0$.
        \item\label{a2:causality} The spectral radii satisfy $\prod^N_{k=1}\rho(\mB_k)<1$.
        \item\label{a3:nonsing} The matrices $\mB_{k}$ and $\mSigma_k$ are nonsingular for each $k=1,\ldots,N$. 
    \end{enumerate}
\end{assumptions}
We first establish the consistency and asympototic distribution of the reduced-form estimator $\hat{\vpi}$. In particular, we state the asymptotic distribution for the reduced-form estimator that includes the coefficient estimators $\{\hat{\mB}_k\}^N_{k=1}$ and the lower-triangular matrix estimators $\{\hat{\mSigma}^{1/2}_k\}^N_{k=1}$ instead of the full $\mSigma_k$, because the structural parameter estimator in \eqref{eq:transforms_estims} uses $\hat{\mL}^{1/2}_k$ instead of $\hat{\mSigma}_k$
\begin{proposition} \label{prop:rf_distr}
    Suppose Assumptions \ref{a1:iid}-\ref{a3:nonsing} hold. Then, if $T\rightarrow \infty$,
    \begin{align}
        \sqrt{T}(\hat{\vpi} - \vpi^0) \ind N(\vzeros, \mXi),
    \end{align}
    where $\mXi$ is defined in Appendix \ref{app:proof}, see Lemma \ref{lem:coeff_distr} and \ref{lem:sigma_distr}.
\end{proposition}
To obtain the asymptotic distribution of $\hat{\vtheta}$, let
\begin{align*}
    \mG_{\vtheta} \equiv \frac{\partial \vg(\vtheta| \vpi)}{\partial \vtheta^\prime}\Bigg|_{\vtheta^0}, \quad \mG_{\vpi} \equiv \frac{\partial \vg(\vtheta| \vpi)}{\partial \vpi^\prime}\Bigg|_{\vpi^{0}}
\end{align*}
be the partial derivatives of $\vg(\vtheta|\vpi)$ with respect to $\vtheta$ and $\vpi$, respectively. Given the asymptotic distribution of the reduced-form estimator in Proposition \ref{prop:rf_distr} and the partial derivatives $\mG_{\vtheta}$ and $\mG_{\vpi}$, we can obtain the asymptotic distribution of the structural parameters.
\begin{theorem} \label{thm:theta_distr}
    Suppose Assumptions \ref{a1:iid}-\ref{a3:nonsing} and identifying condition \eqref{eq:constraints_stacked} hold. If the function $f_k$ is continuously differentiable and satisfies the rank conditions in Theorem \ref{thm:global_param}, if $\mP_k$ does not have $-1$ as an eigenvalue, for each $k=1,\ldots,N$, and if $\mW_{T} \inp \mW$, then
    \begin{align*}
        \sqrt{T}(\hat{\vtheta} - \vtheta^0) = -(\mG_{\vtheta}^\prime\mW \mG_{\vtheta})^{-1}G_{\vtheta}^\prime\mW \mG_{\vpi}\sqrt{T}(\hat{\vpi} - \vpi^{0}) + o_p(1).
    \end{align*}
    Under optimal weight matrix $\mW_0 = (\mG_{\vpi}\mXi\mG_{\vpi}^\prime)^{-1}$ that minimizes the asymptotic variance of $\sqrt{T}(\hat{\vtheta} - \vtheta^0)$, we have
    \begin{align*}
        \sqrt{T}(\hat{\vtheta}-\vtheta^0) \ind \calN(\vzeros, \mG_{\vtheta}^\prime (\mG_{\vpi}\mXi \mG^\prime_{\vpi})^{-1}\mG_{\vtheta}).
    \end{align*}
\end{theorem}
We can use Theorem \ref{thm:theta_distr} to test the validity of the identifying restrictions \eqref{eq:GMM_constr}. Let
\begin{align*}
    J = T\vg(\hat{\vtheta}|\hat{\vpi})^\prime \hat{\mW}_0 \vg(\hat{\vtheta}| \hat{\vpi}),
\end{align*}
where $\hat{\mW}_0$ is a consistent estimator of $\mW_0$. The $J$-statistic follows an asymptotic chi-square distribution with $c-\sum^N_{k=1}n_k (n_k-1)/2$ degrees of freedom, see Proposition \ref{prop:j_test} below.
\begin{proposition} \label{prop:j_test}
    Under the assumption of Theorem \ref{thm:theta_distr}, $J \ind \chi^2_{c-\sum^N_{k=1}n_k(n_k-1)/2}$.
\end{proposition}
We can use the results of Theorem \ref{thm:theta_distr} and Proposition \ref{prop:j_test} to obtain the asymptotic distribution of the structural impulse response functions. Fix a horizon $h$ and a scalar functional of the structural impulse response tensor $\mcalF_{i_1 i_2 \cdots i_N}(h)$ defined as $\psi_h(\vpi, \vtheta) \equiv \vq^\prime \vec(\mcalF_{i_1 i_2 \cdots i_N}(h))$ for some user-specified $\vq$. For instance, taking $\vq$ as a unit vector selects the response of a single variable. We define the following quantities of interest: $\psi^0_h \equiv \psi_h(\vpi^0,\vtheta^0)$ and $\hat{\psi}_h \equiv \psi_h(\hat{\vpi},\hat{\vtheta})$. Furthermore, we define the following partial derivatives, evaluated at the true parameters: $\vg_{\vpi,h} \equiv \partial \psi_h(\vpi, \vtheta)/\partial \vpi$ and $\vg_{\vtheta,h} \equiv \partial \psi_h(\vpi, \vtheta)/\partial \vtheta$, both evaluated at $(\vpi^0, \vtheta^0)$.

\begin{theorem} \label{thm:irf_distr}
     Suppose the assumptions of Theorem \ref{thm:theta_distr} hold, and suppose $\psi_h(\vpi, \vtheta)$ is continuously differentiable in a neighborhood of $(\vpi^0, \vtheta^0)$. Then, for a fixed horizon $h$, if $T\rightarrow\infty$, then
    \begin{align*}
        \sqrt{T}(\hat{\psi}_h - \psi^0_h) \ind \calN(0,\vm_h^\prime \mXi \vm_h),
    \end{align*}
    with $\vm_h = \vg_{\vpi,h} + \mA_{\vtheta}^\prime\vg_{\vtheta,h}$, $\mA_{\vtheta}\equiv -(\mG_{\vtheta}^\prime\mW \mG_{\vtheta})^{-1}\mG_{\vtheta}^\prime\mW \mG_{\vpi}$, and $\mXi$ defined in Proposition \ref{prop:rf_distr}.\footnote{An explicit expression for $\vg_{\vpi,h}$ and $\vg_{\vtheta,h}$ can be obtained by differentiating the Cayley transform and the mode-specific IRF formula, as is done in Theorem 5 of \cite{HAN2025106057} for structural factor models.}
\end{theorem}

\section{Robust inference} \label{sec:robust}
A key assumption of the model is the Kronecker structure of the coefficients. In practice, however, this assumption may only approximately hold. Consider the reduced-form VAR$(1)$, 
\begin{align}\label{eq:sar_vec}
    \vec(\mcalY_t) = \mPhi_T \vec(\mcalY_{t-1}) + \mC_0\vec(\mcalE_t),
\end{align}
where $\mPhi_T$ denotes an $n\times n$ VAR coefficient matrix, and $\mC_0$ an $n\times n$ structural matrix.
Model \eqref{eq:sar} imposes that the propagation dynamics $\mPhi_T$ and contemporaneous structural matrix $\mC_0$ both have a Kronecker structure, and that $\mC_0$ is identified. In practice, Kronecker separability may not need to hold exactly and may be regarded as a parsimonious approximation to a richer transmission mechanism. In this subsection we provide inference methods robust to approximate Kronecker separability for the structural impulse response functions.

We treat the Kronecker assumption in $\mPhi_T$ and $\mC_0$ asymmetrically. If $\mC_0$ is misspecified, the mode-specific identifying restrictions may be misspecified and thus may not pin down a rotation of the true structural shocks. Relaxing Kronecker separability in $\mC_0$ therefore raises a partial identification problem rather than a bias problem, so we leave this issue for further research (see also Remark \ref{rem:sigma-extension}). We rather provide valid inference when the propagation dynamics $\mPhi_T$ depart from exact Kronecker separability, holding the structure in $\mC_0$ fixed.

In particular, we allow the Kronecker structure in $\mPhi_T$ to be locally misspecified. Let $\mathcal{K}\equiv \big\{K(\vb) = \mB_N \otimes \cdots \otimes \mB_1 : 
\vb=(\vec(\mB_1)^\prime,\ldots, \vec(\mB_N)^\prime)^\prime \in \mathcal{B}\big\}$ denote the Kronecker manifold where $\mathcal{B}$ imposes the scale normalizations $||\mB_k||_F=1$ for $k<N$. We consider local misspecification to be departures from $\mathcal{K}$, with a \cite{pitman1949notes} drift,
\begin{align}\label{eq:local-misspec}
    \mPhi_T = \mK(\vb^0) + T^{-1/2}\mD
\end{align}
where $\mD$ is a possibly nonzero matrix. We define vectorizations $\vphi_T \equiv \vec(\mPhi_T), \vk(\vb) \equiv \vec(\mK(\vb))$, and $\vd \equiv \vec(\mD)$. The misspecification is local in the sense that the degree shrinks with the sample size $T$, in particular with the sampling error $\sqrt{T}$ rate. Intuitively, local misspecification here suggests that if $T^{-1/2}\mD$ is small relative to sampling uncertainty, then the Kronecker structure yields an accurate approximation in finite sample. Similar local misspecification has been used for moment specification tests \citep{newey1985generalized, newey1985maximum}, VAR forecasting \citep{schorfheide2005var}, dynamic discrete choice problems \citep{bugni2019inference}, sensitivity analysis \citep{andrews2017measuring, armstrong2021sensitivity}, among other purposes. 

The restricted reduced-form estimator estimates $\vpi$ by maximising the Gaussian log-likelihood \eqref{eq:ll} subject to the Kronecker structure. It is convenient, however, to write the restricted estimator as a minimum-distance projection of an unrestricted estimator onto $\mathcal{K}$. In this way, we can use the asymptotic distribution of the unrestricted estimator to determine the asymptotic distribution of the restricted estimator under local misspecification. Let $\hat{\mPhi}_U \equiv (\sum^T_{t=2}\vy_t \vy_{t-1}^\prime)(\sum^T_{t=2}\vy_{t-1} \vy_{t-1}^\prime)^{-1}$ and $\hat{\mGamma}_0 \equiv (T-1)^{-1}\sum^T_{t=2}\vy_{t-1}\vy_{t-1}^\prime$ denote the unrestricted OLS estimator and the sample second moment, respectively. Moreover, we define $\hat{\vphi}_U = \vec(\hat{\mPhi}_U)$. The following lemma says that the restricted QMLE $\hat{\vb}$ is exactly equivalent to the minimum-distance projection of $\hat{\vphi}_U$ onto $\mathcal{K}.$ 

\begin{lemma}\label{lem:qmle-md}
Suppose $\mC_0$ is fixed. Then, maximising the Gaussian log-likelihood \eqref{eq:ll} over $\vphi \in \mathcal{K}$ is \emph{exactly} equivalent to solving the minimum-distance problem,
\begin{equation}
  \hat{\vb} \;=\; \argmin_{\vb \in \mathcal{B}}
  \bigl( \hat{\vphi}_U - \vk(\vb) \bigr)' \hat{\mW}
  \bigl( \hat{\vphi}_U - \vk(\vb) \bigr),
  \qquad \hat{\mW} = \hat{\mGamma}_0 \otimes \mSigma^{-1} .
  \label{eq:md-problem}
\end{equation}
Under Assumptions \ref{a1:iid}-\ref{a3:nonsing}, $\hat{\mW} \to_p \mW \equiv \mGamma_0 \otimes \mSigma^{-1}
= \mV_\phi^{-1}$, where $\mV_\phi = \mGamma_0^{-1} \otimes \mSigma$ is the asymptotic variance of $\sqrt{T}(\hat{\vphi}_U - \vphi_T)$.\footnote{Since the Gaussian information is block diagonal between $\vphi$ and $\vsigma$, replacing $\mSigma$ by a root-$T$ consistent estimate has no first-order effect on the results in Lemma \ref{lem:qmle-md}.}
\end{lemma}
Lemma \ref{lem:qmle-md} implies the following. First, since $\hat{\vb}$ is a minimum-distance estimator, we can make use of existing results on minimum-distance estimators under local misspecification. In particular, \cite{andrews2017measuring, andrews2020informativeness} and \cite{armstrong2021sensitivity} have been influential on sensitivity analysis for this estimator. We follow \cite{armstrong2021sensitivity} most closely to provide robust confidence intervals for the estimated structural impulse response functions based on $\hat{\vb}$.

Second, the weight matrix $\mW$ in the minimum-distance problem is determined by the asymptotic variance of the reduced-form estimator. Given the QMLE $\hat{\vphi}_U$, $\mW = \mV_{\phi}^{-1}$ minimizes the asymptotic variance of $\hat{\vb}$ within this class of minium-distance estimators. Third, the distribution of the restricted estimator can be decomposed into sampling noise and misspecification bias. To see this, set $r\equiv \sum^N_{k=1}n_k^2 - (N-1)$ and let $\mJ = \left(\partial \vk(\vb)/\partial \vb^\prime\Big|_{\vb=\vb^0}\right)\mR_0$ for some matrix $\mR_0 \in \SR^{\sum_k n_k^2 \times r}$ whose columns span the tangent space of normalization $||\mB_k||_F=1, k<N$, at $\vb^0$. The matrix $\mJ$ has full column rank $r$ at $\vb^0$ by Lemma \ref{lem:kron_to_indiv_consist}.\footnote{The restriction to $\mR_0$ is necessary, because without accounting for arbitrary rescaling the derivative $\partial \vk(\vb)/\partial \vb^\prime\Big|_{\vb=\vb^0}$ is rank deficient by $N-1$. The matrix $\mR_0$ maps local free-coordinate perturbations into the tangent space of the normalization manifold. $\mR_0$ removes the $N-1$ scale-indeterminacy directions and yields a full-rank Jacobian $\mJ$. Then, $(\mJ^\prime\mW\mJ)^{-1}$ is well defined.} Define $\mP_W = \mJ(\mJ^\prime\mW\mJ)^{-1}\mJ^\prime\mW$ as the $\mW$-orthogonal projection onto the tangent space of the Kronecker manifold $\mcalK$. Using that $\sqrt{T}(\hat{\vphi}_U - \vphi_T) \ind \mZ$ where $\mZ \sim \calN(\vzeros, \mV_{\phi})$ by standard OLS results, we obtain under local misspecification \eqref{eq:local-misspec} that $\sqrt{T}(\hat{\vphi}_U - \vk(\vb_0)) \ind \mZ + \vd$. Let $\hat{\vphi}_K \equiv \vk(\hat{\vb})$ denote the restricted estimator. By definition of $\mP_W$, we have that
\begin{align}\label{eq:phi_decomp}
    &\sqrt{T}(\hat{\vphi}_K - \vphi_T) = \mP_W \mZ - (\mI - \mP_W)\vd + o_p(1).
\end{align}
Equation \eqref{eq:phi_decomp} says that the asymptotic distribution of the restricted estimator under local misspecification can be decomposed into two parts. First, the term $\mP_W\mZ$ captures the sampling uncertainty which would be present regardless of misspecification. Second, the term $-(\mI-\mP_W)\vd$ captures the specification bias. If the direction of misspecification $\vd$ does not lie in the column space of $\mJ$, this term is nonzero and we have a first-order specification bias. 

The first-order bias also enters the asymptotic distribution of the structural IRF estimator. Since $\psi_h$ is a smooth function of $(\vphi, \vtheta)$, we can define partial derivatives $g_{\phi, h}\equiv \partial \psi_h /\partial \phi\Big|_{\vphi^0}$ and $g_{\theta, h} = \partial \psi_h /\partial \theta\Big|_{\vtheta^0}$. Note that $g_{\phi, 0}=0$, since the structural impulse response function at horizon zero depends on $\mA_{0,k}$ only. To rule out a partial identification problem, we posit the following assumption.
\begin{assumptions}
    \begin{enumerate}\setcounter{enumi}{3}
        \item\label{a4:robust_identification} For each $k$, the identifying map $f_k$ depends on the contemporaneous coefficients only:
        \begin{align*}
            f_k(\mA^\prime_{0,k}, \mA^\prime_k) = f_k(\mA^\prime_{0,k}).
        \end{align*}
    \end{enumerate}
\end{assumptions}
Assumption \ref{a4:robust_identification} allows for recursive orderings, block exogeneity, and the schemes discussed in the examples in Section \ref{sec:examples}, but excludes long-run restrictions \'a la \cite{blanchard1989} and restrictions on IRFs at $h\geq 1$. Under Assumption \ref{a4:robust_identification}, the moment conditions \eqref{eq:stacked_gmm_moments} depend on $\vpi$ only through $\mSigma$. The theorem below specifies the asymptotic distribution of the structural IRFs estimator $\hat{\psi}_h$ under local misspecification. We restrict ourselves to $\vd\in \mathcal{D}(c) \equiv \{\vd: \vd^\prime \mH \vd \leq c^2\}$ for a user-specified positive definite matrix $\mH$ and scalar $0\leq c < \infty$.\footnote{The matrix $\mH$ in set $\mathcal{D}(c)$ encodes the kind of violations of the Kronecker structure. For example, $\mH=\mI$ corresponds to departures in Frobenius norm and treats all coefficients alike. A block-structured $\mH$ corresponds to mode-specific penalisation of departures: e.g. the within-country monetary block is Kronecker separable while cross-country propagation is only approximately Kronecker separable.} For $\mH=\mV^{-1}_{\phi}$, $\mathcal{D}(c)$ is part of the \cite{cressie1984multinomial} family of divergences. \cite{armstrong2021sensitivity} show that this form of $\mathcal{D}(c)$ is mathematically convenient for optimality and efficiency. We refer to Remark 3.1 in \cite{armstrong2021sensitivity} for a discussion on other choices of norms in $\mathcal{D}(c)$.
\begin{theorem}\label{thm:irf_distr_misspec}
    Suppose Assumptions \ref{a1:iid}-\ref{a4:robust_identification} hold. Let $\mPhi_T$ be locally misspecified as in \eqref{eq:local-misspec} with $\vd\in\mathcal{D}(c)$. Let $\veta \equiv (\vsigma^\prime, \vtheta^\prime)^\prime$. Let $\psi_{h,T} = \psi_h(\vphi_T, \veta^0)$ be the true IRF under the local DGP, and $\vg_{\eta,h} = \partial \psi_h /\partial \veta|_{\veta^0}$. Moreover, suppose that the rank conditions of Theorem \ref{thm:global_param} hold at $\vb^0$. Then, as $T\rightarrow \infty$, 
    \begin{align}
        \sqrt{T}(\hat{\psi}_h - \psi_{h,T}) \ind \calN(b_h(\vd), \sigma^2_h),
    \end{align}
    with $b_h(\vd) = -g_{\phi,h}^\prime(\mI - \mP_W)\vd$ and $\sigma^2_h = \vg_{\phi,h}^\prime \mJ(\mJ^\prime\mV^{-1}_{\phi}\mJ)^{-1}\mJ^\prime \vg_{\phi,h} + \vg^\prime_{\eta,h}\mV_{\eta}\vg_{\eta,h} + 2\vg_{\phi,h}^\prime\mP_W\mV_{\phi \eta}\vg_{\eta,h}$, where $\mV_{\eta}$ is the asymptotic variance of $\hat{\veta}$.
\end{theorem}
The bias term $b_h(\vd)$ may be nonzero for horizons $h>0$ while it is exactly equal to zero for $h=0$, since we assume that the contemporaneous effects are correctly specified and identified. The variance  decomposes into three parts: a propagation term, an identification term, and their interaction (which is zero under Gaussian errors). More importantly, the variance does not depend on $\vd$. Therefore, we can construct robust confidence intervals by centering only.

\subsection{Robust confidence intervals}
We now construct robust confidence intervals. Since $\vd$ cannot be estimated, we use the worst-case IRF bias to compute critical values. Let $\vl_h \equiv (\mI-\mP_W)^\prime g_{\phi, h}$ such that $b_h(\vd) = -\vl_h^\prime \vd$. For the set of ellipsoids $\mcalD(c)$, the worst-case absolute bias is defined as $B_h(c) \equiv \sup_{\vd\in \mcalD(c)}|b_h(\vd)|$. By the Cauchy-Schwarz inequality, we have that $B_h(c) = c\sqrt{\vl_h^\prime\mH^{-1}\vl_h}$. For $\mH=\mI$, we have that $B_h(c) = c||(\mI-\mP_W)^\prime g_{\phi,h}||_2$. We refer to $B_h(c)$ as the misspecification sensitivity measure, since a large $B_h(c)$ implies that the IRF is highly sensitive to directions that are excluded by the Kronecker structure, while small $B_h(c)$ implies that the IRF is robust to violations of the Kronecker structure in the propagation mechanisms. For $\mH = \mV_{\phi}^{-1}$, we have that 
\begin{align}\label{eq:worstcasebias_eff}
    B_h(c) = c\sqrt{g_{\phi,h}^\prime (\mI - \mP_W)\mV_{\phi}(\mI - \mP_W)^\prime g_{\phi,h} },
\end{align}
which corresponds to the worst-case bias as $c$ times the sampling standard deviation of the part of $\hat{\vphi}_U$ that the Kronecker assumption ignores. Moreover, under $\mH = \mV_{\phi}^{-1}$, \eqref{eq:worstcasebias_eff} follows the form of the test statistic for Kronecker separability in \cite{chen2021autoregressive}, which converges to a noncentral $\chi^2$ distribution with $n^2 - r$ degrees of freedom and noncentrality measure equal to $\vd^\prime (\mI-\mP_W)^\prime\mV^{-1}_{\phi}(\mI-\mP_W)\vd \leq c^2$. Hence, in this testing interpretation, $c$ controls the power of the test, and thus, $c$ must be specified a-priori to avoid post-selection inference issues.

A simple approach to valid robust confidence intervals would be to directly add the worst case bias $B_h(c)$ to the confidence interval,
\begin{align}\label{eq:robustci_add}
    \mathrm{CI}^{\mathrm{add}}_h(c) = \left[\hat{\psi}_h - \frac{z_{1-\alpha/2}\hat{\sigma}_h + \hat{B}_h(c)}{\sqrt{T}}, \hat{\psi}_h + \frac{z_{1-\alpha/2}\hat{\sigma}_h + \hat{B}_h(c)}{\sqrt{T}}\right],
\end{align}
where $\hat{B}_h(c)$ denotes the plug-in estimator of $B_h(c)$. The term $\hat{B}_h(c)/\sqrt{T}$ captures the Kronecker separability uncertainty. The confidence interval $\mathrm{CI}^{\mathrm{add}}_h(c)$ is easy to implement but conservative, since it penalizes each direction of possible misspecification. To tighten the robust confidence interval, we follow \cite{armstrong2021sensitivity} and account for the misspecification through the critical value instead. 

Let $\bar{b}_h(c) = B_h(c)/\sigma_h$ (abbreviated by $\bar{b}$) and define
\begin{align*}
    cv_\alpha(\bar{b}) \equiv \argmin_q \left\{\inf_{|b|\leq \bar{b}}P(|Z+b|\leq q)\geq 1-\alpha\right\},
\end{align*}
for $Z\sim \calN(0,1)$. Critical value $cv_{\alpha}(\bar{b})$ minimizes the length of the confidence interval subject to a bound on the worst-case bias. Since $P(|Z+b|\leq q)$ is strictly decreasing in $|b|$, the infimum is attained at the bound $\bar{b}$, and thus, $cv_{\alpha}(\bar{b})$ is the unique solution to $\Phi(q - \bar{b}) + \Phi(q+\bar{b}) -1 = 1-\alpha$. That is, $cv_{\alpha}(\bar{b})$ is the $1-\alpha$ quantile of a folded Gaussian variable $|Z+\bar{b}|$. The \textit{bias-aware} confidence intervals can then be constructed as
\begin{align} \label{eq:biasaware_ci}
    \mathrm{CI}^{\mathrm{BA}}_h(c) = \left[\hat{\psi}_h - cv_{\alpha}(\hat{\bar{b}}_{h}(c))\frac{\hat{\sigma}_h}{\sqrt{T}}, \hat{\psi}_h + cv_{\alpha}(\hat{\bar{b}}_{h}(c))\frac{\hat{\sigma}_h}{\sqrt{T}}\right],
\end{align}
for a given $c$. Note that for $c=0$, we have that $B_h(0)=0$ and $\bar{b}_h(0)=0$ such that $cv_{\alpha}(0) = z_{1-\alpha/2}$. In other words, under correct specification the bias-aware confidence intervals reduce to the ordinary confidence intervals. Moreover, it holds that $cv_{\alpha}(\bar{b}) \leq z_{1-\alpha/2} + \bar{b}$, such that $\mathrm{CI}^{\mathrm{BA}}_h(c) \subseteq \mathrm{CI}^{\mathrm{add}}_h(c)$. Theorem \ref{thm:robust_baci_cov} below shows the uniform validity for the bias-aware confidence intervals for a class of local misspecification. 
\begin{theorem}\label{thm:robust_baci_cov}
    Suppose Assumptions \ref{a1:iid}-\ref{a4:robust_identification} hold. Moreover, let $\calP_T$ denote the set of data generating processes of the form \eqref{eq:local-misspec} for $\vd\in\mathcal{D}(c)$ and a fixed $\vb^0 \in \mathcal{B}$,
\begin{align*}
    \calP_T(c) = \Big\{P_{T,d}: \vphi_T = \vk(\vb^0) + \frac{\vd}{\sqrt{T}}, \quad \vd^\prime \mH\vd\leq c^2\Big\}.
\end{align*}
    Then, for a fixed $c\geq 0$ and $\alpha\in (0,1)$, it holds that
    \begin{align*}
        \liminf_{T\rightarrow\infty}\inf_{\vd\in \mathcal{D}(c)} P_{T,d}\left(\psi_{h,T}\in \mathrm{CI}^{\mathrm{BA}}_h(c)\right) \geq 1-\alpha.
    \end{align*}
\end{theorem}
Theorem \ref{thm:robust_baci_cov} is uniform over the class of misspecification in \eqref{eq:local-misspec}, and therefore validity also for the least favourable directions of misspecification. This is important in our context as departures from the Kronecker structure are not known a priori. Several remarks on Theorem \ref{thm:robust_baci_cov} apply.
\begin{remark}
    The procedure requires a practitioner to fix the value of $c$. This may feel arbitrary. We therefore recommend to use a grid of values, and use our methods as a sensitivity measure. If inference drastically changes for small values of $c$, it suggests that a Kronecker structure may be inappropriate.
\end{remark}

\begin{remark}\label{rem:sigma-extension}
    For a richer model that has a sufficient number of identifying restrictions to identify $\mSigma$, independently of the Kronecker separability, we could analogously build robust confidence intervals by constructing the tangent-space projection for $\mSigma$ and deriving the joint bias for $\mSigma$ and $\mPhi_T$. Without these restrictions, relaxing Kronecker separability of $\mSigma$ implies set-identification, and confidence intervals must be replaced by confidence sets. We leave both issues as an extension of the present work. 
\end{remark}

\section{Simulations} \label{sec:simulation}
We evaluate the finite-sample performance of the MD estimator 
using a Monte-Carlo simulation study. We fix the number of modes to $N=2$ and generate the data according to model \eqref{eq:sar} with one lag, and
\begin{align*}
    \mA_{0,1}\mcalY_t \mA_{0,2} = \mA_{1,1}\mcalY_{t-1} \mA_{1,2} + \mcalE_t, \quad t=1,\ldots,T,
\end{align*}
where the errors $\vec(\mcalE_t)$ are generated from a multivariate standard normal distribution. We consider the sizes $(n_1,n_2) = (2,3), (5,5), (6,9)$ and $T\in\{100, 250, 500, 1000\}$. For each $(n_1, n_2)$, we impose upper triangularity for both $\mA_{0,1}$ and $\mA_{0,2}$. In particular, the diagonal elements of $\mA_{0,j}, j=1,2$, are generated i.i.d. $1 + 0.1 \calU(0,1)$ while the upper off-diagonal elements are generated i.i.d. $0.15\phi \calU(0,1)$, where $\phi \in \SR$ controls the signal of the cross-equation effects. For large $\phi$, the identification of the contemporaneous effects is strong. We consider $\phi \in \{1, 0.5, 0.05\}$, which we refer to as \textit{baseline}, $\textit{weak}$, and \textit{very weak} identification settings. Moreover, we draw $\mA_{1,j}$ using that $\mA_{1,j} = \mA_{0,j}\mB_{1,j}$. We draw the entries of the reduced-form parameters $\mB_{1,1}$ independently from a standard normal distribution. To normalize the scale, we rescale $\mB_{1,1}$ such that $||\mB_{1,1}||_F=1$. We draw the entries of $\mB_{1,2}$ also independently from a standard normal distribution and rescale $\mB_{1,2}$ such that $\rho(\mB_{1,1})\rho(\mB_{1,2}) = \rho$ for $\rho\in\{0.3, 0.5, 0.8\}$. The value of $\rho$ controls the level of stationarity of the simulated data. We use the same generated matrices across all 1,000 Monte Carlo repetitions. For the MD estimator, we use weighting matrix $\mW = \mI$ for estimation. The main results are reported below. Appendix \ref{app:sims} provides additional simulation results.

\subsection{RMSE results} \label{sec:sim_rmse}
 In each Monte Carlo repetition, we compute the root mean square error (RMSE) of $||\hat{\mA}_{0,2}\otimes\hat{\mA}_{0,1} - \mA_{0,2}\otimes\mA_{0,1}||_2$, the RMSE of the estimation impulse response function with 8 horizons for a shock in $\mcalE_{12,t}$. We ignore the structural zeros and average the results over the variables. Table~\ref{tab:sim_rmse} shows the RMSE results for the MD estimator. 
 
 We make the following observations. First, as the sample size increases, the RMSE and its standard deviation decrease across the dimension sizes $(n_1, n_2)$ and values of $\phi$. These patterns conform with the $\sqrt{T}$-consistency established in Theorem~\ref{thm:theta_distr}-\ref{thm:irf_distr}. The finite-sample improvements are in particular important for larger $(n_1,n_2)$ and small $T$, because our Kronecker structure reduces the dimensionality of the system compared to SVAR which makes estimation under small $T$ reliable, without having to resort to additional data-driven shrinkage methods. For example, under $(n_1,n_2)=(5,5)$ there are 80 free parameters in model \eqref{eq:sar}, instead of 300 free parameters in a SVAR$(1)$ model with triangular identification.  Table \ref{tab:sim_rmse} shows that finite-sample for $T=250$ is already close to the performance for $T=1000$, implying that the model \eqref{eq:sar} can be applied reliably to a large structural system of variables.

\begin{table}[!ht]
\centering
\resizebox{\textwidth}{!}{%
\begin{tabular}{llllclllclll}
\toprule
                                  & \multicolumn{11}{c}{Average RMSE}                                               \\
                                  \midrule
$T$\textbackslash{}($n_1$, $n_2$) & \multicolumn{3}{c}{(2,3)} & & \multicolumn{3}{c}{(5,5)} & & \multicolumn{3}{c}{(6,9)} \\
\cmidrule(l){2-4} \cmidrule(l){6-8} \cmidrule(l){10-12} 
                                  & $\phi=1$          & $\phi=0.5$         & $\phi=0.05$   &       & $\phi=1$          & $\phi=0.5$         & $\phi=0.05$       &  &  $\phi=1$        & $\phi=0.5$         & $\phi=0.05$ \\
                                  \cmidrule(l){2-4} \cmidrule(l){6-8} \cmidrule(l){10-12}
100                               & 21.913  & 20.894  & 21.715 & & 7.769 & 5.921 & 5.539 & & 15.266 & 9.194 & 7.759\\
250                               &  7.951  & 7.843 & 7.563 & & 3.705 & 2.251 & 2.002 & & 13.687 & 5.535 & 4.583\\
500                               &  3.841 & 3.742 & 3.708 & & 2.508 & 1.183 & 0.991 & & 12.304 & 4.937 & 3.916\\
1000                              &  1.899 & 1.881 & 1.841 & & 1.997 & 0.659 & 0.481 & & 11.454 & 4.175 & 3.326\\
\midrule
   & \multicolumn{11}{c}{St. dev. RMSE}                                               \\
                                  \midrule
$T$\textbackslash{}($n_1$, $n_2$) & \multicolumn{3}{c}{(2,3)} & & \multicolumn{3}{c}{(5,5)} & &\multicolumn{3}{c}{(6,9)} \\
\cmidrule(l){2-4} \cmidrule(l){6-8} \cmidrule(l){10-12} 
                                  & $\phi=1$          & $\phi=0.5$         & $\phi=0.05$   &       & $\phi=1$          & $\phi=0.5$         & $\phi=0.05$      &   &  $\phi=1$          & $\phi=0.5$         & $\phi=0.05$ \\
                                  \cmidrule(l){2-4} \cmidrule(l){6-8} \cmidrule(l){10-12}
100                               &  1.366  &  1.260 & 1.330 & & 0.468 & 0.374 & 0.332 & & 0.298 & 0.239 & 0.290\\
250                               &  0.772  & 0.752 & 0.735 & & 0.394 & 0.215 & 0.198 & & 0.462 & 0.183 & 0.207\\
500                               &  0.527 & 0.500 & 0.517 & & 0.403 & 0.168 & 0.138 & & 0.185 & 0.160 & 0.093\\
1000                              &  0.377 & 0.366 & 0.370 & & 0.438 &  0.152 & 0.095 & & 0.374 & 0.180 & 0.103\\
\bottomrule
\end{tabular}
}
\caption{This table reports the average root mean square error (RMSE) and its standard deviation for the structural MD estimator for stationarity level $\rho=0.5$, across different values of the sample size $T$ and dimension sizes $(n_1, n_2)$.  All reported values have been multiplied by $10^{2}$.}
\label{tab:sim_rmse}
\end{table}

\begin{figure}[!ht]
    \centering
    \includegraphics[width=\linewidth]{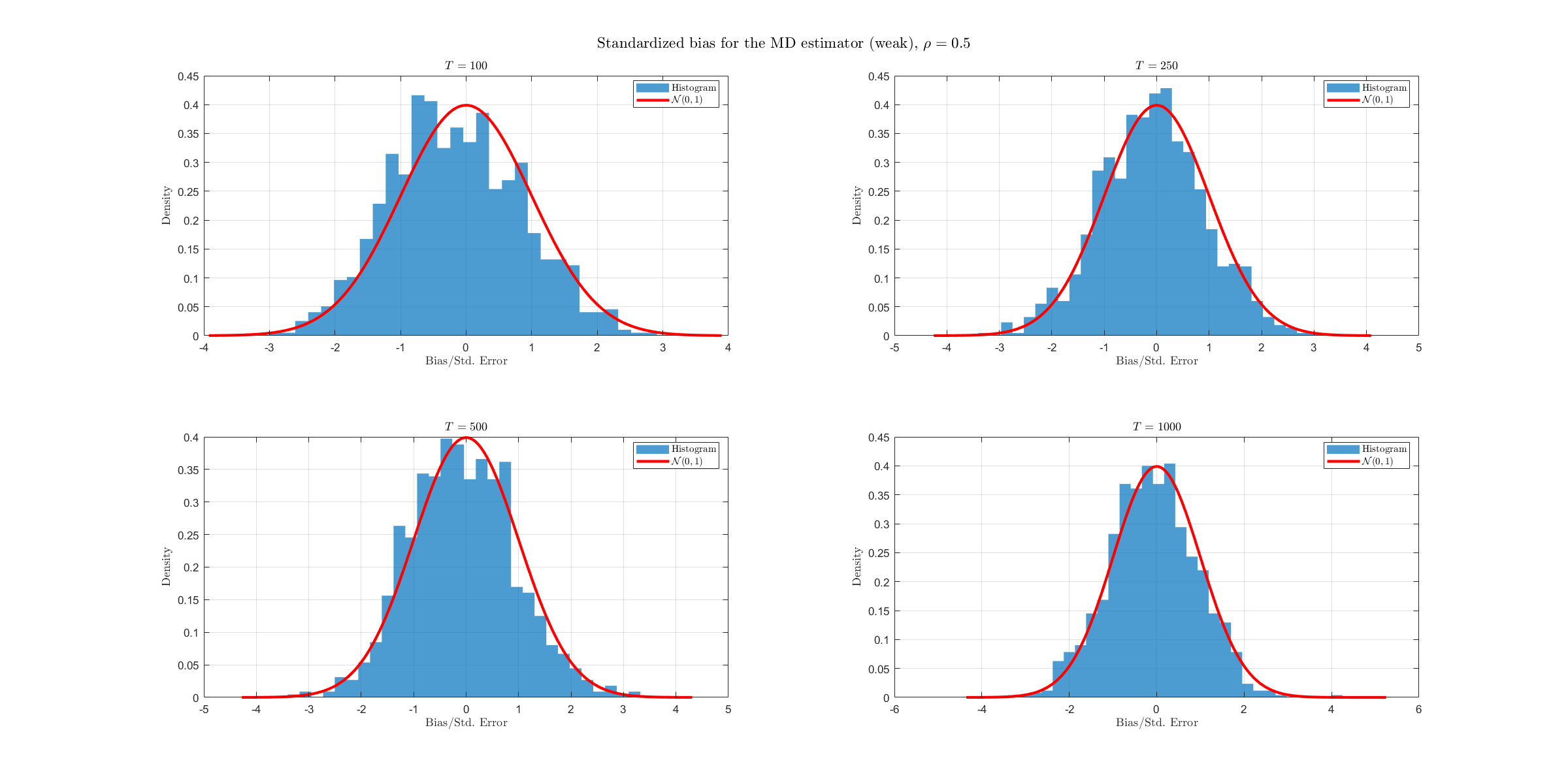}
    \caption{The figure displays the distribution of the contemporaneous structural IRFs with $(n_1, n_2) = (2,3)$, $\rho=0.5$, $\phi=0.5$, and different values of $T$. The red solid curves correspond to the standard normal density functions.}
    \label{fig:sim_dist_MD_23}
\end{figure}

\begin{figure}[!ht]
    \centering
    \includegraphics[width=\linewidth]{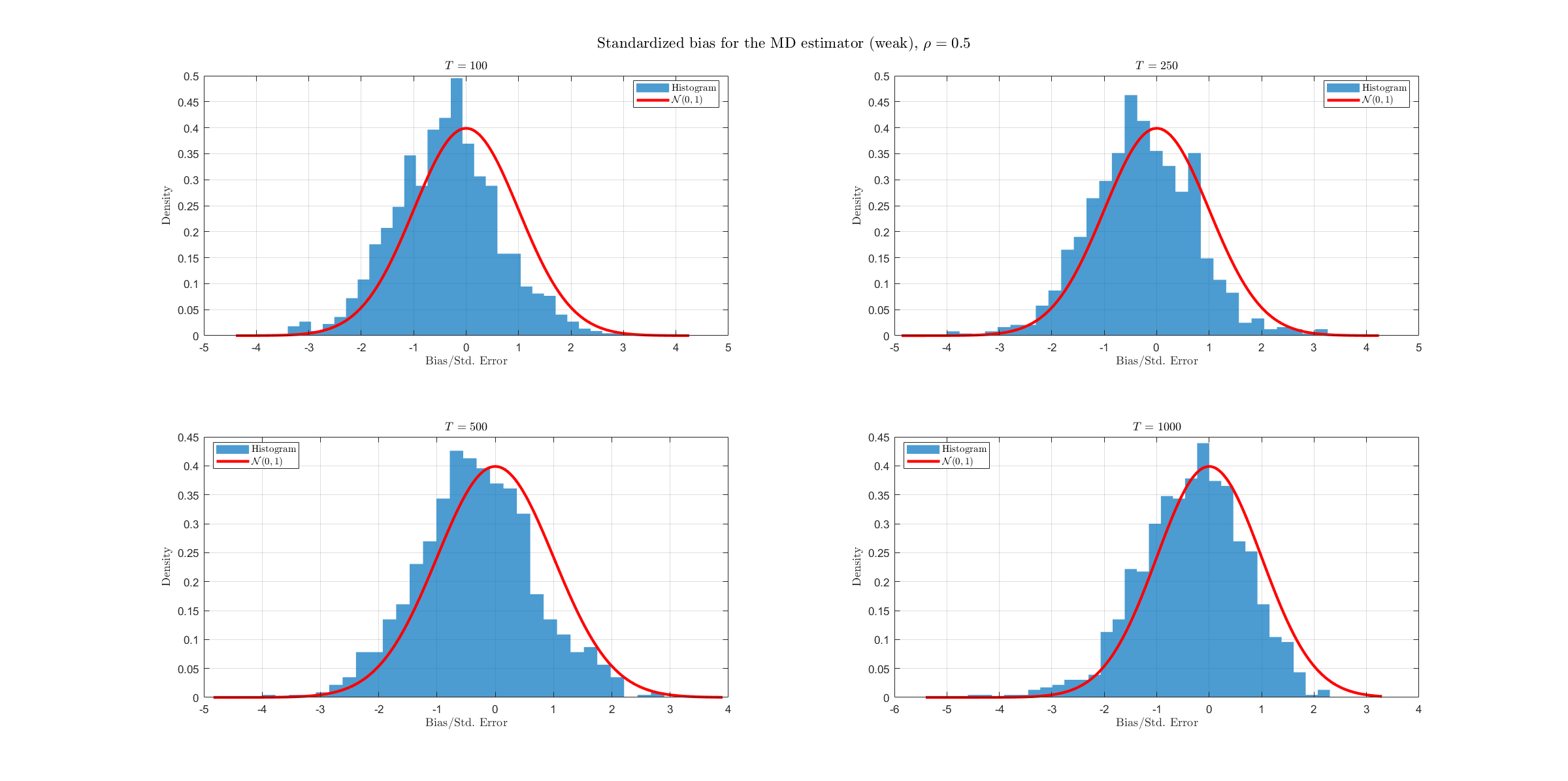}
    \caption{The figure displays the distribution of the contemporaneous structural IRFs with $(n_1, n_2) = (5,5)$, $\rho=0.5$, $\phi=0.5$, and different values of $T$. The red solid curves correspond to the standard normal density functions.}
    \label{fig:sim_dist_MD_55}
\end{figure}

We further evaluate the finite-sample performance of the asymptotic distribution of the MD estimator (Theorem \ref{thm:theta_distr}). Figure \ref{fig:sim_dist_MD_23} displays the distribution of standardized biases (i.e. the ratio of the estimated biases and their standard error) for the MD-estimated contemporaneous IRFs at element $(1,3)$ to a unit shock in element $(1,2)$, for sizes $(n_1, n_2) = (2,3)$, $\rho=0.5$, $\phi=0.5$, and different values of the sample size. The standardized biases appear to be approximately normally distributed, and get closer to a standard normal distribution as the sample size increases. When the size of the dimensions is larger, $(n_1, n_2) = (5,5)$, we see more finite-sample deviations for small $T$ and a a larger $T$ is needed before the asymptotic distribution is accurate (see Figure \ref{fig:sim_dist_MD_55}). We are estimating a larger number of parameters and therefore $T$ needs to be larger in finite-sample. If the dimensions are very large or diverging, we may want to impose sparsity on the parameters. The current estimator is not valid under this asymptotic regime, but can be explored in further research.

\subsection{Overidentification results}
\begin{figure}[!ht]
 \begin{subfigure}{0.49\textwidth}
     \includegraphics[width=\textwidth]{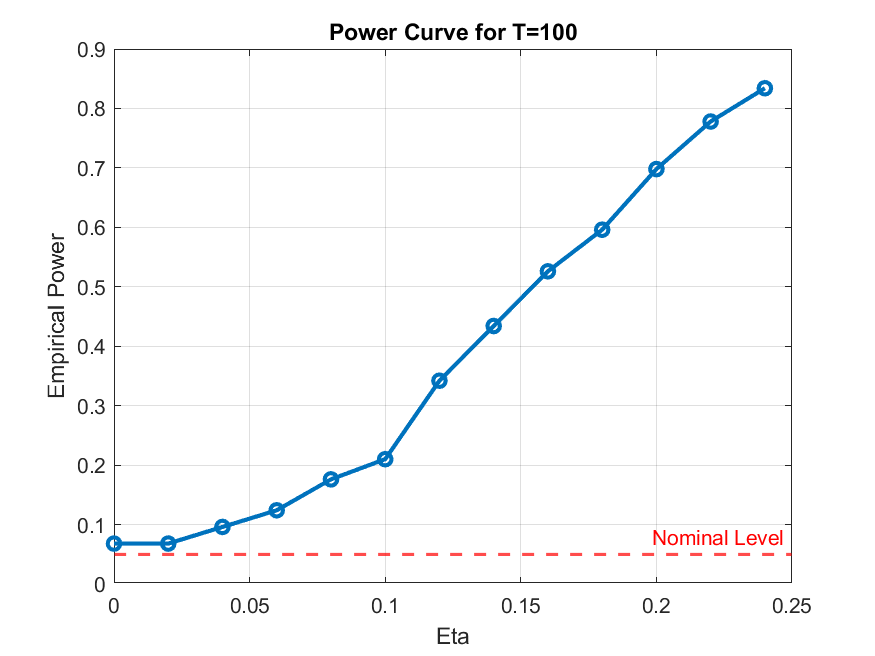}
 \end{subfigure}
 \hfill
 \begin{subfigure}{0.49\textwidth}
     \includegraphics[width=\textwidth]{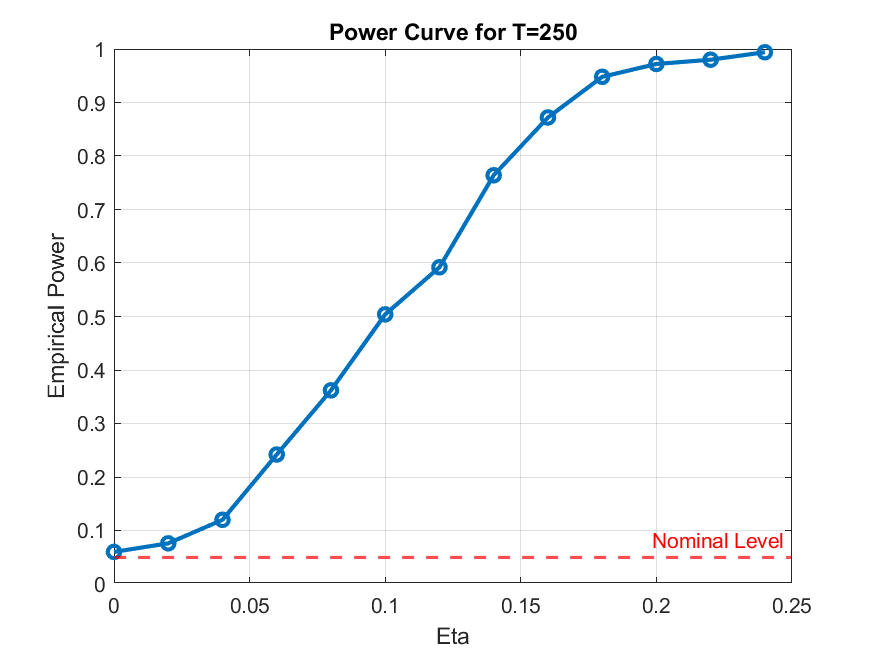}
 \end{subfigure}
 
 \medskip
 \begin{subfigure}{0.49\textwidth}
     \includegraphics[width=\textwidth]{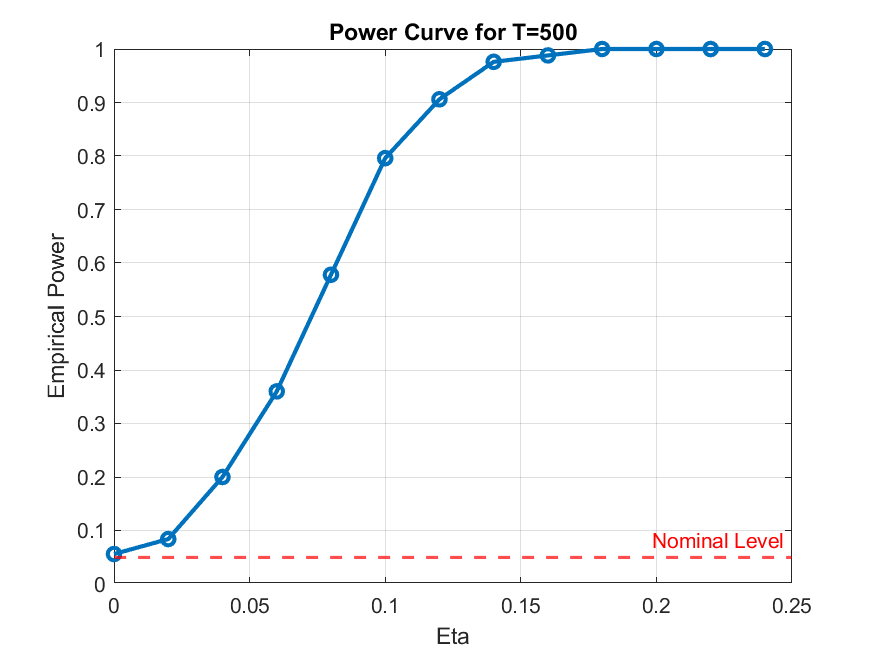}
 \end{subfigure}
 \hfill
 \begin{subfigure}{0.49\textwidth}
     \includegraphics[width=\textwidth]{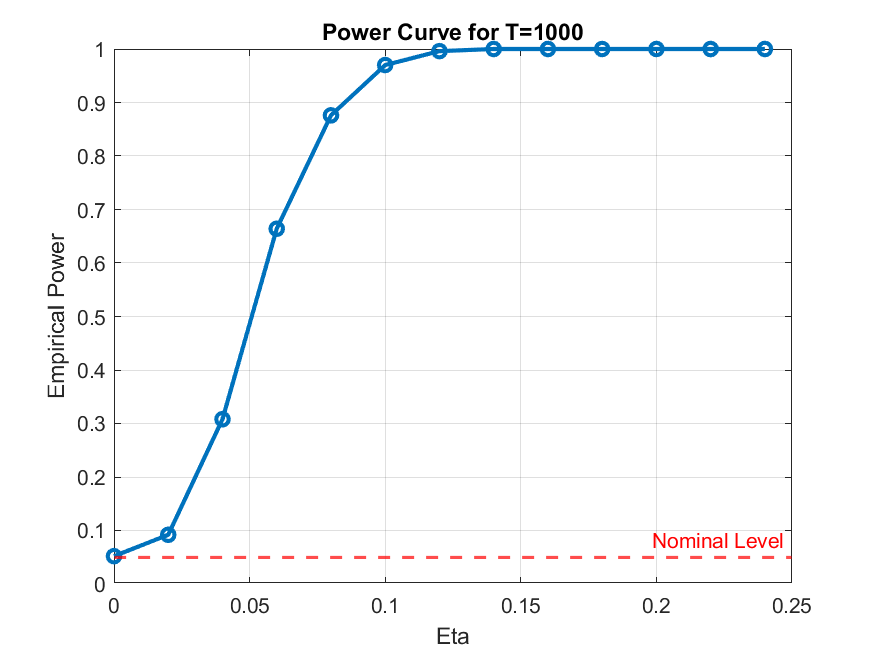}
 \end{subfigure}

 \caption{This figure reports the rejection rates of the specification test for identifying restrictions. The value of $\eta$ measures how far away the alternative is from the null hypothesis. The significance level is 5\%.}
 \label{fig:sims_power}

\end{figure}

Finally, we illustrate the validity of the overidentification test in finite samples. We fix $(n_1, n_2) = (2,3)$ and we generate samples according to model \eqref{eq:sar} with one lag, and with contemporaneous coefficients
 \begin{align*}
     \mA_{0,1} = \begin{pmatrix}
         1 & 0.5 \\ 0 & 0.8
     \end{pmatrix};\quad \mA_{0,2} = \begin{pmatrix}
         1 & \eta & 0.2\\
         0 & 1 & 0.4\\
         0 & 0 & 0.9
     \end{pmatrix},
 \end{align*}
 and the lag coefficients and reduced-form coefficients are drawn and standardized as before. In addition to triangular identification, we impose a zero restriction on the place of $\eta$. The value of $\eta$ determines whether there is overidentification. The null hypothesis $H_0$ of overidentification corresponds to $\eta = 0$. We choose a grid $\eta \in [0, 0.25]$ for $\eta$ and compute the average rejection rate of the test. We assume a significance level of $0.05$. The results are shown in Figure \ref{fig:sims_power} for different values of $T$. We make the following main observations in Figure \ref{fig:sims_power}. First, the test has correct size in finite-sample.The test slightly over-rejects the null for $T=100$, but the rejection rate approaches the nominal rate as $T$ increases. Second, as $\eta$ departs from zero, the test correctly rejects the null at a higher rate, even for small $T$. As $T$ increases, the power increases as shown in Proposition \ref{prop:j_test}: the asymptotically valid chi-square critical values are more appropriate as $T$ increases.

\subsection{Local misspecification}
\begin{figure}[!ht]
    \centering
    \includegraphics[width=\linewidth]{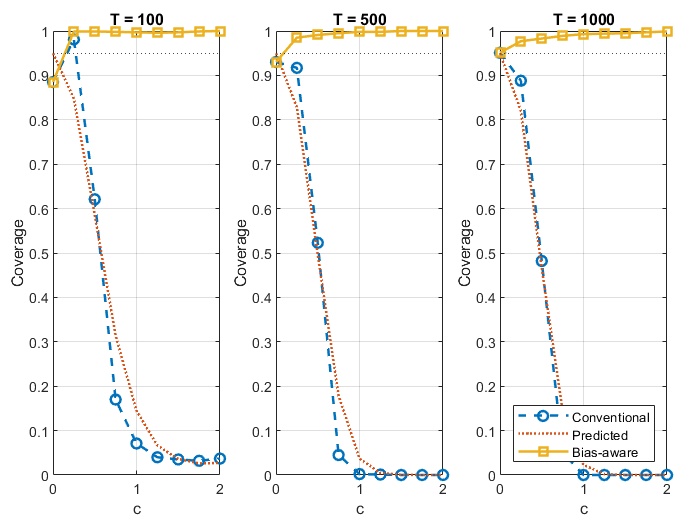}
    \caption{This figure reports the coverage results for the robust bias-aware confidence intervals (Section \ref{sec:robust}), conventional confidence intervals (Section \ref{sec:estim and props}) and its predicted coverage, for horizon $h=3$ and different values of $T$.}
    \label{fig:robust_conv_pred}
\end{figure}

Finally, we explore the coverage properties of the robust confidence intervals described in Section \ref{sec:robust}. We generate data that is locally misspecified. In particular, we fix $(n_1,n_2) = (2,3)$ and employ the following DGP for the coefficients:
\begin{align*}
    \mPhi_T = \mB_{2} \otimes \mB_{1} + c\frac{\mD}{(T-1)^{1/2}},
\end{align*}
where $\mB_j, j=1,2$ are generated as in Section \ref{sec:sim_rmse}, and where $\mD$ is chosen to be the least-favorable direction for the response $(1,1)$ to shock $(1,2)$ at horizon $h=3$,
\begin{align*}
    \vec(\mD) = \frac{(\mI - \mP_W)^\prime g_{\phi, 3}}{\sqrt{g_{\phi,3}^\prime (\mI - \mP_W)(\mI - \mP_W)^\prime g_{\phi,3}}}.
\end{align*}
Here, $\mP_W$ and $g_{\phi, h}$ are defined in Section \ref{sec:robust}. Note that $||\mD||_F=1$ such that the Kronecker scale remains normalized. The value of $c\in \SR$ determines the size of local misspecification. We consider $c\in\{0,0.25, 0.50,\ldots, 2\}$, where $c=0$ corresponds to correct specification. We set $\mH = \mV_{\phi}^{-1}$ We impose the triangular identification, as before, satisfying Assumption \ref{a4:robust_identification}. Furthermore, we know that the standardized estimation error $\Bar{b}$ can be approximated by $Z + \Bar{b}$ with $Z\sim \calN(0,1)$ under local misspecification. Thus, under local misspecification, a convential $95\%$ confidence interval is predicted to have 
\begin{align*}
    P(|Z+\Bar{b}|\leq 1.96) = \Phi(1.96 - \Bar{b}) + \Phi(1.96+\Bar{b}) - 1.
\end{align*}
We report the empirical coverage of the robust confidence intervals $CI^{\mathrm{BA}}_h(c)$ (see \eqref{eq:biasaware_ci}) and the conventional confidence intervals, together with the predicted coverage. Figure \ref{fig:robust_conv_pred} shows the results.

When $c=0$, the conventional and robust confidence intervals coincide and approach nominal coverage as $T$ increases. As $c$ increases, the coverage of the conventional confidence intervals drops sharply towards zero, as also predicted by the theory. The coverage of the robust confidence intervals is conservative for $c>0$ in finite sample, but the coverage approaches $95\%$ as $T$ increases. In short, the results show that robust confidence intervals are better suited for inference in applications where the Kronecker structure in coefficients may be disputed.

\section{Application to Monetary Spillovers} \label{sec:empirics}
In this section we apply our proposed methods to study the effects of monetary policy shocks on macroeconomic variables across countries. We use quarterly observations of short-term interest rate (IR), GDP, industrial production (IP), and inflation (CPI), across United States (US), Germany (DE), France (FR), United Kingdom (UK), and Canada (CA), obtained from the OECD database. The data is available from 1990-Q1 and we end the sample period at 2019-Q4 to exclude the COVID-19 period. The five countries are selected based on earlier evidence of synchronized business cycles and monetary spillovers \citep{monfort2003economic, dees2007exploring, canova2009estimating, canova2013panel}. We follow the data transformations described by \cite{chen2021autoregressive} to ensure approximate stationarity in our sample. We represent the data at each time $t$ as an $5\times 4$ matrix $\mcalY_t$, where the rows correspond to the countries and the columns to the variables. For the purpose of illustration, we assume a lag order of one for model \eqref{eq:sar}.

We order the variables in a recursive way. The variables are ordered according to a typical assumption that short-term interest rates shocks can affect the other variables contemporaneously, output shocks can affect industrial production and inflation contemporaneously, and industrial production shocks can only affect inflation contemporaneously. We order the countries according to trade openness, where higher trade with other countries suggests a more open economy. This ordering has been adopted in structural VARs that contained different countries \citep[see, e.g., ][]{schnucker2016restrictions}. In this ordering, the US can have contemporaneous impact on all other countries but cannot be impacted by the variables of the other countries. Canada is ordered last, implying that its variables can be impacted by all countries in the system, but Canada itself does not influence the variables of the foreign economies. The identification strategy can be summarized as follows,
\begin{align*}
\begin{array}{c}
\mathrm{US}\\
\mathrm{DE}\\
\mathrm{FR}\\
\mathrm{UK}\\
\mathrm{CA}
\end{array}
\left(
    \begin{array}{c@{\quad}ccccc}
\times &  &  &  &  \\
\times & \times  &  &  & \scalebox{1.8}{$0$}\\
\times & \times  & \times &  &\\
 \times & \times  & \times  &\times  & \\
\times &\times  &\times  & \times  & \times
\end{array}
\right)= \mA_{0,1}, \qquad   \begin{array}{c}
\mathrm{IR}\\
\mathrm{GDP}\\
\mathrm{IP}\\
\mathrm{CPI}\\
\end{array}
\left(
    \begin{array}{c@{\quad}cccc}
\times &  &  &  \\
\times & \times  &  & \scalebox{1.8}{$0$} \\
\times & \times  & \times  &\\
 \times & \times  & \times  &\times   
\end{array}
\right)=\mA_{0,2}.
\end{align*}

For estimation, we estimate the reduced-form parameters $\mB_{1,1}, \mB_{1,2}, \mSigma_{1},$ and $\mSigma_2$  by maximum-likelihood estimation subject to normalization constraints $||\hat{\mB}_{1,1}||_F=||\hat{\mSigma}_1||_F=1$. To obtain the MD estimator and the structural IRFs, we set weight matrix $\mW=\mI$ because we are under an exact identification scheme, and solve \eqref{eq:gmm_uncon}. In the following, we first report estimated IRFs and historical decompositions in Section \ref{sec:emp_irf}. To verify the results, we present robust confidence intervals for the IRFs in Section \ref{sec:emp_rob}. 

\subsection{Estimation results} \label{sec:emp_irf}
\begin{figure}[!ht]
    \centering
    \includegraphics[width=\linewidth]{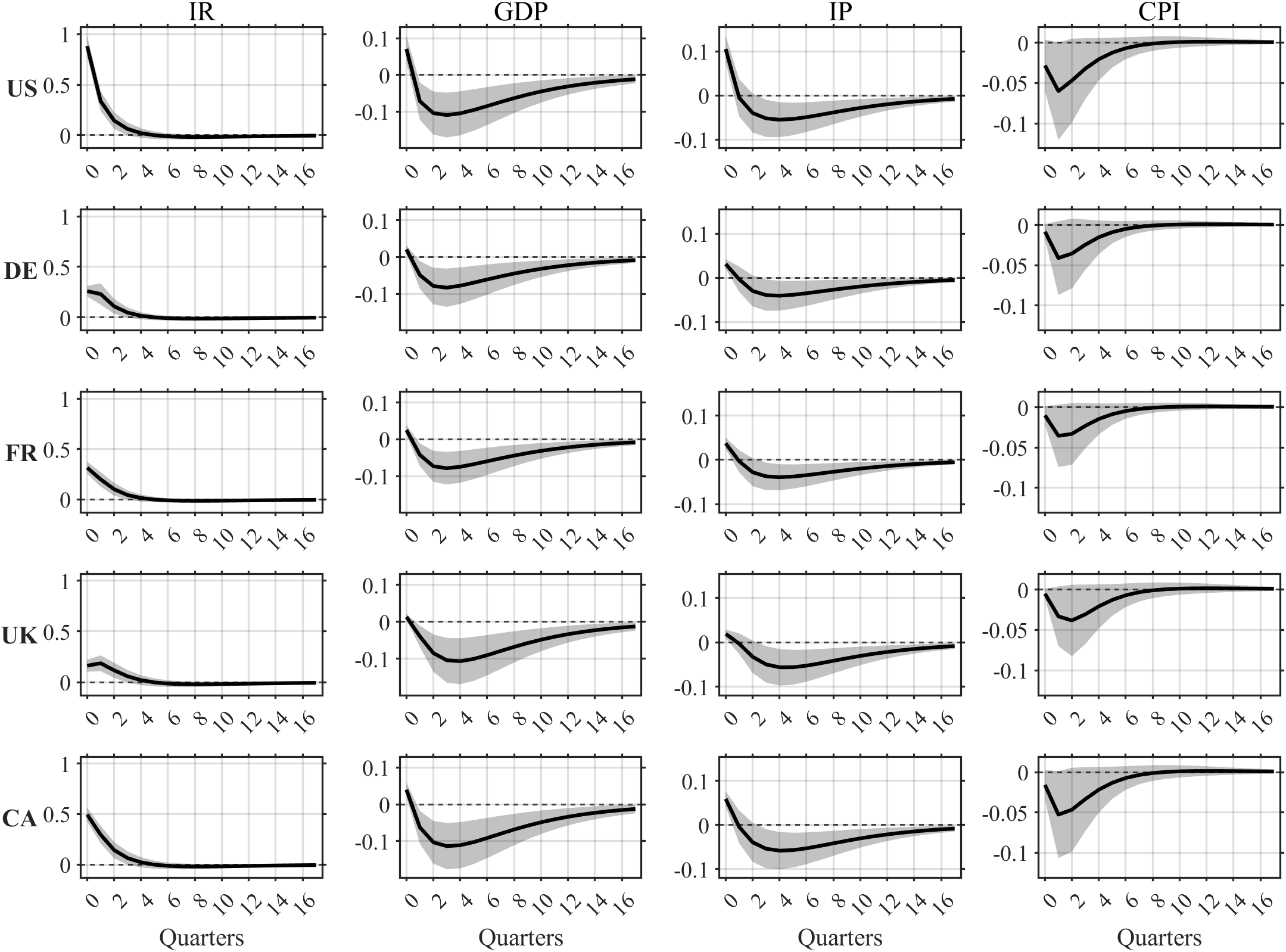}
    \caption{This figure reports the estimated structural impulse response functions to a unit US monetary policy shock. The solid black curves represent the MD estimates, and the gray shaded area represent the 68 percent confidence intervals.}
    \label{fig:appms_sirf_shock11}
\end{figure}

Figure \ref{fig:appms_sirf_shock11} shows the structural impulse response functions for a unit monetary policy/short-term interest rate shock in the US. The rows correspond to the countries' shocks, while the columns correspond to the variables. The solid line represents the MD estimates and the gray shaded area the 68\% confidence intervals. From the figure, we see that a positive US monetary policy shock has a significant positive contemporaneous effect on the interest rates in foreign countries and the effect last for around 4 quarters, with Canada showing the strongest effect among the foreign economies. Moreover, the confidence intervals indicate that there is a significant negative effect on GDP and industrial production after 2 quarters until 3 years later across economies. We observe heterogeneity across the impulse responses. For example, the US shock has significant positive impact on GDP in the US and CA, whereas the contemporaneous effects in DE, FR, and UK are close to zero. 

\begin{figure}[!ht]
\centering
\begin{subfigure}[b]{0.9\textwidth}
  \includegraphics[width=1\linewidth]{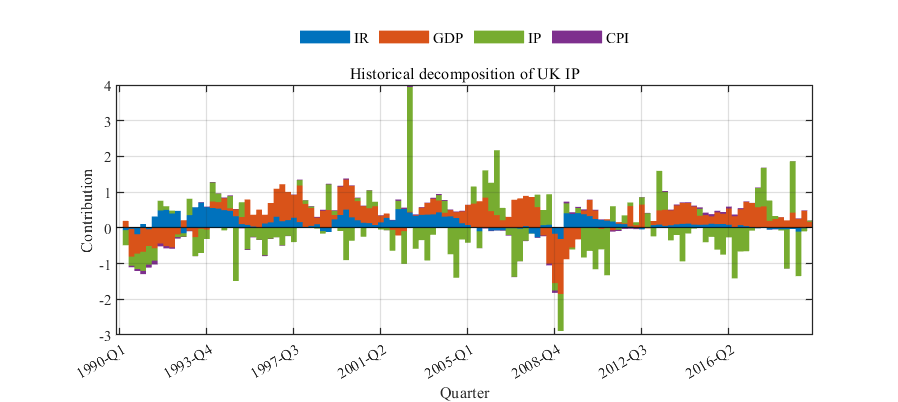}
  \caption{Variables}
  \label{fig:appms_hdecmp_var} 
\end{subfigure}

\medskip 
\begin{subfigure}[b]{0.9\textwidth}
  \includegraphics[width=1\linewidth]{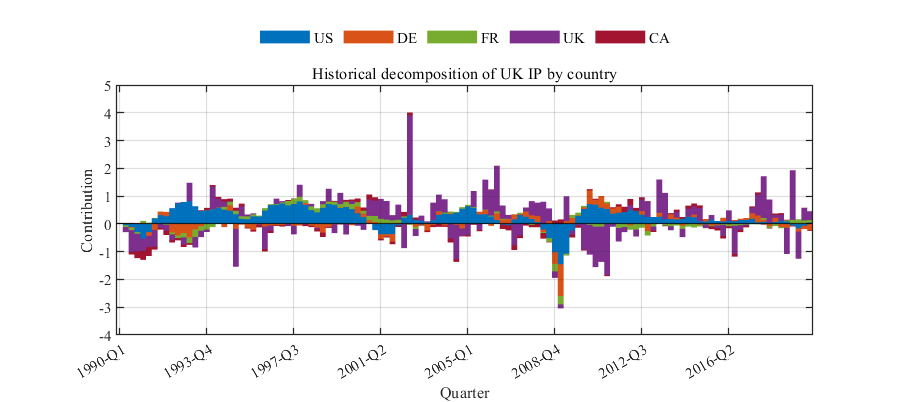}
  \caption{Countries}
  \label{fig:appms_hdecmp_country}
\end{subfigure}

\caption[Historical Decomposition]{%
This figure presents the estimated historical decomposition to UK shocks in industrial production by (a) variables and (b) countries.}
\label{fig:appms_hdecmp}
\end{figure}

Historical decompositions provide an alternative interpretation of structural shocks. We compute two historical decompositions. First, we decompose a structural shock in terms of variables, in which we aggregate the explained variation over the countries. For example, the share of a UK industrial production shock at a given quarter that is explained by interest rate is a sum of interest rate variables across countries. Second, we decompose a structural shock in terms of countries, in which we aggregate explained variation over the variables. In this way, we can obtain insight in, for instance, how important foreign activity is for IP shocks in the UK. Figure \ref{fig:appms_hdecmp} presents the estimated historical decompositions for UK industrial production shocks.
Figure \ref{fig:appms_hdecmp_var} shows that interest rates play an important (positive) role of explaining production shocks. In crises such as 2008-2009, lagged industrial productions and GDP play an important role. Domestic and foreign inflation has little to no effect on UK industrial production shocks. Figure \ref{fig:appms_hdecmp_country} shows that UK industrial production is mostly driven by domestic and US variables throughout our sample period. During crises, for example 2008-2009, the spillovers of other countries seems to become stronger. Post-2008 the impact of US seems to decrease, suggesting that the UK industrial production is accounted for by domestic policy during the pre-Brexit period.

\begin{figure}[!ht]
    \centering
    \includegraphics[width=0.8\linewidth]{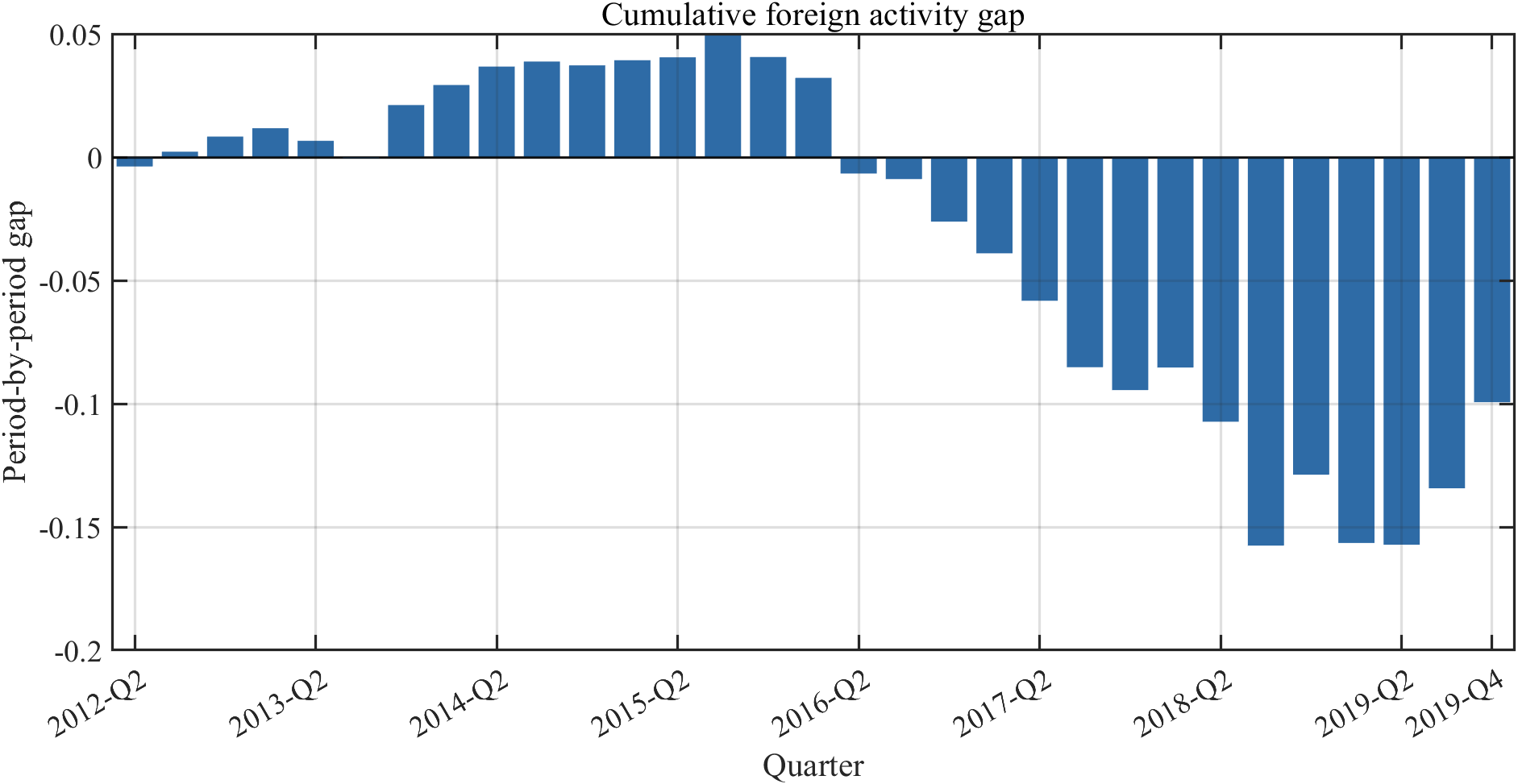}
    \caption{This figure represents the historical decomposition to a shock in industrial production, estimated by the MD estimator.}
    \label{fig:appms_counter}
\end{figure}

Furthermore, our model allows for a causal analysis of monetary policy. We perform a counterfactual analysis on US monetary policy and its impact on foreign activity. To do so, we set the US interest rate shocks to zero for the last 25\% of the sample period (2012-Q2 to 2019-Q4), continue the iterations in \eqref{eq:sar}, and measure the gap between the observed foreign activity and counterfactual foreign activity. The foreign activity is measured as average of IP and GDP. Figure \ref{fig:appms_counter} shows that there is positive gap in the beginning of the counterfactual episode, suggesting that observed U.S. monetary policy shocks raised foreign activity relative to the counterfactual. After 2015, the gap turns negative and becomes increasingly large, showing a large cumulative drag on foreign activity. The results of the counterfactual exercise suggest that U.S. monetary shocks have time-varying spillover effects.

\subsection{Robust inference results} \label{sec:emp_rob}
\begin{figure}[!ht]
    \centering
    \includegraphics[width=\linewidth]{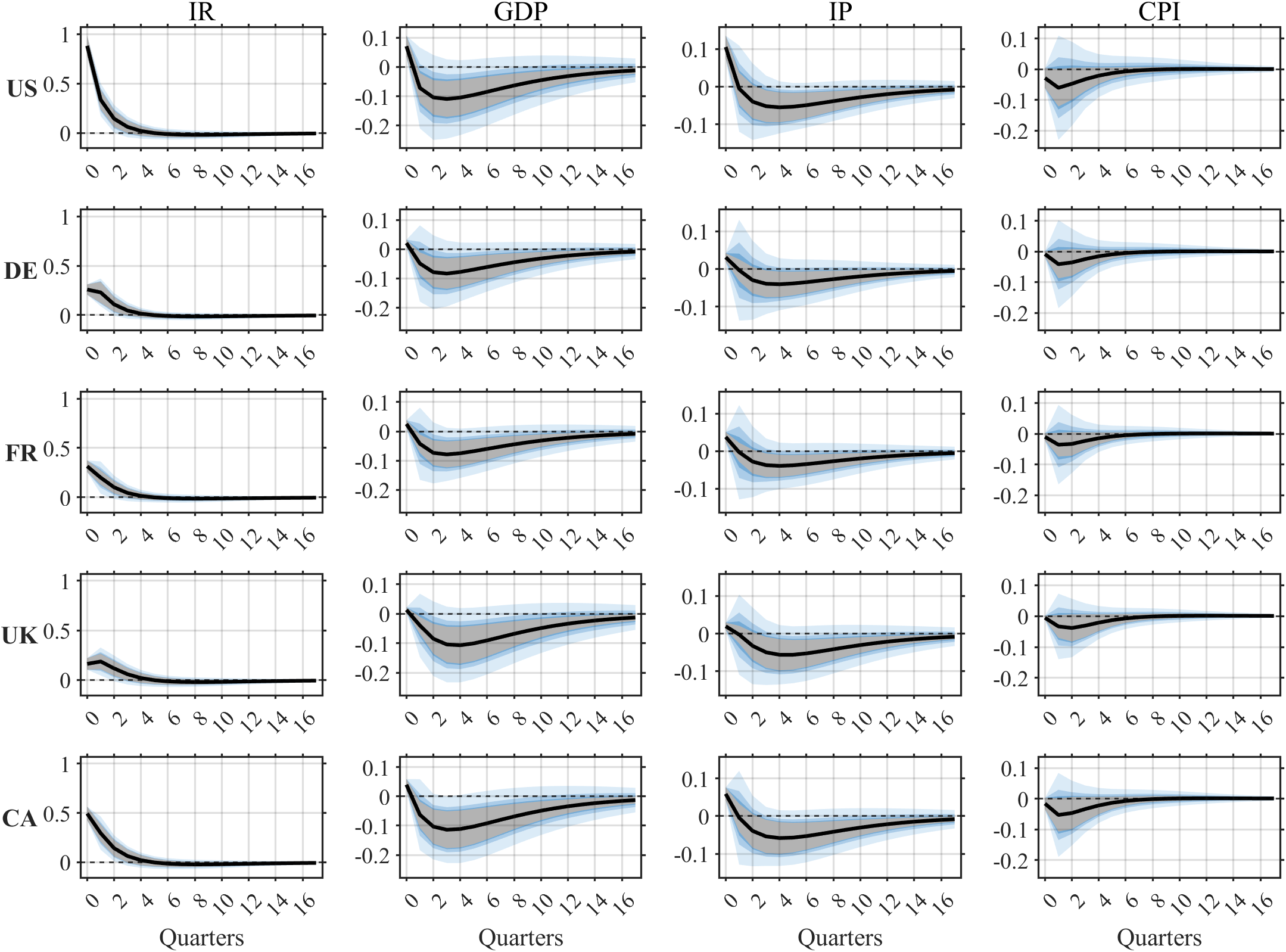}
    \caption{This figure reports the estimated structural impulse response functions to a unit US monetary policy shock. The solid black curves represent the MD estimates. The gray shaded area represent the 68\% confidence intervals. The blue, light blue, very light blue, represent the 68\% robust bias-aware confidence intervals for $c=0.25, 0.5, 1$, respectively.}
    \label{fig:appms_sirf_robust}
\end{figure}

We challenge our results by applying our robust bias-aware confidence intervals (Section \ref{sec:robust}) to a unit US interest rate shock. We consider three values for sensitivity parameter $c$, namely $c\in\{0.25, 0.5, 1\}$. The results are shown in Figure \ref{fig:appms_sirf_robust}. We can see that the robust confidence intervals for the IRFs of IR, foreign and domestic, are similar to the conventional confidence intervals, for all values of $c$. This suggests that the Kronecker structure is accurate for the IR variable. The robust confidence intervals for GDP, IP, and CPI are slightly wider than the conventional confidence intervals. However, for GDP and IP, the IRFs remain significantly negative for quarters 4 to 10, up to $c=0.5$. In this way, the robust confidence intervals can provide a sensitivity analysis on the IRFs.

\clearpage
\newpage
\section{Conclusion} \label{sec:conclusion}
This paper develops a framework for identification, estimation, and inference of structural autoregressions for multidimensional data. We restrict the coefficients to follow a Kronecker structure to reduce the number of parameters, obtain interpretable identification, and have a factor structure for the propagation of shocks. We show that the our minimum-distance estimator is consistent and asymptotically normal, and establish the asymptotic distribution of structural impulse response function. In addition, we accommodate potential misspecification of Kronecker separability by providing robust confidence intervals for the structural impulse response function under local misspecification.  Our Monte Carlo study shows that our asymptotic distributions  have a reasonable finite-sample performance, both under correct specification and local misspecification. We apply our methods on a country panel dataset to analyze dynamic and spillover effects of US monetary policy shocks. We provide further discussions on Bayesian estimation, bootstrap inference, identification with external instruments, sign and set identification, computational details, and proofs in the appendix.

\begin{spacing}{1.85}
\bibliographystyle{chicago}
\bibliography{bib}
\end{spacing}

\appendix
\clearpage
\newpage
\setcounter{page}{1}
\renewcommand{\thepage}{Appendix p. \arabic{page}}
\centerline{\LARGE Appendix to:}
\begin{center}
{\LARGE
Identification, estimation, and robust inference\\[1ex]
of structural autoregressions for multidimensional data
}
\end{center}
\bigskip
\centerline{\large \textit{Bernhard van der Sluis$^{1}$}}
\bigskip
\centerline{$^1$\ \textit{Econometric Institute, Erasmus University Rotterdam}}
\bigskip
\noindent{\Large\textbf{Contents}}\\[2ex]
Appendix~\ref{app:methods}:\ \nameref{app:methods}\hfill\pageref{app:methods}\\[1ex]
Appendix~\ref{app:proof}:\ \nameref{app:proof}\hfill\pageref{app:proof}\\[1ex]
Appendix~\ref{app:sims}:\ \nameref{app:sims}\hfill\pageref{app:sims}\\[1ex]

\newpage
\clearpage
\section{Appendix to Methodology} \label{app:methods}

\subsection{Illustration: Number of parameters} \label{sec:app_noparams}
\begin{figure}[!h]
    \centering
    \includegraphics[width=\linewidth]{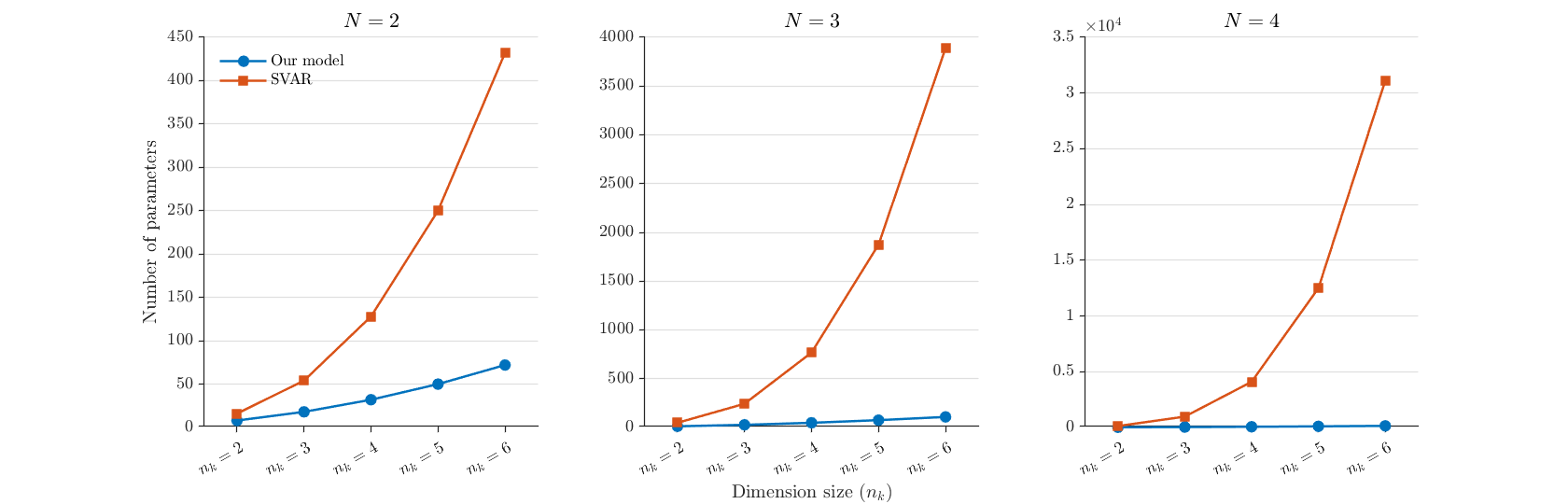}
    \caption{The number of parameters as a function of $N$ and $n_k$ for SVAR and model \eqref{eq:sar}.}
    \label{fig:num_of_params_svar}
\end{figure}

The figures below show the number of parameters for different specification of number of modes $N$ and their sizes $\{n_k\}^N_{k=1}$, for an unrestricted SVAR and our model \eqref{eq:sar}. We set $N\in\{2,3,4\}$ and $n_1 =\ldots = n_N$ with $n_1\in\{2,3,4,5,6\}$. We assume there are no lags, meaning the figure below shows the most favorable case for the SVAR model. Figure \ref{fig:num_of_params_svar} shows that the number of parameters explodes for the SVAR model, while it remains limited for model \eqref{eq:sar}. Our model pays a price in terms of flexibility, but heterogeneity is observed through its specification.

\subsection{Special case: matrix time series} \label{app:mar}
The notation and model derivations simplify for $N=2$. Let us illustrate the derivation of the structural impulse response functions for $N=2$. To keep notation simple, we assume there is no intercept term $\widetilde{\mC}$ and a single lag ($p=1$). In this case, the reduced form of model \ref{eq:sar} can be written as 
\begin{align}\label{eq:rf_compactmar1}
    \mcalY_t = \mB_{1,1} \mcalY_{t-1}\mB_{1,2}^\prime + \mA^{-1}_{0,1}\mcalU_t\mA_{0,2}^{-1\prime}. 
\end{align}
Therefore, model \eqref{eq:sar} with $N=2$ admits a causal and stationary solution if the spectral radius of $\mB_{1,2}\otimes \mB_{1,1}$ is strictly less than one.In other words, we would need to assume that $\rho(\mB_{1,1}) \cdot \rho(\mB_{1,2}) < 1$.
Under this assumption, the process $\{\mcalY_t\}$ is stationary and stable for $N=2$ modes. We need this for the following causal representation
\begin{align*}
    \vec(\mcalY_t) &= \sum^{\infty}_{s=0}\left(\mB_{1,2}^s\otimes \mB_{1,1}^s\right)\vec(\mcalU_{t-s})\\
    &= \sum^{\infty}_{s=0}\left(\mB_{1,2}^s\otimes \mB_{1,1}^s\right)\left(\mA_{0,2}^{-1}\otimes \mA_{0,1}^{-1}\right)\vec(\mcalE_{t-s}),
\end{align*}
with convention that $\mB_{1,2}^s \otimes \mB_{1,1}^s = \mI_{n_1n_2}$ for $s=0$. Let $\mcalE_{t-s}$ be a matrix with nonzero element $[\mcalE_{t-s}]_{ij}$ and zero elsewhere. Then the structural impulse response function
 of $\mcalY_t$ to shock $[\mcalE_{t-s}]_{ij}$ can be computed as
 \begin{align}
\frac{\partial \vec(\mcalY_t)}{\partial \vec(\mcalE_{t-s})'}
&=
\left(\mB_{1,2}^s\otimes \mB_{1,1}^s\right)\left(\mA_{0,2}^{-1}\otimes \mA_{0,1}^{-1}\right)\vec(\mcalE_{t-s}).
\label{eq:sr_irf_vec_mar}
 \end{align}
 In matrix form, we can write
\begin{align}
     \mF_{ij}(s) \equiv \frac{\partial \mcalY_t}{\partial [\mcalE_{t-s}]_{ij}} = \left[\mB_{1,1}^s \mA_{0,1}^{-1}\right]_i\left[\mB_{1,2}^s\mA_{0,2}^{-1}\right]_j^\prime,
\label{eq:sr_irf_matrix_mar}
\end{align}
The impulse response functions $\mF_{ij}(s)$ are $n_1 \times n_2$ matrices, with each element reflecting the response of an individual unit.

\subsection{Identification with external instruments}
In the paper, we globally identify the structural shocks in the structural autoregression using zero restrictions, up to sign normalizations. An alternative approach is to use external instruments to partially identify structural shocks. This approach is typically referred to as Proxy-SVAR, and has been widely used in macroeconomics, see for example \citetapndx{Stock2008proxy}, \citetapndx{StockWatson2012}, \citetapndx{MertensRavn2013}, and \citetapndx{arias2021proxyvar}. In this section we provide a discussion on the proxy-approach in our framework for multidimensional data, though we will not develop a complete theory of estimation and inference.

Recall that the relation between the vectorized reduced-form innovations and the structural shocks is given by $\vu_t = \mC_0 \ve_t$. Suppose we are interested in structural shock $e_{\iota,t}$ for some index $\iota$ and let $\vc_{\iota}$ denote the corresponding column of $\mC_0$. To identify the dynamic effects of $e_{\iota,t}$, the idea is to estimate $\vc_{\iota}$ with external instruments $\mZ_t = (Z_{1t,\ldots,Z_{kt}})^\prime\SR^k, t=1,\ldots,T$. The instruments are assumed to satisfy a relevance condition, $\E(\mZ_t \e_{\iota,t}) = \valpha\neq \vzeros$, and an exogeneity condition, $\E(\mZ_t e_{j,t}) = \vzeros$ for $j\neq \iota$. It follows that the instruments satisfy
\begin{align*}
    \E(\vu_t \mZ_t^\prime) = \E(\mC_0 \ve_t \mZ_t^\prime) = \vc_{\iota}\valpha^\prime.
\end{align*}
This moment condition identifies $\vc_{\iota}$, up to normalization. Moreover, if $k>1$, we have overidentification. 

In our model, $\vc_{\iota}$ has a Kronecker structure, which has several consequences. To see these, let us consider the case for $N=2$. Suppose we are interested in a structural shock indexed by pair $\iota=(i,j)$ for $i=1,\ldots,n_1$ and $j=1,\ldots,n_2$. Our model assumes that $\mC_0 = \mC_{0,2}\otimes \mC_{0,1}$. Let $\vc_{1,i}$ denote the $i$-th column of $\mC_{0,1}$ and $\vc_{2,j}$ denote the $j$-th column of $\mC_{0,2}$. Under relevance and exogeneity, the moment condition can be written as
\begin{align*}
    \E(\vu_t \mZ_t^\prime) = (\vc_{2,j}\otimes \vc_{1,i})\valpha^\prime.
\end{align*}
Hence, the instruments identify the direction $\vc_{\iota}$ up to scale, as in the conventional Proxy-SVAR. The Kronecker structure, however, restricts the structural direction that is being identified to a set of vectors that can be written as a product of two mode-specific directions. This restriction has several consequences.

First, with multiple dimensions, different exogeneity conditions may be appropriate depending on the structural object we want to identify. Suppose, for illustration, that $N=2$ where the first dimension indexes countries and the second indexes variables. If the object of interest is $e_{11,t}$ and we impose $\E(\mZ_t e_{ij,t})=0$ for all $(i,j)\neq (1,1)$, then both mode-specific directions $\vc_{1,1}$ and $\vc_{2,1}$ are identified, up to scale normalizations. However, if it is difficult to find or argue that a proxy is exogoneous to every other structural error within country 1, then one may want to allow for country-specific exogeneity, $\E(Z_{\ell,t}e_{1j,t}) = \alpha_{\ell,j}\neq0$ for all $j=1,\ldots,n_2$, and $\E(Z_{\ell,t}e_{ij,t})=0$ for all $i\neq 1$. Under country-specific exogeneity, it follows that
\begin{align*}
    \E(\mcalU_t Z_{\ell,t}) &= \sum^{n_2}_{j=1}\alpha_{\ell,j}\vc_{1,1}\vc_{2,j}^\prime\\
    &= \vc_{1,1}\left(\sum^{n_2}_{j=1}\alpha_{\ell,j}\vc_{2,j}\right)^\prime.
\end{align*}
Thus, under weak exogeneity, the shock $e_{11,t}$ is no longer identified because the second-mode component is a mixture of variable-specific directions. Nevertheless, country-specific direction $\vc_{1,1}$ is common to all admissible shocks and can still be identified up to scale. This suggests a hierarchy of proxy assumptions, with exogeneity versus mode-specific exogeneity. This also implies that an instrument that is not valid for one exact country-variable shock may still contain information about a shock with one country dimension.

Second, the Kronecker structure restricts the admissible rotations when the instruments are correlated with several structural shocks. In an unrestricted SVAR, instruments can identify the space spanned by the correlated structural shocks, but they need not identify a unique rotation, and therefore it requires additional zero and sign restrictions \citepapndx{arias2021proxyvar}. The Kronecker structure restricts the admissible rotations since it imposes a rank-one direction. 

Third, our model reduces the number of parameters dramatically (Figure \ref{fig:num_of_params_svar}), wherein parameters are common within a mode. This suggests a potential pooling argument as the number of units within a mode increases. Similarly, in the panel data literature, grouped heterogeneity is used to classify individuals into groups that share common parameters. The common parameters can then be estimated more efficiently as the number of individuals within a group diverge \citepapndx{bonhomme2015, bonhomme2022discretizing, cheng2026}. We may use this observation to alleviate weak-instruments in our proxy-context. 

To see this, consider again a single structural shock $e_{ij,t}$ and a scalar instrument $Z_t$ satisfying relevance and exogeneity. Let us denote the moment condition by $\mG \equiv \E(\mcalU_t Z_t) = \alpha \vc_{1,i}\vc_{2,j}^\prime$. The corresponding sample moment $\hat{\mG}$ can be written as
\begin{align*}
    \hat{\mG} = \alpha \vc_{1,i}\vc_{2,j}^\prime + (\hat{\mG}-\mG).
\end{align*}
Suppose relevance is local-to-zero weak, such that $\alpha=\alpha_T= T^{-1/2}\delta$ for some $\delta\neq 0$. For fixed $(n_1, n_2)$, the signal $(\alpha_T\vc_{1,i}\vc_{2,j}^\prime)$ and the sampling error are of the same asymptotic order, so that the weak-instrument problem is not resolved by a Kronecker structure. If $(n_1, n_2)$ grow with $T$, however, this may be different. Assuming that $(\vc_1, \vc_2)$ do not shrink to zero as the dimensions increase, $\mG$ will be of asymptotic order $|\delta|\sqrt{n_1n_2/T}$. Thus, common mode-specific directions are observed across an increasing number of units and thus $\mG$ may need not to vanish. This suggests a pooling argument, under suitable rates. At the same time, one would need to control the sampling error as the dimensions grow and the dependence across unit-specific moments. We leave a formal joint large-$T$, large-$n_k$, weak-instrument framework for further research.

\subsection{Sign and set identification}
Throughout the paper we have assumed that we have a sufficient number of zero restrictions such that the structural shocks are point identified. If we impose insufficiently many zero restrictions, there remains a set of admissible rotations of $\mP\in O(n)$, and thus, the shocks are set identified. Typically, one would impose economically motivated sign restrictions to shrink the admissible set of rotations. 

Sign restrictions could also be imposed in our framework for multidimensional data, but it requires additional care, because they are affected by the sign indeterminacy of the Kronecker product. In particular, a sign restriction imposed on one mode may imply a set of sign restrictions on the product, i.e., on the multidimensional impulse responses. 

\subsection{Bayesian Inference} \label{sec:bayes}

In practice, Bayesian inference is often used for SVARs. It allows the practitioner to include prior information and does not require large $T$ for valid inference. Alternatively to the MD estimation, our model can be estimated using existing Bayesian techniques for SVARs and large SVARs. In particular, the Kronecker structure in our model separates the large SVAR in lower dimensional modes. It allows for a different prior for each dimension, and both priors are efficiently updated by iterating between modes using a Gibbs sampling step.

In particular, we provide an efficient algorithm to independently draw from any posterior over a Kronecker orthogonal parameterization, conditional on the zero restrictions. For efficiency, we work with a family of prior distributions that have a conjugate posterior. The conjugate normal-inverse-Wishart family of distributions for the SVAR \citepapndx{arias2018inference} can be used, but we need to take care of the Kronecker structure in the parameters. The idea is to independently draw  orthogonal reduced form parameters from the posterior distribution, conditional on the restrictions. We map the independent draws into the structural parameterization and weight the draws such that the draws are indeed from the posterior distribution over the structural parameters.

\subsubsection{Orthogonal Parameterization} \label{sec:orth_param}
We propose the Kronecker orthogonal reduced-form parameterization, where each mode follows the orthogonal parameterization, separately. Following \citetapndx{arias2018inference}, the parameterization of the reduced-form can be written as follows,
\begin{align} \label{eq:orth_parma}
    &\vec(\mcalY_t)^\prime = \vec(\mcalY_{t-1})^\prime \mUpsilon + \vec(\mcalE_t)^\prime\mP^\prime \mH(\mSigma),\\
    & \mUpsilon = \mB_N \otimes \cdots \otimes \mB_1,\\
    & \mP = \mP_N \otimes \cdots \otimes \mP_1,\\
    & \mSigma = \mSigma_N \otimes \cdots \otimes \mSigma_1,
\end{align}
where $(\mB_k, \mSigma_k, \mP_k)$ are defined as in Section \ref{sec:methods}, and the transformation $\mH(\mR) \equiv \mH(\mR_N \otimes \cdots \otimes \mR_1)$ denotes any decomposition of a matrix $\mR$ with a Kronecker structure such that $\vh(\mR_k)'\vh(\mR_k) = \mR_k$. We follow \citetapndx{arias2018inference} and take the Cholesky decomposition for $\vh_k$, $k=1,\ldots,N$. Equation \eqref{eq:orth_parma} is equivalent to the orthogonal parameterization in \citetapndx{arias2018inference}, where $(\mUpsilon, \mSigma,\mP)$ each follow a Kronecker structure.

Let $g_k$ be a mapping from the structural parameters to the reduced form parameters for mode $k$,
\begin{align*}
    g_k(\mA_{0,k}, \mA_k) = \left(\underbrace{\mA^{-1}_{0,k}\mA_k}_{\mB_k},\quad
    \underbrace{\left(\mA_{0,k} \mA_{0,k}'\right)^{-1}}_{\mSigma_k},
    \quad \underbrace{\mA_{0,k}'h\left(\left(\mA_{0,k} \mA_{0,k}'\right)^{-1}\right)}_{\mP_k}, \right)
\end{align*}

The function $g_k$ is invertible for each $k$, where the inverse is defined as
\begin{align*}
    g_k^{-1}\left(\mB_k, \mSigma_k, \mP_k\right) = \left(\underbrace{\mP_k \vh(\mSigma_k)^{-1}}_{\mA_{0,k}}, \quad \underbrace{\mP_k \vh(\mSigma_k)^{-1}\mB_k}_{\mA_k}\right)
\end{align*}
Therefore, given the reduced-form parameters and a decomposition $\mH$, each value of $\mP_k \in \calO(n_k)$ implies a particular choice of structural parameters for mode $k$. 

\subsubsection{Prior and posterior distribution}
Since the orthogonal parameterization in \eqref{eq:orth_parma} follows the same form as the reduced-form model in \citetapndx{arias2018inference}, the normal-inverse-Wishart family is also conjugate for the reduced-form in \eqref{eq:orth_parma}. The Kronecker structure has three important implications. First, each mode has their own independent inverse-Wishart prior for $\mSigma_k$, instead of one for the entire covariance matrix $\mSigma$. Second, the tensor structure of the data implies that a normal-inverse-Wishart distribution over the reduced-form parameters is defined by $N\times 4$ parameters: a degrees of freedom for each mode, $\nu_k\geq n_k$; a symmetric and positive definite matrix $\mUpsilon_k$ with size $n_k\times n_k$; a matrix $\mPsi_k$ with size $n_k p\times n_k$; a symmetric and positive definite matrix $\mOmega_k$ with size $n_k p\times n_k p$. Let $\nu=\sum^N_{k=1}\nu_k$, $\mUpsilon = \bigotimes^{1}_{k=N}\mUpsilon_k$, $\mPsi = \bigotimes^{1}_{k=N}\mPsi_k$, and $\mOmega = \bigotimes^{1}_{k=N}\mOmega_k$. We denote the distribution of the \emph{Kronecker normal-inverse-Wishart} distribution by KNIW and its density by
\begin{align*}
    &\mathrm{KNIW}(\mPhi, \mSigma) \\
    &\propto |\det\left(\mSigma\right)|^{-\frac{\nu + n + 1}{2}}e^{-\frac{1}{2}\tr\left(\mUpsilon\mSigma^{-1}\right)}
    |\det\left(\mSigma\right)|^{-\frac{m}{2}} e^{-\frac{1}{2}\vec\left(\mPhi-\mPsi\right)'\left(\mSigma\otimes \mOmega\right)^{-1}\vec\left(\mPhi-\mPsi\right)}\\
    &=\prod^N_{k=1}|\det(\mSigma_k)|^{-\frac{\nu_k + n_k +1}{2}} e^{-\frac{1}{2}\tr\left(\mUpsilon_k\mSigma_k^{-1}\right)} |\det(\mSigma_k)|^{-\frac{n_k p}{2}} e^{-\frac{1}{2}\vec\left(\mB_k-\mPsi_k\right)'\left(\mSigma_k\otimes \mOmega_k\right)^{-1}\vec\left(\mB_k-\mPsi_k\right)}.
\end{align*}
To simplify the notation for the posterior, we define for each mode $k$ the effective regressors. Let $\mSigma_{-k} = \bigotimes_{\ell \neq k}\mSigma_\ell$ with dimension $n_{-k} = n/n_k$, and let $\mB_{j,-k} = \bigotimes_{\ell \neq k}\mB_\ell$. Recall that $\vy_t = \vec(\mY_t)$ where $\mcalY_t$ is an $n_1\times\cdots\times n_N$ tensor. Let $\mY_{k,t}$ denote its mode-$k$ matricization, i.e., the $n_k\times n_{-k}$ matrix satisfying $\vec(\mcalY_t) = (\mI_{n_{-k}}\otimes \mY_{k,t})\vec(\mI_{n_{-k}})$ up to a permutation of columns. Define the $n_{-k}\times n_k$ effective regressor matrix $\tilde{\mX}_{k,t}$ as $\mY_{k,t-1} \mA_{-k}^\prime$. The mode-$k$ regression then reads $\mY_{k,t}^\prime = \mX_{k,t}^\prime\mB_k + \mU_{k,t}$, where the rows of $\mU_{k,t}$ have covariance $\mSigma_k$ and are correlated across rows through $\mSigma_{-k}$. After whitening by $\mSigma_{-k}^{-1/2}$, we obtain $\mZ_{k,t} = \mSigma_{- k}^{-1/2}\mX_{k,t}^\prime$ and $\mY_{k,t}^w = \mSigma_{\setminus k}^{-1/2}\mY_{k,t}^\prime$, yielding $T_{\mathrm{eff}} = n_{\setminus k}\cdot T$ whitened observations with independent rows. The conditional posterior distribution of $(\mB_k, \mSigma_k)$ given all other factors is KNIW with the parameters,
\begin{align*}
    &\tilde{\nu}_k = T_{\mathrm{eff}} + \nu_k\\
    &\tilde{\mOmega}_k = \left(\mOmega_k^{-1} + \sum^T_{t=1}(\mZ_{k,t})^\prime\mZ_{k,t}\right)^{-1}\\
    &\tilde{\mPsi}_k = \tilde{\mOmega}_k\left(\mOmega_k^{-1}\mPsi_k + \sum^T_{t=1}(\mZ_{k,t})^\prime\mY_{k,t}^w\right)\\
    &\tilde{\mUpsilon}_k = \left(\sum^T_{t=1}(\mY_{k,t}^w)^\prime\mY_{k,t}^w\right) + \mUpsilon_k + \mPsi_k^{\top}\mOmega_k^{-1}\mPsi_k - \tilde{\mPsi}_k^{\top}\tilde{\mOmega}_k^{-1}\tilde{\mPsi}_k
\end{align*}
Since $\mZ_{k,t}^w$ and $\mY_{k,t}^w$ depend on the parameters of the other modes, we use Gibbs sampling to draw from the posterior distribution. In particular, we iteratively draw from the conditional posterior distributions, cycling through $k=1,\ldots,N$ and updating each $(\mB_k, \mSigma_k)$ given the current values of all other factors. To resolve the scale indeterminacy inherent in the Kronecker factorization, we normalize $\|\mB_k\|_F = 1$ and $\|\mSigma_k\|_F = 1$ for $k < N$ after each sweep, absorbing the scale factors into the last mode.

\subsubsection{Bayesian sampling}
We can draw independently from the KNIW distribution. However, to identify the SMAR, we need to impose zero (and sign) restrictions, see Section \ref{sec:identification}. The zero restrictions are linear restrictions on the columns of orthogonal matrix $\mP$, and therefore on the columns of $\mP_k$ for $k=1,\ldots,N$, conditional on the reduced-form parameters. Using this observation, we can draw independently from the set of structural parameters that satisfy zero restrictions. For mode $k$, let $z_{k,j}$ denote the number of zero restrictions on the $j$th column of $\mP_k$, and define $o_{k,j} = z_{k,j} + j - 1$ for $j = 1,\ldots, n_k$. One caveat is the sign indeterminacy: although drawing $\mP_k$ separately for each $k$ works, their Kronecker product may not deliver the sign consistent with the data. Hence, we need a sign normalization for $\mP_k$, employed through a matrix $\mD_r$. Drawing from the KNIW posterior and constructing each $\mP_k$ separately yields a set of structural parameters satisfying the zero restrictions. The following algorithm describes the procedure.

\begin{enumerate}
    \item Draw $(\mB_k, \mSigma_k)$ for $k=1,\ldots,N$ from the KNIW conditional posteriors via Gibbs sampling, and normalize $\|\mB_k\|_F = 1$, $\|\mSigma_k\|_F = 1$ for $k < N$.
    \item For each mode $k=1,\ldots,N$ and each column $j=1,\ldots,n_k$:
    \begin{enumerate}
        \item Draw $\vbeta_{k,j} \in \mathbb{R}^{n_k - o_{k,j}}$ independently from a standard normal distribution, and set $\vomega_{k,j} = \vbeta_{k,j}/\|\vbeta_{k,j}\|$.
        \item Let $\mK_{k,j}$ be any $n_k \times (n_k - o_{k,j})$ matrix whose columns form an orthonormal basis for the null space of the $o_{k,j} \times n_k$ matrix
        \begin{align*}
            \mM_{k,j} = \begin{bmatrix} \vp_{k,1} & \cdots & \vp_{k,j-1} & \bigl(\mQ_{k,j} f\bigl(g^{-1}_h(\mB_k, \mSigma_k, \mI_{n_k})^\prime\bigr)\bigr)^\prime \end{bmatrix}^{\prime},
        \end{align*}
        and set $\vp_{k,j} = \mK_{k,j}\,\vomega_{k,j}$.
    \end{enumerate}
    \item Set $\mP_k = [\vp_{k,1}\;\cdots\;\vp_{k,n_k}]$ for each mode and $\mP = \bigotimes^{1}_{k=N}\mP_k$.
    \item Compute the structural parameters $(\mA_{0,k}, \mA_{k}) = g^{-1}_h(\mB_k, \mSigma_k, \mP_k)$ for $k=1,\ldots,N$.
    \item Sign normalization: $(\mA_{0,k}, \mA_{k}) = (\mA_{0,k} \mD_k, \mA_{k} \mD_k)$ 
    \item Retain the draw with importance weight 
    \begin{align*}
        \prod_{k=1}^{N} w_{k},
    \end{align*}
    where $w_k$ denotes the ratio of volume elements as in \citetapndx{arias2018inference}, evaluated per mode.
    \item Return to Step~1 until the required number of draws has been obtained.
    \item Re-sample with replacement using the normalized importance weights.
\end{enumerate}

\subsection{Bootstrap inference}
Theorem \ref{thm:irf_distr} establishes the asymptotic distribution of the structural IRF estimator. In practice, to avoid estimating the (numerical) derivatives in the asymptotic distribution, we may prefer to use a bootstrap algorithm that replicates the asymptotic distribution of the IRF estimator. We present a residual-based bootstrap algorithm for this purpose. This bootstrap algorithm repeatedly resamples the reduced-form residual matrices with replacement and applies the MD estimator (Section \ref{sec:estim and props}) in each iteration. In each iteration we obtain the reduced-form and structural parameter estimates, and construct the bootstrap estimate for the IRFs. The quantiles of the bootstrap IRF estimates give the bootstrap confidence bands for the true IRF. The bootstrap algorithm is outlined below.
\begin{enumerate}
    \item Estimate the reduced-form parameters $\hat{\vpi}$ from the data $\{\mcalY_t\}$. Obtain residuals $\hat{\mcalU}_t$ for $t=1,\ldots,T$, and $\hat{\vtheta}$ by minimizing \eqref{eq:GMM_constr}.
    \item Generate a bootstrap sample $\{\mcalY^*_t\}$ with $\hat{\vpi}$, where the error term is randomly sampled with replacement from $\{\hat{\mcalU}_t\}^T_{t=1}$. 
    \item Estimate reduced-form parameters from $\hat{\vpi}_b$ from $\{\mcalY^*_t\}$.
    \item Solve the minimization problem \eqref{eq:GMM_constr} for $\hat{\vpi}_b$ and $\{\mcalY^*_t\}$. Obtain $\hat{\vtheta}_b$. 
    \item Repeat Steps 2-5 $B$ times. Obtain $\{\hat{\vtheta}_b\}^B_{b=1}$.
    \item Obtain $\alpha/2$ and $1-\alpha/2$ quantiles from $\{\hat{\vtheta}_b\}^B_{b=1}$.
\end{enumerate}
The validity of the algorithm rests on the time-series behavior of the data. Additionally, since we consider tensor data, we need to account for dependence within each mode and possibly dependence across dimensions. To capture the latter, we resample the full residual $\hat{\mcalU}_t$ instead of individual elements. This allow arbitrary cross-sectional dependence within the tensor, as dependence is preserved. We leave a theoretical justification for this bootstrap algorithm for further research.

\begin{remark}
    Our bootstrap algorithm does not account for heteroskedasticity, serial correlation, or matrix-like cross-sectional dependence. In essence, we assume here that the reduced form accounts for these. For example, by choosing the appropriate the number of lags $p$ we expect the serial dependence in the errors to be negligible. However, we acknowledge that the bootstrap algorithm can be adopted to account for more complex patterns. For heteroskedasticity, for example, one might consider a wild component to the bootstrap residuals. We leave such extensions for further research.
\end{remark}

\newpage
\section{Proofs} \label{app:proof}
\subsection{Proofs for impulse response functions}

\begin{proof}[Proof of Proposition \ref{prop:causality}]
    It is well known that a VAR$(1)$ process is stationary and stable if the spectral radius of the coefficient matrix is less than 1. The coefficient matrix for our reduced-form model is given by $\mB_{1,N}\otimes \ldots \otimes \mB_{1,1}$. Using that\\ $\rho(\mB_{1,N}\otimes \ldots \otimes \mB_{1,1}) = \rho(\mB_{1,N})\rho(\mB_{1,N-1})\cdots\rho(\mB_{1,1})$ \citepapndx[see, e.g.,][]{chen2021autoregressive, horn2012matrix}, we obtain the desired result.
\end{proof}

\subsection{Proof for global identification}
\begin{proof}[Proof of Proposition \ref{prop:global_irfs}]
    We show that the sign indeterminancy of $\{(\mA_{0,k}, \mA_{1,k})\}_k$ across modes $k=1,\ldots,N$ does not affect the uniqueness of impulse response function $\mcalF_{i_1,i_2,\ldots,i_N}(s)$ under normalization $\SZ$. Under $\SZ$, the parameters are up to a sign flip, that is, coefficient matrices $\bigotimes^N_{k=1}\mA_{l,k}$ is equivalent to $\prod^N_{k=1}\delta_{k} \bigotimes^N_{k=1}\mA_{l,k}$ for any $\delta_{k}\in \{-1,1\}$ such that $\prod^N_{k=1}\delta_{k}=1$. The structural IRFs under coefficients $\prod^N_{k=1}\delta_{l,k} \bigotimes^N_{k=1}\mB_{l,k}$ are given by
    \begin{align*}
        \mcalF_{i_1,i_2,\ldots,i_N}(s) = \mcalE_{t-s}\times^N_{k=1}\left(\mB_{k}^s\delta_k\mA_{0,k}^{-1}\right)
        &= \mcalE_{t-s}\prod^N_{k=1}\delta_k\times^N_{k=1}\left(\mB_{k}^s\mA_{0,k}^{-1}\right)\\
        &=\mcalE_{t-s}\times^N_{k=1}\left(\mB_{k}^s\mA_{0,k}^{-1}\right),
    \end{align*}
    which is the same value as the structural IRF with coefficients $\{(\mA_{0,k}, \mA_{1,k})\}_k$. Therefore, the value of $\mcalF_{i_1,i_2,\ldots,i_N}$ is unique under normalization $\SZ$ when $\vtheta$ is globally identified.
\end{proof}

\begin{proof}[Proof of Theorem \ref{thm:global_param}]
    We prove the theorem by contradiction: if model \eqref{eq:sar} is not identified at $\{(\mA_{0,k}, \mA_k)\}^N_{k=1}$, then there exists a $j$ for some $k$ such that $\mM_{j,k}(f_k(\mA^\prime_{0,k}, \mA^\prime_k))$ has rank strictly less than $n_k$. We know that if the model is not identified at $\{(\mA_{0,k}, \mA_k)\}^N_{k=1}$, there must exist a $\mP_k \in O(n_k)$ such that $\mP_k$ is not diagonal and $\left(\mA^\prime_{0,k}\mP_k, \mA^\prime_k \mP_k\right) \in \SZ$. Let $j$ denote the the index of the first column of $\mP_k$ with a non-zero off-diagonal element. Furthermore, let $p_{k,jj}$ be the $(j,j)$-th element in $\mP_k$

    Consider the vector $\vv_{j,k} = \mP_k \ve_{j} - p_{k,jj}\ve_j$, where $\ve_j$ is the $j$-th column of the identity matrix $\mI_{n_k}$. By construction, we have that $\vv_j \neq 0$. It rests us to show that $\mM_{j,k}(f_k(\mA^\prime_{0,k}, \mA^\prime_k))\vv_j=\vzeros$. We have that
    \begin{align*}
        \mM_{j,k}(f_k(\mA^\prime_{0,k}, \mA^\prime_k))\vv_j &= \mM_{j,k}(f_k(\mA^\prime_{0,k}, \mA^\prime_k)) \mP_k \ve_j - \mM_{j,k}(f_k(\mA^\prime_{0,k}, \mA^\prime_k))p_{k,jj}\ve_j\\
        &= \begin{pmatrix}
            \mQ_{j,k}f_k(\mA^\prime_{0,k}, \mA^\prime_k)\mP_{k}\ve_j\\
            [\mI_j \quad \vzeros_{j\times(n_k-j)}]\mP_k\ve_j
        \end{pmatrix} - \begin{pmatrix}
            \mQ_{j,k}f_k(\mA^\prime_{0,k}, \mA^\prime_k)\ve_j\\
            [\mI_j \quad \vzeros_{j\times(n_k-j)}]\ve_j
        \end{pmatrix}p_{k,jj}\\
        &= \begin{pmatrix}
            \vzeros_{z_k\times 1}\\
            [\mI_j \quad \vzeros_{j\times(n_k-j)}]\mP_k\ve_j
        \end{pmatrix} - \begin{pmatrix}
            \vzeros_{z_k\times 1}\\
            [\mI_j \quad \vzeros_{j\times(n_k-j)}]\ve_j
        \end{pmatrix}p_{k,jj}\\
        &= \begin{pmatrix}
            \vzeros_{z_k\times 1}\\
            p_{k,1j}\\
            \vdots\\
            p_{k,jj}
        \end{pmatrix} - \begin{pmatrix}
            \vzeros_{z_k\times 1}\\
            0\\
            \vdots\\
            0\\
            p_{k,jj}
        \end{pmatrix}=\vzeros,
    \end{align*}
    where the third equality follow from admissibilty of $f_k(\cdot)$ and the last equality from orthogonality of $\mP_k$ \citepapndx[see][Theorem 1]{rubio2010structural}. Hence, we can conclude that the rank of $\mM_{j,k}(f_k(\mA^\prime_{0,k}, \mA^\prime_k))$ is strictly less than $n_k$, which contradicts the full-rank assumption. 
\end{proof}

\subsection{Proofs for reduced-form results}
To reiterate, the reduced-form parameters are $\vpi = (\vb^\prime, \vsigma)$, where $\vb = \left(\vec(\mB_1)^\prime, \ldots, \vec(\mB_N)^\prime\right)^\prime$ and $\vsigma \coloneq \left(\vec(\mSigma^{1/2}_1)^\prime,\ldots,\vec(\mSigma^{1/2}_N)^\prime\right)^\prime$. We estimate $\vpi$ by maximum likelihood estimation. The log-likelihood $\ell(\vpi)$ is defined in \eqref{eq:ll}. The $\hat{\mSigma}_k^{1/2}$ is computed from $\hat{\mSigma}_k$ using the Cholesky decomposition. For the full computation details, see Appendix \ref{app:computation}. 

We know that the VAR$(1)$ representation of the reduced-form model is given by $\vec(\mcalY_t) = \mPhi\vec(\mcalY_{t-1}) + \vec(\mcalU_t)$, where $\mPhi = \mB_N \otimes \cdots \otimes \mB_1$ and $\mSigma = \E(\vec(\mcalU_t)\vec(\mcalU_t)^\prime) = \mSigma_N \otimes \cdots \otimes \mSigma_1$. Let $\mOmega \coloneq \mSigma^{-1} = \mOmega_N \otimes \cdots \otimes \mOmega_1$ with $\mOmega_k = \mSigma^{-1}_k$. Using the VAR$(1)$ form of the model, we can rewrite the log-likelihood \eqref{eq:ll} at parameters $(\mPhi,\mOmega)$ as
\begin{align*}
    \ell(\mPhi, \mOmega) = -\frac{T-1}{2}h(\mOmega, \mS(\mPhi)),
\end{align*}
where 
\begin{align*}
    &h(\mOmega, \mS) = -\log|\mOmega| + tr(\mOmega\mS);\\
    &\mS(\mPhi) = \frac{1}{T-1}(\vy - \mPhi \vx)(\vy - \mPhi\vx)^\prime;\\
    &\vy = \left(\vec(\mcalY_{2})^\prime, \ldots, \vec(\mcalY_{T})^\prime\right)^\prime;\\
    &\vx = \left(\vec(\mcalY_{1})^\prime, \ldots, \vec(\mcalY_{T-1})^\prime\right)^\prime
\end{align*}
The maximum-likelihood estimator $\hat{\vpi}$ maximizes $\ell(\mPhi, \mOmega)$ subject to the Kronecker structure and normalization constraints. We need an auxiliary lemma that says consistency of a Kronecker product implies consistency of the individual factors.
\begin{lemma} \label{lem:kron_to_indiv_consist}
     Let $\mPsi = \bigotimes^N_{k=1}\mC_k$ for $\mC_k \in \SR^{n_k \times n_k}/\{\vzeros\}$, $k=1,\ldots,N$. Assume $\mC_k$ are normalized such that $||\mC_k||_F=1$ for $k=1,\ldots,N-1$. Let $\hat{\mC}_k$ satisfy the same normalization and set $\hat{\mPsi}=\bigotimes^N_{k=1}\hat{\mC}_k$. Then, it holds that
    \begin{enumerate}[label=(\roman*)]
        \item if $\hat{\mPsi} \inp \mPsi$ and the same sign normalization is used on $\hat{\mC}_k$ and $\mC_k$, then $\hat{\mC}_k \inp \mC_k$, for $k=1\ldots,N$.
        \item  if $\mC_k$ is nonsingular, for each $k=1,\ldots,N$, then
        there exist some constant $c>0$ and a neighborhood of $\left(\mC_1,\ldots,\mC_N\right)$ such that
        \begin{align*}
        ||\hat{\mPsi}-\mPsi||_F \geq c\sum^N_{k=1}||\hat{\mC}_k - \mC_k||_F.
    \end{align*}
    \end{enumerate}
\end{lemma}

\begin{proof}[Proof of Lemma \ref{lem:kron_to_indiv_consist}]
    We first prove result (i). Since the Frobenius norm of a Kronecker product is the product of Frobenius norms, we have that
    \begin{align*}
        ||\hat{\mPsi}||_F = \prod^N_{k=1}||\hat{\mC}_k||_F.
    \end{align*}
    The first $N-1$ norms are normalized to one, so it holds that
    \begin{align*}
        ||\hat{\mPsi}||_F = ||\hat{\mC}_N||_F.
    \end{align*}
    Moreover, if $\hat{\mPsi}\inp \mPsi$, this implies that $||\hat{\mPsi}||_F$ is bounded, and hence, $||\hat{\mC}_N||$ is bounded. Therefore, any subsequence of $(\hat{\mC}_1, \ldots, \hat{\mC}_N)$ has a further subsequence for which $\hat{\mC}_k \inp \mC^*_k$. By the continuous mapping theorem, it must hold that $\mC^*_N\otimes \cdots \otimes \mC^*_1 = \mC_N\otimes \cdots \otimes \mC_1$. Furthermore, the norm and sign normalizations give uniqueness, since $\mC_k\neq \vzeros$, $\mC^*_k$ and $\mC_k$ may only differ through rescaling. Therefore, every converging subsequence has the unique limit $(\mC_1,..., \mC_N)$, so that $\hat{\mC}_k \inp \mC_k$ for every $k=1,\ldots,N$. This shows the result in (i).
    
    To prove result (ii), we define the map $\mF(\mC_1,\ldots,\mC_N) = \bigotimes^N_{k=1}\mC_k$. Furthermore, let $\mDelta_k \equiv \hat{\mC}_k - \mC_k$. By definition, we have that
    \begin{align*}
        \mPsi &= \mF(\mC_1, \ldots,\mC_N)\\
        \hat{\mPsi} &= \mF(\mC_1 + \mDelta_1,\ldots,\mC_N + \mDelta_N).
    \end{align*}
    Expanding the product of sums in $\hat{\mPsi}$ we get that
    \begin{align*}
        \hat{\mPsi} = \bigotimes^N_{k=1}\mC_k + \sum^N_{k=1}\mC_N\otimes \cdots \otimes \mDelta_k \otimes \cdots \otimes \mC_1 + \mR,
    \end{align*}
    where $\mR$ denotes a remainder term that holds all the products where two or more $\mDelta_k$'s appear. Hence, we have that
    \begin{align*}
        \hat{\mPsi} - \mPsi = \sum^N_{k=1}\mC_N\otimes \cdots \otimes \mDelta_k \otimes \cdots \otimes \mC_1 + \mR.
    \end{align*}
    The first term is linear in the $\mDelta_k$'s. We can show that the kernel of the first term only consists of rescalings $\mDelta_k = \alpha_k \mC_k$ with $\sum^N_{k=1}\alpha_k=0$. To do so, suppose $\sum^N_{k=1}\mC_N\otimes \cdots \otimes \mDelta_k = \vzeros$. Since $\mC_k$ is nonsingular for all $k=1,\ldots,N$, we can premultiply the left by $\bigotimes^N_{k=1} \mC^{-1}_k$ to obtain
    \begin{align*}
        \sum^N_{k=1} \mI\otimes \cdots \otimes \mE_k \otimes \cdots \otimes \mI = \vzeros,
    \end{align*}
    where $\mE_k \equiv \mC_k^{-1}\mDelta_k$. First, every off-diagonal element of $\mE_k$ must be zero. Second, along the diagonal of $\mE_k$ the entries must be equal and must cancel across modes for the diagonal equation to be equal to zero. Therefore, it must hold that $\mE_k = \alpha_k \mI$ for $\sum^N_{k=1}\alpha_k=0$. This implies that the linear map $\sum^N_{k=1}\mC_N\otimes \cdots \otimes \mDelta_k \otimes \cdots \otimes \mC_1$ is not injective on the entire space. 

    We now show that the directions that make the linear map not injective are eliminated after restricting to the tangent space of the normalization constraints. We define the tangent space at $(\mC_1, \ldots, \mC_N)$ as $\mathcal{T} \equiv \{(\tilde{\mDelta}_1, \ldots,\tilde{\mDelta}_N):\langle\mC_k, \tilde{\mDelta}_k\rangle_F = 0, k<N\}$. If $(\tilde{\mDelta}_1,\ldots,\tilde{\mDelta}_N)\in\mathcal{T}\cap \mathrm{kernel}\left(\sum^N_{k=1}\mC_N\otimes \cdots \otimes \mDelta_k \otimes \cdots \otimes \mC_1\right)$, then the kernel implies that $\tilde{\mDelta}_k = \alpha_k \mC_k$ and $\mathcal{T}$ implies that $\langle\mC_k, \tilde{\mDelta}_k\rangle_F = 0$. Therefore, it must hold that
    \begin{align*}
        0=\langle\mC_k, \alpha_k \mC_k\rangle_F = \alpha_k||\mC_k||_F^2,
    \end{align*}
    which holds if and only if $\alpha_k = 0$ for $k=1,\ldots,N-1$ under the assumption that $||\mC_k||_F=1$. Since in the kernel it holds that $\sum^N_{k=1}\alpha_k=0$, it must hold that $\alpha_N=0$. Therefore,
    \begin{align*}
         \mathcal{T}\cap \mathrm{kernel}\left(\sum^N_{k=1}\mC_N\otimes \cdots \otimes \mDelta_k \otimes \cdots \otimes \mC_1\right) = \{0\}.
    \end{align*}

 Thus, the linear map $\sum^N_{k=1}\mC_N\otimes \cdots \otimes \mDelta_k \otimes \cdots \otimes \mC_1$ is injective on the tangent space to the normalization constraints at $(\mC_1,\ldots, \mC_N)$. Injectivity implies that for each $(\tilde{\mDelta}_1,\ldots,\tilde{\mDelta}_N)\in\mathcal{T}$,
    \begin{align*}
        ||\sum^N_{k=1}\mC_N\otimes \cdots \otimes \tilde{\mDelta}_k \otimes \cdots \otimes \mC_1||_F \geq c_0 \sum^N_{k=1}||\tilde{\mDelta}_k||_F,
    \end{align*}
    for some constant $c_0>0$. Since $||\mDelta_k + \mC_k||_F=||\mC_k||_F=1$, $k<N$, we know that
    \begin{align*}
        \langle \mC_k, \mDelta_k\rangle_F = -||\mDelta_k||^2_F/2.
    \end{align*}
    Let $r_k \equiv  \langle \mC_k, \mDelta_k\rangle_F$. Set $\tilde{\mDelta}_k = \mDelta_k - r_k\mC_k$ for $k=1,\ldots,N-1$, and $\tilde{\mDelta}_N=\mDelta_N$. By definition, it holds that $(\tilde{\mDelta}_1,\ldots,\tilde{\mDelta}_N)\in\mathcal{T}$. It follows that
    \begin{align*}
        \sum^N_{k=1}||\mDelta_k - \tilde{\mDelta}_k||_F &= \sum^{N-1}_{k=1}|r_k|||\mC_k||_F\\
        &=\sum^{N-1}_{k=1}||\mDelta_k||^2_F\\
        &\leq C_1 \left(\sum^N_{k=1}||\mDelta_k||_F\right)^2
    \end{align*}
    by the triangle inequality for some constant $C_1>0$. This also implies that $\sum^N_{k=1}||\tilde{\mDelta}_k||_F \geq \sum^N_{k=1}||\mDelta_k||_F - C_1\left(\sum^N_{k=1}||\mDelta_k||_F\right)^2$ By the linearity of $\sum^N_{k=1}\mC_N\otimes \cdots \otimes \mDelta_k \otimes \cdots \otimes \mC_1$ we get that
    \begin{align*}
        &||\sum^N_{k=1}\mC_N\otimes \cdots \otimes \mDelta_k \otimes \cdots \otimes \mC_1||_F\\
        &=||\sum^N_{k=1}\mC_N\otimes \cdots \otimes \tilde{\mDelta}_k \otimes \cdots \otimes \mC_1 + \sum^N_{k=1}\mC_N\otimes \cdots \otimes (\mDelta_k - \tilde{\mDelta}_k) \otimes \cdots \otimes \mC_1||_F\\
        &\geq c_0\sum^N_{k=1}||\tilde{\Delta}_k||_F + ||\sum^N_{k=1}\mC_N\otimes \cdots \otimes (\mDelta_k - \tilde{\mDelta}_k) \otimes \cdots \otimes \mC_1||_F\\
        &\geq c_0\left(\sum^N_{k=1}||\mDelta_k||_F - C_1\left(\sum^N_{k=1}||\mDelta_k||_F\right)^2\right) - C_1\left(\sum^N_{k=1}||\mDelta_k||_F\right)^2\\
        &\geq c_0\sum^N_{k=1}||\mDelta_k||_F - C_2 (\sum^N_{k=1}||\mDelta_k||_F)^2.
    \end{align*}
    
    For the remainder term $\mR$, we have that it contains at least two differences, say $\mDelta_i$ and $\mDelta_j$. Since the fixed $\mC_k$ are bounded and $N$ is finite, 
    \begin{align*}
        ||\mR||_F \leq C_3\left(\sum^N_{k=1}||\mDelta_k||_F\right)^2
    \end{align*}
    in a sufficiently small neighborhood, for some constant $C_3$. Combining this result with the lower bound on the linear term, we obtain
    \begin{align*}
        ||\hat{\mPsi}-\mPsi||_F \geq c_0\sum^N_{k=1}||\hat{\mC}_k - \mC_k||_F - C\left(\sum^N_{k=1}||\mDelta_k||_F\right)^2.
    \end{align*}
    Let $\varepsilon = \sum^N_{k=1}||\mDelta_k||_F$. Then
    \begin{align*}
         ||\hat{\mPsi}-\mPsi||_F \geq \varepsilon(c_0 - C\varepsilon).
    \end{align*}
    Choose the neighborhood small enough that $\varepsilon < c_0/(2C)$ so that
    \begin{align*}
         ||\hat{\mPsi}-\mPsi||_F \geq c \sum^N_{k=1}||\hat{\mC}_k - \mC_k||_F,
    \end{align*}
    where $c=c_0/2$. This proves the result.
\end{proof}

\subsubsection{Step 1: asymptotic distribution of coefficients}
In the first step, we establish the asymptotic distribution of the coefficient matrices $\mA_k$, $k=1,\ldots,N$. To do so, we need to generalize Lemma 8 in \citetapndx{chen2021autoregressive}.

\begin{lemma}\label{lem:rf_uni_bound}
    Consider the VAR$(1)$ representation of model \eqref{eq:rf-sar} with true coefficients $\mPhi = \mA_N \otimes \cdots \otimes \mA_1$. Suppose Assumptions \ref{a1:iid}-\ref{a3:nonsing} hold. Then, for any sequence ${c_T}$ with $c_T\rightarrow\infty$,
    \begin{align*}
        \SP\Bigg[\inf_{\sqrt{T}||\Bar{\mPhi} - \mPhi||_F > c_T}\sum^T_{t=2}||\vec(\mcalY_t) - \Bar{\mPhi}\vec(\mcalY_{t-1})||^2 \leq \sum^T_{t=2}||\vec(\mcalU_{t})||^2\Bigg] \rightarrow 0.
    \end{align*}
\end{lemma}

\begin{proof}[Proof Lemma \ref{lem:rf_uni_bound}]
    From Proposition \ref{prop:causality}, it follows that the autocovariance matrices are given by
    \begin{align*}
        \cov(\vec(\mcalY_t), \vec(\mcalY_{t-l})) = \sum^{\infty}_{m=0}\left(\mB_N^{l+m}\otimes \cdots \otimes \mB_1^{l+m}\right)\left(\mSigma_N\otimes \cdots \otimes \mSigma_1\right)\left(\mB_N^m \otimes \cdots \otimes \mB_1^m\right)^\prime, l\geq 0,
    \end{align*}
    and are absolutely summable. Let $\mGamma_0 \equiv  \cov(\vec(\mcalY_t), \vec(\mcalY_{t}))$. The ergodic theorem implies that 
    \begin{align*}
        \frac{1}{T}\sum^T_{t=2}\vec(\mcalY_{t-1})\vec(\mcalY_{t-1})^\prime \rightarrow \mGamma_0 \quad a.s
    \end{align*}
    Therefore, for any constant $c>0$,
    \begin{align*}
        \sup_{\sqrt{T}||\Bar{\mPhi} - \mPhi||_F \leq c}\left|\sum^T_{t=2}\tr\left((\Bar{\mPhi}-\mPhi)\vec(\mcalY_{t-1})\vec(\mcalY_{t-1})^\prime(\Bar{\mPhi}-\mPhi)^\prime\right) - T\tr\left((\Bar{\mPhi}-\mPhi)\mGamma_0(\Bar{\mPhi}-\mPhi)^\prime\right)\right| \rightarrow 0 \quad a.s.
    \end{align*}
    The supremum over $\Bar{\mPhi}$ implies that there exists a sequence $\{\bar{c}_T\}$ such that $\bar{c}_T \rightarrow \infty$, $\bar{c}_T \leq c_T$, and 
    \begin{align*}
        \sup_{\sqrt{T}||\Bar{\mPhi} - \mPhi||_F \leq \bar{c}_T}\left|\sum^T_{t=2}\tr\left((\Bar{\mPhi}-\mPhi)\vec(\mcalY_{t-1})\vec(\mcalY_{t-1})^\prime(\Bar{\mPhi}-\mPhi)^\prime\right) - T\tr\left((\Bar{\mPhi}-\mPhi)\mGamma_0(\Bar{\mPhi}-\mPhi)^\prime\right)\right| \inp 0.
    \end{align*}
    Following the arguments in \citetapndx{chen2021autoregressive}, we obtain that, on the boundary $\sqrt{T}||\Bar{\mPhi} - \mPhi||_F = \bar{c}_T$, we have that
    \begin{align*}
        &\sum^T_{t=2}||\vec(\mcalY_t) - \Bar{\mPhi}\vec(\mcalY_{t-1})||^2 - \sum^T_{t=2}||\vec(\mcalU_{t})||^2\\
        &= -2 \sum^T_{t=2}\tr\left((\Bar{\mPhi}-\mPhi)\vec(\mcalY_{t-1})\vec(\mcalU_t)^\prime\right) + \sum^T_{t=2}\tr\left((\Bar{\mPhi}-\mPhi)\vec(\mcalY_{t-1})\vec(\mcalY_{t-1})^\prime(\Bar{\mPhi}-\mPhi)^\prime\right)\\
        &= O_p(\bar{c}_T) + \sum^T_{t=2}\tr\left((\Bar{\mPhi}-\mPhi)\vec(\mcalY_{t-1})\vec(\mcalY_{t-1})^\prime(\Bar{\mPhi}-\mPhi)^\prime\right)\\
        &\geq O_p(\bar{c}_T) + \lambda_{\min}(\mGamma_0)\bar{c}_T^2,
    \end{align*}
    where $\lambda_{\min}(\mGamma_0)$ denotes the minimum eigenvalue of $\mGamma_0$. Under Assumption \ref{a3:nonsing}, $\lambda_{\min}(\mGamma_0)$ is strictly positive. Therefore, we have that for $\bar{c}_T\rightarrow\infty$,
    \begin{align*}
        \SP\Bigg[\inf_{\sqrt{T}||\Bar{\mPhi}-\mPhi||_F = \bar{c}_T}\sum^T_{t=2}||\vec(\mcalY_{t}) - \Bar{\mPhi}\vec(\mcalY_{t-1})||^2 \leq \sum^T_{t=2}||\vec(\mcalU_{t})||^2\Bigg] \rightarrow 0.
    \end{align*}
    Convexity of $\sum^T_{t=2}||\vec(\mcalY_{t}) - \Bar{\mPhi}\vec(\mcalY_{t-1})||^2$ with respect to $\Bar{\mPhi}$ yields the desired result.
\end{proof}

\begin{lemma} \label{lem:coeff_consist}
    Suppose the assumptions in \ref{lem:rf_uni_bound} hold. Then, if $T\rightarrow\infty$,
    \begin{align*}
        \hat{\mB}_k = \mB_k + O_p(T^{-1/2}), \quad \text{for }k=1,\ldots,N.
    \end{align*}
\end{lemma}

\begin{proof}[Proof of Lemma \ref{lem:coeff_consist}]
    For a given $\mOmega$, $h(\mOmega, \mS)$ is minimized at $\mS = \Tilde{\mS} = \mS(\Tilde{\mPhi})$ with $\Tilde{\mPhi} = \vy \vx^\prime(\vx\vx^\prime)^{-1}$. The corresponding minimum value of $h$ is $h(\mOmega, \Tilde{\mS})$, for a given $\mOmega$. Let $\Tilde{\mOmega} = \Tilde{\mS}^{-1}$. To show the root-$T$ consistency of the coefficient estimator, we first need to establish consistency of the variance estimators. Lemma \ref{lem:kron_to_indiv_consist} implies that in order to establish that $\hat{\mSigma}_k = \mSigma_k + o_p(1)$  it suffices to show that for any constant $c>0$ we have
    \begin{align}
        \SP\left(\inf_{||\bar{\mOmega} - \mOmega||_F > c}h(\bar{\mOmega}, \Tilde{\mS})\leq h(\mOmega, \mS(\mPhi))\right) \rightarrow 0.
    \end{align}
    The results follows from a multivariate Taylor expansion of $h$ and the ergodic theorem. We refer to page 50 in the supplement of \citetapndx{chen2021autoregressive} for the technical details.

    Now we have established that $\hat{\mSigma}_k = \mSigma_k + o_p(1)$ for all $k=1,\ldots,N$, we can establish $\hat{\mB}_k = \mB_k + O_p(T^{-1/2})$ for all $k$. Let $\SH^{\mSigma}$ denote the set of all positive definite matrices of the tensor product form:
    \begin{align*}
        \SH^\Sigma = \{\mSigma: \mSigma = \mSigma_N\otimes \cdots \otimes \mSigma_1 \text{ for some }n_k\times n_k \text{ positive definite matrices }\mSigma_k, k=1,\ldots,N\}.
    \end{align*}
    Following Lemma \ref{lem:kron_to_indiv_consist}, in order to show $\hat{\mB}_k = \mB_k + O_p(T^{-1/2})$ for all $k$ it suffices to show that for any sequence $c_T\rightarrow \infty$,
    \begin{align}
        \SP\left(\inf_{\sqrt{T}||\bar{\mPhi} - \mPhi||\geq c_T}\inf_{\bar{\mOmega}\in \SH^\Sigma}h(\bar{\mOmega}, \mS(\bar{\mPhi}))\leq \inf_{\bar{\mOmega}\in\SH^\Sigma}h(\bar{\mOmega}, \mS(\mPhi))\right)\rightarrow 0.
    \end{align}
    By the same technical arguments of \citetapndx[page 51-52]{chen2021autoregressive}, this result can be established directly.
\end{proof}

\begin{lemma}\label{lem:coeff_distr}
     Suppose the assumptions in Lemma \ref{lem:rf_uni_bound} hold. Define $\vb = (\vec(\mB_1)^\prime, \ldots, \vec(\mB_N)^\prime)^\prime$, and $\hat{\vb}$ similarly. Then, if $T\rightarrow\infty$,
    \begin{align*}
        \sqrt{T}\left(\hat{\vb} - \vb\right) \ind N(0,\mXi_1),
    \end{align*}
    where
    \begin{align}
        \mXi_1 = \mH^{-1} \E(\mD_t\prime\mOmega \mD_t)\mH^{-1},
    \end{align}
    with $\mH = \E(\mD_t\prime\mOmega \mD_t) + \mGamma\mGamma^\prime$, $\mD_t = \partial \mPhi \vy_{t-1}/\partial \vb^\prime$, and $\mGamma = (\vgamma_1,\ldots,\vgamma_{N-1})$ with columns $\vgamma_k= \vec(0,\ldots,\vec(\mB_k)^\prime,\ldots,0)^\prime$.
\end{lemma}

\begin{proof}[Proof of Lemma \ref{lem:coeff_distr}]
The gradient condition for $\hat{\mB}_k$ \eqref{eq:rf_mle_gradient} at the MLE is given by
    \begin{align}
\hat{\mB}_k\left(\sum^T_{t=2}\mY_{t-1(k)}\hat{\mM}_k\mY_{t-1(k)}^\prime\right) -  \sum^T_{t=2}\mY_{t(k)}\hat{\mOmega}_{-k}\hat{\mB}_{-k}\mY_{t-1(k)}^\prime = \mathbf{0}{n_k\times n_k}.
\end{align}
Since this condition holds for all $k$, we have that the score of the log-likelihood at the MLE is zero. Let $\dot{\mL}(\vb, \mOmega)$ denote the score function at some coefficient value $\vb$ and covariance matrix $\mSigma$. We have that the score $\dot{\mL}(\hat{\vb}, \hat{\mOmega}) = \vzeros$ at the MLE. A first-order Taylor expansion in $\va$ gives that
\begin{align} \label{eq:lem4_foc_taylor}
    \vzeros = \dot{\mL}(\hat{\vb}, \hat{\mOmega}) = \dot{\mL}(\vb, \mOmega) + \frac{\partial \dot{\mL}(\vb, \mOmega)}{\partial \vb}(\hat{\vb} - \vb) + r_T,
\end{align}
where $r_T = o_p(T^{1/2})$ is a remainder term. The rate of the remainder term is determined by the rate in Lemma \ref{lem:coeff_consist}. By defining $\mD_t \equiv \partial \mPhi \vy_{t-1}/\partial \vb^\prime$, we can write $\dot{\mL}(\vb, \mOmega) = \sum^T_{t=2}\mD_t \mOmega \ve_t$ and 
\begin{align*}
    \frac{\partial \dot{\mL}(\vb, \mOmega)}{\partial \vb} &= -\sum^T_{t=2}\mD_t \mOmega \mD_t + \sum^T_{t=2}\frac{\partial \mD_t}{\partial \vb^\prime}\mOmega \vu_t\\
    &= -\sum^T_{t=2}\mD_t \mOmega \mD_t + o_p(T^{1/2}),
\end{align*}
where the second equality follows from Lemma \ref{lem:coeff_consist}. Similarly, \citetapndx{chen2021autoregressive} substitute the true process into the first-order conditions and conclude that the leading term is linear in the estimation error and the higher order terms are absorbed in the $o_p(T^{1/2})$ remainder. 

Plugging the terms back in \eqref{eq:lem4_foc_taylor}, we get that 
\begin{align*}
    T^{-1}\sum^T_{t=2}\mD_t^\prime\mOmega \mD_t (\hat{\vb} - \vb) = T^{-1}\sum^T_{t=2}\mD_t^\prime\mOmega \vu_t  -  + o_p(T^{-1/2}).
\end{align*}
First, by the martingale central limit theorem  \citepapndx{hall2014martingale}, we have that
\begin{align*}
    T^{-1/2}\sum^T_{t=2}\mD_t^\prime\mOmega \vu_t \ind \calN(\vzeros, \E(\mD_t\prime\mOmega \mD_t)),
\end{align*}
where $\mOmega\mSigma\mOmega = \mOmega$ since $\mOmega = \mSigma^{-1}$. Second, since $\mcalY_t$ is strictly stationary and ergodic with i.i.d. innovations, the ergodic theorem implies that
\begin{align*}
    T^{-1}\sum^T_{t=2}\mD_t^\prime \mOmega \mD_t \inp \E(\mD_t ^\prime\mOmega\mD_t) \quad \text{a.s}.
\end{align*}
However, $\E(\mD_t ^\prime\mOmega\mD_t)$ is not full rank due to the $N-1$ arbitrary rescalings of the Kronecker factors. This is analogous to the reduced-rank result in \citetapndx{chen2021autoregressive} when $N=2$. Define matrix $\mGamma \equiv (\vgamma_1,\ldots,\vgamma_{N-1})$ where each column $\vgamma_k \equiv (0, \ldots,\vec(\mB_k)^\prime,\ldots,0)^\prime$. By Lemma \ref{lem:coeff_consist}, it holds that $\mGamma^\prime (\hat{\vb}-\vb) = o_p(T^{-1/2})$. Therefore, we can define $\mH = \E(\mD_t ^\prime\mOmega\mD_t) + \mGamma \mGamma^\prime$, and obtain that
\begin{align*}
    \sqrt{T}(\hat{\vb} - \vb) \ind \calN(\vzeros,\mH^{-1} \E(\mD_t\prime\mOmega \mD_t)\mH^{-1}),
\end{align*}
where we know that $\mH$ is nonsingular due to Lemma \ref{lem:kron_to_indiv_consist}.

\end{proof}

\subsubsection{Step 2: asymptotic distribution of variance estimates}
In the second step, we establish the asymptotic distribution of the estimators of the Cholesky-decomposed variance matrices $\mSigma_k$, $k=1,\ldots,N$. To do so, we need an auxiliary lemma.

\begin{lemma}\label{lem:sigma_unif_bound}
    Consider the VAR$(1)$ representation of model \eqref{eq:rf-sar} with true coefficients $\mPhi$ and true covariance matrix $\mSigma = \mSigma_N \otimes \cdots \otimes \mSigma_1$. Define $\mOmega \equiv \mSigma^{-1} = \mSigma_N^{-1} \otimes \cdots \otimes \mSigma^{-1}_1$. Let $\tilde{\mS} \equiv (\vy - \tilde{\mPhi}
    \vx)(\vy - \tilde{\mPhi}\vx)^\prime/(T-1)$ with $\tilde{\mPhi} \equiv \vy\vx^\prime(\vx\vx^\prime)^{-1}$. Suppose Assumptions \ref{a1:iid}-\ref{a3:nonsing} hold. Then, for any sequence ${c_T}$ with $c_T\rightarrow\infty$,
    \begin{align*}
        \SP\Bigg[\inf_{\sqrt{T}||\Bar{\Omega} - \mOmega||_F \geq c_T}h(\bar{\mOmega}, \tilde{\mS}) \leq h(\mOmega, S(\mPhi))\Bigg]\rightarrow 0.
    \end{align*}
\end{lemma}
\begin{proof}[Proof of Lemma \ref{lem:sigma_unif_bound}]
    Let $\tilde{\mOmega} \equiv \tilde{\mS}^{-1}$. By Proposition (ii) in \citetapndx{chen2021autoregressive}, the first-order Taylor expansion of $h(\bar{\mOmega}, \tilde{\mS})$ around $\mOmega$ gives
    \begin{align}\label{eq:auxlem_sigma_unif}
        h(\bar{\mOmega}, \tilde{\mS}) - h(\tilde{\mOmega}, \tilde{\mS}) = \tr\left((\mSigma - \tilde{\mS})(\bar{\mOmega}-\mOmega)\right) + \frac{1}{2}\left(\vec(\bar{\mOmega} - \mOmega)\right)^\prime\left(\mOmega^{-1}\otimes \mOmega^{-1}\right)\vec(\bar{\mOmega}-\mOmega).
    \end{align}
For the first term on the right-hand side in \eqref{eq:auxlem_sigma_unif} we have that $\mSigma - \tilde{\mS} = \mSigma - \mS(\mPhi) + \mS(\mPhi) - \tilde{\mS}$. By Assumption \ref{a1:iid}, we know that $\vu_t$ has finite fourth moments, such that each term in  
\begin{align*}
    \mSigma - \mS(\mPhi) = \frac{1}{T-1}\sum^T_{t=2}(\mSigma - \vu_t \vu_t^\prime)
\end{align*},
has finite variance, and thus, by the central limit theorem we have that $\sqrt{T}\vec(\mS(\mPhi) - \mSigma) = O_p(1)$. In other words, we have that
\begin{align*}
    \mS(\mPhi) - \mSigma = O_p(T^{-1/2}).
\end{align*}
Moreover, for the term $\mS(\mPhi) - \tilde{\mS}$ we can apply standard VAR theory for the unrestricted estimator $\tilde{\mS}$ that $\tilde{\mPhi} - \mPhi = O_p(T^{-1/2})$, and thus,
\begin{align*}
    \tilde{\mS} - \mS(\mPhi) = -(\tilde{\mPhi} - \mPhi)\left(\frac{\vx_t\vx_t^\prime}{T-1}\right)(\tilde{\mPhi}-\mPhi)^\prime = O_p(T^{-1}),
\end{align*}
since no multicollinearity implies $\vx_t\vx_t^\prime/(T-1) = O_p(1)$. Therefore, we have that $\mSigma - \tilde{\mS} = O_p(T^{-1/2})$. On the boundary $\sqrt{T}||\bar{\mOmega} - \mOmega||_F = c_T$, we obtain that
\begin{align*}
    |\tr\left((\mSigma - \tilde{\mS})(\bar{\mOmega}-\mOmega)\right)| = O_p(c_T/T).
\end{align*}
For the second quadratic term on the right-hand side in \eqref{eq:auxlem_sigma_unif}, we use nonsingularity of $\mOmega$ to obtain that
\begin{align*}
    \lambda_{min}(\mOmega^{-1} \otimes \mOmega^{-1}) = \lambda_{min}^2(\mOmega^{-1})>0.
\end{align*}
Therefore, on the boundary $||\bar{\mOmega} - \mOmega||_F=c_T$ we obtain that
\begin{align*}
    \frac{1}{2}\left(\vec(\bar{\mOmega} - \mOmega)\right)^\prime\left(\mOmega^{-1}\otimes \mOmega^{-1}\right)\vec(\bar{\mOmega}-\mOmega) = \frac{1}{2}\lambda^2_{min}(\mOmega^{-1})c^2_T = O_p(\frac{c^2_T}{T}).
\end{align*}
The deterministic curvature (second term in \eqref{eq:auxlem_sigma_unif}) dominates the stochastic term (first term in \eqref{eq:auxlem_sigma_unif}). Furthermore, since $\tilde{\mS} - \mS(\mPhi) = O_p(T^{-1})$, we have that 
\begin{align*}
    h(\mOmega, \tilde{\mS}) - h(\mOmega, \mS(\mPhi)) = \tr\left(\mOmega(\tilde{\mS} - \mS(\mPhi))\right) = O_p(T^{-1}).
\end{align*}
This term is negligible relative to $c^2_T/T$ for $c_T\rightarrow \infty$. It follows that, with $c_T\rightarrow\infty$, we have that
\begin{align*}
    \SP\Bigg[\inf_{\sqrt{T}||\Bar{\Omega} - \mOmega||_F = c_T}h(\bar{\mOmega}, \tilde{\mS} \leq h(\mOmega, S(\mPhi))) \rightarrow 0.
\end{align*}
By the convexity of $h(\bar{\mOmega}, \tilde{\mS})$ as a function of $\bar{\mOmega}$ \citealpapndx[Proposition 9]{chen2021autoregressive}, we can conclude the proof.
\end{proof}

\begin{lemma}\label{lem:sigma_consist}
    Let the assumptions in \ref{lem:rf_uni_bound} hold. Then, if $T\rightarrow\infty$,
    \begin{align*}
        \hat{\mSigma}_k = \mSigma_k + O_p(T^{-1/2}), \quad \text{for }k=1,\ldots,N
    \end{align*}
\end{lemma}

\begin{proof}[Proof of Lemma \ref{lem:sigma_consist}]
    Let $\SH_{\mSigma} \equiv \{\mSigma: \mSigma = \mSigma_N \otimes \cdots \otimes \mSigma_1 \text{ is an }n\times n \text{ matrix, with } ||\mSigma_k||_F=1 \text{ for }k=1,\ldots, N-1; \mSigma_k \text{positive definite for all }k=1,\ldots,N \}$. Let $\{c_T\}$ be any sequence such that $c_T\rightarrow\infty$ and $c_T/\sqrt{T}\rightarrow 0$. For $\mSigma, \bar{\mSigma}\in \SH_{\mSigma}$ and $T\sum^N_{k=1}||\bar{\mOmega}_k - \mOmega_k||^2_F \geq c^2_T$, then $||\bar{\mOmega}-\mOmega||_F \geq C c_T$ for some constant $C>0$. Lemma \ref{lem:sigma_unif_bound} yields that
    \begin{align*}
        \SP\Bigg[\min_{T\sum^N_{k=1}||\Bar{\Omega}_k - \mOmega_k||_F \geq c^2_T}h(\bar{\mOmega}, \tilde{\mS} \leq h(\mOmega, S(\mPhi))) \rightarrow 0,
    \end{align*}
    for $\bar{\mSigma}\in\SH_{\mSigma}$. Therefore, we obtain that
    \begin{align}\label{eq:aux_sigma_const}
        \SP\Bigg[T\sum^N_{k=1}||\hat{\Omega}_k - \mOmega_k||_F \leq c^2_T\Bigg] \rightarrow 1,
    \end{align}
    also requiring that $\hat{\mSigma}\in\SH_{\mSigma}$. Equation \eqref{eq:aux_sigma_const} holds for all sequences $\{c_T\}$ such that $c_T\rightarrow\infty$ and $c_T/\sqrt{T} \rightarrow 0$, and hence, it holds that
    \begin{align*}
        \hat{\mOmega}_k = \mOmega_k + O_p(T^{-1/2}),
    \end{align*}
    for all $k=1,\ldots,N$. Using that $\hat{\mSigma}_k - \mSigma_k = \hat{\mSigma}_k\left(\mSigma^{-1}_k - \hat{\mSigma}^{-1}_k\right)\mSigma_k$, the desired result holds.
\end{proof}

\begin{lemma}\label{lem:sigma_distr}
Let $\mL_k \mL_k^\prime = \mSigma_k$ with $\mL_k$ a $n_k\times n_k$ denote a lower triangular matrix with positive diagonal elements, for all $k$. Let $\vell = \left(\vech(\mL_1)^\prime,\ldots, \vech(\mL_N)^\prime\right)^\prime$. Let $\mD_m$ denote the duplication matrix such that $\vec(\mA) = \mD_m\vech(\mA)$ for a symmetric matrix $\mA\in\mathbb{R}^{m\times m}$, and let $\mD^+_m = (\mD_m^\prime\mD_m)^{-1}\mD_m^\prime$ denote its pseudoinverse. Let the assumptions in \ref{lem:rf_uni_bound} hold. Then it holds that
    \begin{align*}
        \sqrt{T}(\hat{\vell} - \vell) \ind \calN(\vzeros, \mXi_2),
    \end{align*}
    where $\mXi_2 = \mH \mV \mH^\prime$, $\mH = \diag\left(\mD^+_{n_k}(\mI\otimes \mL_k)\mG_{n_k}(\mL^{-1}_k \otimes \mL^{-1}_k)\mD_{n_k}\right)$, and $\mV =  \tilde{\mW}^{-1}_1 \mW_2 \tilde{\mW}^{-1}_1$. 
\end{lemma}

\begin{proof}[Proof of Lemma \ref{lem:sigma_distr}]
    The MLE $\hat{\mSigma}_k$ satisfies the first-order condition in \eqref{eq:rf_mle_gradient} such that 
    \begin{align*}
        \hat{\mSigma}_k = \frac{n_k}{n(T-1)}\sum^T_{t=2}\hat{\mU}_{t(k)}\hat{\mOmega}_{-k}\hat{\mU}^\prime_{t(k)}.
    \end{align*}
    We define the auxiliary estimator $\tilde{\mSigma}_k$ that replaces the estimated residuals $\hat{\mU}_{t(k)}$ with the infeasible true errors $\mU_{t(k)}$,
    \begin{align*}
       \tilde{\mSigma}_k = \frac{n_k}{n(T-1)}\sum^T_{t=2}\mU_{t(k)}\hat{\mOmega}_{-k}\mU^\prime_{t(k)}.
    \end{align*}
    For all $k=1,\ldots,N$, we can write
    \begin{align*}
        \hat{\mSigma}_k - \mSigma_k &= \hat{\mSigma}_k - \tilde{\mSigma}_k + \tilde{\mSigma}_k - \mSigma_k\\
        &= o_p(T^{-1/2}) + \tilde{\mSigma}_k - \mSigma_k,
    \end{align*}
    where the final equality follows from Lemma \ref{lem:coeff_consist}. We now consider $\tilde{\mSigma}_k - \mSigma_k$. By Lemma \ref{lem:sigma_consist} we obtain that for all $j=1,\ldots,N$
    \begin{align*}
        \hat{\mOmega}_j = \mOmega_j - \mOmega_j\left(\hat{\mSigma}_j - \mSigma_j\right)\mOmega_j + o_p(T^{-1/2}), 
    \end{align*}
    which implies that
    \begin{align}\label{eq:omega-k_const}
        \hat{\mOmega}_{-k} = \mOmega_{-k} - \sum_{j\neq k}\mOmega_N\otimes \cdots \otimes \mOmega_j\left(\hat{\mSigma}_j - \mSigma_j\right)\mOmega_j\otimes \cdots \mOmega_1 + o_p(T^{-1/2}).
    \end{align}
    Using the result in \eqref{eq:omega-k_const}, we get that
    \begin{align*}
        \tilde{\mSigma}_k - \mSigma_k &= \frac{n_k}{n(T-1)}\sum^T_{t=2}\left(\mU_{t(k)}\hat{\mOmega}_{-k}\mU_{t(k)}^\prime - \frac{n}{n_k}\mSigma_k\right)\\
        &= \frac{n_k}{n(T-1)}\sum^T_{t=2}\left(\mU_{t(k)}\mOmega_{-k}\mU_{t(k)}^\prime - \frac{n}{n_k}\mSigma_k\right)\\
        &\quad -\mSigma_k\sum_{j\neq k}\frac{1}{n_j}\tr\left(\mOmega_j(\hat{\mSigma}_j - \mSigma_j)\right) + o_p(T^{-1/2}),
    \end{align*}
    for all $k$. In other words, we get the following equations,
    \begin{align*}
        &\hat{\mSigma}_k - \mSigma_k + \mSigma_k\sum_{j \neq k}\frac{1}{n_j}\tr\left(\mOmega_j(\hat{\mSigma}_j - \mSigma_j)\right) \\
        &= \frac{n_k}{n(T-1)}\sum^T_{t=2}\left(\mU_{t(k)}\mOmega_{-k}\mU_{t(k)}^\prime - \frac{n}{n_k}\mSigma_k\right) + o_P(T^{-1/2}), \qquad \text{for all }k=1,\ldots,N.
    \end{align*}
    By vectorizing the equations we get the following equations
    \begin{align*}
        &\vec(\hat{\mSigma}_k - \mSigma_k) + \sum_{j \neq k}\frac{1}{n_j}\vec(\mSigma_k)\vec(\mOmega_j^\prime)^\prime\vec(\hat{\mSigma}_j - \mSigma_j) \\
        &= \vec\left(\frac{n_k}{n(T-1)}\sum^T_{t=2}\left(\mU_{t(k)}\mOmega_{-k}\mU_{t(k)}^\prime - \frac{n}{n_k}\mSigma_k\right)\right) + o_p(T^{-1/2}), \qquad \text{for all }k=1,\ldots,N.
    \end{align*}
    We stack this system of equations into vectors and matrices to obtain the following system,
    \begin{align*}
        \mW_1 \begin{pmatrix}
            \vec(\hat{\mSigma}_1 - \mSigma_1)\\
            \vdots\\
            \vec(\hat{\mSigma}_N - \mSigma_N)
        \end{pmatrix}
        = \begin{pmatrix}
            \vec\left(\frac{n_1}{n(T-1)}\sum^T_{t=2}\left(\mU_{t(1)}\mOmega_{-1}\mU_{t(1)}^\prime - \frac{n}{n_1}\mSigma_1\right)\right)\\
            \vdots\\
            \vec\left(\frac{n_N}{n(T-1)}\sum^T_{t=2}\left(\mU_{t(N)}\mOmega_{-N}\mU_{t(N)}^\prime - \frac{n}{n_N}\mSigma_N\right)\right)
        \end{pmatrix}
        + o_P(T^{-1/2}),
    \end{align*}
    where $\mW_1$ contains blocks
    \begin{align}
        \Big[\mW_1\Big]_{kj} = \begin{cases}
            \mI_{n^2_k}, \qquad &j=k\\
            \frac{1}{n_j}\vec(\mSigma_k)\vec(\mOmega_j^\prime)^\prime, & j\neq k.
        \end{cases}
    \end{align}
    The matrix $\mW_1$ is not invertible due to the arbitrary $N-1$ Kronecker product rescalings. Therefore, we define $\tilde{\mW}_1 = \mW_1 + \mGamma_{\Sigma}\mGamma_{\Sigma}^\prime$ with $\mGamma_{\Sigma} = (0,\ldots,\vec(\mSigma_k)^\prime,\ldots,0)$. Furthermore, the central limit theorem implies that
    \begin{align}
        &\sqrt{T}\begin{pmatrix}
            \vec\left(\frac{n_1}{n(T-1)}\sum^T_{t=2}\left(\mU_{t(1)}\mOmega_{-1}\mU_{t(1)}^\prime - \frac{n}{n_1}\mSigma_1\right)\right)\\
            \vdots\\
            \vec\left(\frac{n_N}{n(T-1)}\sum^T_{t=2}\left(\mU_{t(N)}\mOmega_{-N}\mU_{t(N)}^\prime - \frac{n}{n_N}\mSigma_N\right)\right)
        \end{pmatrix} \\
        & \ind \calN(\vzeros, \mW_2),
    \end{align}
    where $\mW_2 = \E(\vq_t \vq_t^\prime)$ with $\vq_t = (\vq_{1t}^\prime,\ldots,\vq_{Nt}^\prime)^\prime$ and
    \begin{align*}
        \vq_{kt}\equiv\frac{n_k}{n} \vec\left( \mU_{t(k)}\mOmega_{-k}\mU_{t(k)}' - \frac{n}{n_k}\mSigma_k \right)
    \end{align*}
    By Assumption \ref{a1:iid}, $\mW_2$ exists. We obtain that
    \begin{align*}
        \sqrt{T}\begin{pmatrix}
            \vec(\hat{\mSigma}_1-\mSigma_1)\\
            \vdots\\
            \vec(\hat{\mSigma}_N-\mSigma_N)
        \end{pmatrix}
        = \widetilde{\mW}^{-1}_1 T^{-1/2}\sum^T_{t=2}\vq_t + o_p(1).
    \end{align*}
    Therefore, as $T\rightarrow\infty$,
    \begin{align}
        \sqrt{T}\begin{pmatrix}
            \vec(\hat{\mSigma}_1 - \mSigma_1)\\
            \vdots\\
            \vec(\hat{\mSigma}_N - \mSigma_N)
        \end{pmatrix} \ind \calN\left(\vzeros, \tilde{\mW}_1^{-1}\mW_2\tilde{\mW}_1^{-1}\right).
    \end{align}
    Recall that $\mL_k\mL_k^\prime = \mSigma_k$ for all $k$. Differentiation of the Cholesky decomposition says that, for all $k$,
    \begin{align*}
        \vec(d\mL_k) = (\mI\otimes \mL_k)\mZ_{n_k}(\mL^{-1}_k \otimes \mL^{-1}_k)\vec(\mSigma_k),
    \end{align*}
    where $\mZ_{n_k}$ is an $n^2_k\times n^2_k$ matrix such that
    \begin{align*}
        \mZ_{n_k}\vec(\mA) = \vec(\mJ(\mA)), \quad \text{with }\mJ(\mA)\equiv \begin{cases}
            A_{ij}, & \text{ for }i>j;\\
            \frac{1}{2}A_{ij}, & \text{ for }i=j;\\
            0, &\text{ for } i<j.
        \end{cases}
    \end{align*}
    We obtain that 
    \begin{align}
        \sqrt{\hat{\vell}-\vell} = \sqrt{T}\diag(\mD^+_{n_k}(\mI\otimes \mL_k)\mZ_{n_k}(\mL^{-1}_k \otimes \mL^{-1}_k)\mD_{n_k})\begin{pmatrix}
            \vec(\hat{\mSigma}_1 - \mSigma_1)\\
            \vdots\\
            \vec(\hat{\mSigma}_N - \mSigma_N)
        \end{pmatrix},
    \end{align}
    and thus, we obtain the desired result.
\end{proof}

\begin{proof}[Proof of Proposition \ref{prop:rf_distr}]
    By Lemma \ref{lem:coeff_distr}, we know that
    \begin{align*}
        \sqrt{T}(\hat{\vb}-\vb) = \mH^{-1} T^{-1/2}\sum^T_{t=2}\mD_t^\prime\mOmega \vu_t + o_p(1).
    \end{align*}
    Moreover, by Lemma \ref{lem:sigma_distr}, we know that
    \begin{align*}
        \sqrt{T}(\hat{\vell}-\vell) = \mH_{\ell}\widetilde{\mW}^{-1}_1T^{-1/2}\sum^T_{t=2}\vq_t + o_p(1),
    \end{align*}
    where $\mH_{\ell}\equiv\diag(\mD^+_{n_k}(\mI\otimes \mL_k)\mZ_{n_k}(\mL^{-1}_k \otimes \mL^{-1}_k)\mD_{n_k})$. Under Assumption \ref{a1:iid}-\ref{a3:nonsing}, the multivariate martingale CLT yields
    \begin{align*}
        \frac{1}{\sqrt{T}}\sum^T_{t=2}\begin{pmatrix}
            \mD_t^\prime\mOmega\vu_t\\ \vq_t
        \end{pmatrix}
        \ind \calN\left(\vzeros, \begin{pmatrix}
            \E(\mD^\prime_t\mOmega \mD_t) & \E(\mD_t^\prime\mOmega\vu_t \vq_t^\prime)\\
            \E(\mD_t^\prime\mOmega\vu_t \vq_t^\prime)^\prime & \mW_2
        \end{pmatrix}\right).
    \end{align*}
    We define 
    \begin{align}
    \mXi_{12}\equiv \mH^{-1}\E(\mD_t^\prime\mOmega\vu_t\vq_t^\prime)\widetilde{\mW}^{-1\prime}_1\mH_{\ell}^\prime.
    \end{align}
    Then, by Lemmas \ref{lem:coeff_distr} and \ref{lem:sigma_distr}, we have that
    \begin{align*}
    \sqrt{T}(\hat{\vpi} - \vpi^0) \ind \calN\left(\vzeros, \begin{pmatrix}
            \mXi_1 & \mXi_{12}\\
           \mXi_{12}^\prime & \mXi_2
        \end{pmatrix}
        \right).
        \end{align*}
\end{proof}

\subsection{Proofs for structural-form results}
\begin{proof}[Proof of Theorem \ref{thm:theta_distr}]
    The proof follows the proof of Theorem 4 in \citetapndx{HAN2025106057} closely. Proposition \ref{prop:rf_distr} implies that $Q_T(\vtheta|\hat{\vpi}) \inp Q_T(\vtheta|\vpi^0)$ uniformly over $\vtheta in \mTheta$. Moreover, by the continuous mapping theorem, we have that $\partial g(\vtheta|\hat{\vpi})/\partial \vtheta^\prime \inp \partial g(\vtheta|\vpi^0)/\partial \vtheta^\prime$. For some $\bar{\vtheta}$ between $\vtheta^0$ and $\hat{\vtheta}$, the mean value theorem implies
    \begin{align*}
        \sqrt{T}(\hat{\vtheta}-\vtheta^0) = -\left(\frac{\partial^2 Q_T(\bar{\vtheta}|\hat{\vtheta})}{\partial \vtheta \partial \vtheta^\prime}\right)^{-1}\left(\frac{\sqrt{T}\partial Q_T(\vtheta^0|\hat{\vpi})}{\partial \vtheta}\right).
    \end{align*}
    Using that $\vg(\vtheta^0|\vpi^0)=\vzeros$, we can write $\sqrt{T}\partial Q_T(\vtheta^0|\hat{\vpi})/\partial \vtheta$ as
    \begin{align*}
        \frac{\sqrt{T}\partial Q_T(\vtheta^0|\hat{\vpi})}{\partial \vtheta} = 2\left(\frac{\partial \vg(\vtheta^0, \hat{\vpi})^\prime}{\partial \vtheta}\right)\mW_T \sqrt{T}\left(\vg(\vtheta^0|\hat{\vpi}) - \vg(\vtheta^0|\vpi^0)\right).
    \end{align*}
    By definition of $\mG_{\vtheta}$ and $\mG_{\vpi}$, and $\mW_T \inp \mW$, we have that 
    \begin{align*}
        \frac{\sqrt{T}\partial Q_T(\vtheta^0|\hat{\vpi})}{\partial \vtheta} = 2\mG^\prime_{\vtheta}\mW\mG_{\vpi}\sqrt{T}(\hat{\vpi} - \vpi^0) + o_p(1),
    \end{align*}
    and
    \begin{align*}
        \frac{\partial^2 Q_T(\bar{\vtheta}|\hat{\vpi})}{\partial \vtheta \partial \vtheta^\prime} = 2\mG_{\vtheta}^\prime\mW\mG_{\vtheta} + o_p(1).
    \end{align*}
    Combining the results above concludes the proof.
\end{proof}

\begin{proof}[Proof of Proposition \ref{prop:j_test}]
    The result follows directly from the arguments in the proof of  \citetapndx{HAN2025106057} for their Proposition 3.
\end{proof}

\begin{proof}[Proof of Theorem \ref{thm:irf_distr}]
Combining the results of Proposition \ref{prop:rf_distr} and Theorem \ref{thm:theta_distr} gives us that
\begin{align*}
    \begin{pmatrix}
        \sqrt{T}(\hat{\vpi} - \vpi^0)\\
        \sqrt{T}(\hat{\vtheta} - \vtheta^0)
    \end{pmatrix} = \begin{pmatrix}
        \mI\\
        \mA_{\vtheta}
    \end{pmatrix}
    \sqrt{T}(\hat{\vpi} - \vpi^0) + o_p(1)
\end{align*}
Since $\psi_h$ is continuously differentiable, a first-order Taylor expansion around $(\vpi^0, \vtheta^0)$ gives us that
\begin{align*}
    \sqrt{T}(\hat{\psi}_h - \psi^0) &= \vg^\prime_{\vpi,h}\sqrt{T}(\hat{\vpi}-\vpi^0) + \vg^\prime_{\vtheta,h}\sqrt{T}(\hat{\vtheta}-\vtheta^0) + o_p(1)\\
    &= \vg^\prime_{\vpi,h}\sqrt{T}(\hat{\vpi}-\vpi^0) + \vg^\prime_{\vtheta,h}\mA_{\vtheta}\sqrt{T} (\hat{\vpi} - \vpi^0) + o_p(1)\\
    &= (\vg^\prime_{\vpi,h} + \vg^\prime_{\vtheta,h}\mA_{\vtheta}) \sqrt{T} (\hat{\vpi} - \vpi^0) + o_p(1)\\
    &= \vm^\prime_h\sqrt{T} (\hat{\vpi} - \vpi^0) + o_p(1),\\
\end{align*}
where $\vm_h = \vg_{\vpi,h} + \mA^\prime_{\vtheta}\vg_{\vtheta,h}$. By Proposition \ref{prop:rf_distr} and the continuous mapping theorem, we have that
\begin{align*}
    \vm^\prime_h\sqrt{T}(\hat{\vpi} - \vpi^0) \ind \calN(\vzeros, \vm_h^\prime \mXi \vm_h).
\end{align*}
This concludes the proof. 

\end{proof}

\subsection{Proofs for Section \ref{sec:robust}} \label{app:app:robust-proofs}
In this subsection, we provide the proofs for the results in Section \ref{sec:robust}. Before we prove the results for the restricted estimator (under Kronecker product), we need the asymptotic distribution of the unrestricted estimator under our assumptions. Lemma \ref{lem:aux_unrestr_gaus} below says that the unrestricted estimator is consistent and asymptotically normal. 
\begin{lemma}\label{lem:aux_unrestr_gaus}
    Suppose Assumptions \ref{a1:iid}-\ref{a3:nonsing} hold at $\vb^0$. Then it holds that
    \begin{enumerate}[label=\roman*]
        \item $\hat{\mGamma}_0 \inp \mGamma_0 \equiv \E\left(\vy_{t}\vy_t^\prime\right)$;
        \item $\sqrt{T}(\hat{\vphi}_U - \vphi_T) \ind \mZ$, with $\mZ \sim \calN(\vzeros, \mV_{\phi})$;
        \item $\sqrt{T}(\hat{\vphi}_U - \vphi_0) \ind \mZ + \vd$.
    \end{enumerate}
\end{lemma}
\begin{proof}[Proof of Lemma \ref{lem:aux_unrestr_gaus}]
    Plugging the true GDP $\vy_t = \mPhi_T\vy_{t-1} + \vu_t$ into $\hat{\mPhi}_U$ yields
    \begin{align*}
        \hat{\mPhi}_U - \mPhi_T &= \left(\sum^T_{t=2}\vu_t\vy_{t-1}^\prime\right)\left(\sum^T_{t=2}\vy_{t-1}\vy_{t-1}^\prime\right)^{-1}\\
        &= (T-1)^{-1}\left(\sum^T_{t=2}\vu_t\vy_{t-1}^\prime\right) \hat{\mGamma}_0^{-1},
    \end{align*}
    where the second equation follows from the definition of $\hat{\mGamma}_0$. By vectorizing both sides and multiplying by $T^{1/2}$, we get that
    \begin{align*}
        \sqrt{T}(\hat{\vphi}_U - \vphi_T) = \left(\hat{\mGamma}^{-1}_0 \otimes \mI_n\right)\frac{1}{\sqrt{T}}\sum^T_{t=2}(\vy_{t-1}\otimes \mI_n)\vu_t + o_p(1).
    \end{align*}
    By Assumption \ref{a1:iid}, we have that $\E(\vu_t|\vy_{t-1}, \vy_{t-2},\cdots)=\vzeros$ and limiting variance \begin{align*}\E[\left(\vy_{t-1}\otimes\mI_n\right)\vu_t\vu_t^\prime\left(\vy_{t-1}\otimes\mI_n\right)] = \mGamma_0 \otimes \mSigma.\end{align*} Therefore, by law of large numbers, we have that (i) $\hat{\mGamma}_0 \inp \mGamma_0$ holds. Furthermore, by the martingale CLT, we have that (ii) $\sqrt{T}(\hat{\vphi}_U - \vphi_T)\ind \mZ$ for $\mZ\sim \calN(\vzeros, \mV_{\phi})$ holds. Since $\sqrt{T}(\vphi_T - \vphi_0) = \vd$, it follows that (iii) $\sqrt{T}(\hat{\vphi}_U - \phi_0) \ind \mZ +\vd$ holds.
\end{proof}

\begin{proof}[Proof of Lemma \ref{lem:qmle-md}]
If $\mSigma$ is fixed, then for every $\vphi$,
\begin{align*}
  &\sum_{t=2}^{T} \left(\vy_t - \mPhi \vy_{t-1}\right)^\prime \mSigma^{-1} \left(\vy_t - \mPhi \vy_{t-1}\right)\\
  &= \sum^T_{t=2}\left(\vy_t - \hat{\mPhi}_U\vy_{t-1} + \hat{\mPhi}_U\vy_{t-1} - \mPhi\vy_{t-1}\right)^\prime\mSigma^{-1}\left(\vy_t - \hat{\mPhi}_U\vy_{t-1} + \hat{\mPhi}_U\vy_{t-1} - \mPhi\vy_{t-1}\right)\\
  &=\sum^T_{t=2}\left(\hat{\vu}_t + (\hat{\mPhi}_U - \mPhi)\vy_{t-1}\right)^\prime\mSigma^{-1}\left(\hat{\vu}_t + (\hat{\mPhi}_U - \mPhi)\vy_{t-1}\right)\\
  &= \sum^T_{t=2}\hat{\vu}_{t}^\prime\mSigma^{-1}\vu_{t} + 2\sum^T_{t=2}\hat{\vu}_t^\prime\mSigma^{-1}(\hat{\mPhi}_U - \mPhi)\vy_{t-1} + \sum^T_{t=2}\vy_{t-1}^\prime (\hat{\mPhi}_U - \mPhi)^\prime \mSigma^{-1}(\hat{\mPhi}_U - \mPhi)\vy_{t-1},
  \end{align*}
  where we have defined $\hat{\vu}_t = \vy_t - \hat{\mPhi}_U \vy_{t-1}$. By definition of OLS, the middle term in the last equality equals zero. Therefore, we have that
  \begin{align*}
      &\sum_{t=2}^{T} \left(\vy_t - \mPhi \vy_{t-1}\right)^\prime \mSigma^{-1} \left(\vy_t - \mPhi \vy_{t-1}\right)\\
      &= \sum^T_{t=2}\hat{\vu}_{t}^\prime\mSigma^{-1}\vu_{t}  + \sum^T_{t=2}\vy_{t-1}^\prime (\hat{\mPhi}_U - \mPhi)^\prime \mSigma^{-1}(\hat{\mPhi}_U - \mPhi)\vy_{t-1}\\
      &= \sum^T_{t=2}\hat{\vu}_{t}^\prime\mSigma^{-1}\vu_{t} + (T-1)\tr\left[(\hat{\mPhi}_U - \mPhi)^\prime\mSigma^{-1}(\hat{\mPhi}_U - \mPhi)\hat{\mGamma}_0\right].
  \end{align*}
By the identity $\tr\left(\mA^\prime\mB\mA\mC\right) = \vec(\mA)^\prime(\mC^\prime\otimes \mB)\vec(\mA)$ and the symmetric of $\hat{\mGamma}_0$, we obtain that
\begin{equation*}
    \sum_{t=2}^{T} \left(\vy_t - \mPhi \vy_{t-1}\right)^\prime \mSigma^{-1} \left(\vy_t - \mPhi \vy_{t-1}\right) = \sum^T_{t=2}\hat{\vu}_{t}^\prime\mSigma^{-1}\vu_{t} + (T-1)(\hat{\vphi}_U - \vphi)^\prime\left(\hat{\mGamma}_0 \otimes \mSigma^{-1}\right)(\hat{\vphi}_U - \vphi).
\end{equation*}
The first term does not depend on $\mPhi$. Consequently,  maximizing the Gaussian log-likelihood for $\vphi\in\mathcal{K}$ is exactly equivalent to solving the minimum-distance problem \eqref{eq:md-problem}.  Lemma \ref{lem:aux_unrestr_gaus} gives us that $\hat{\mGamma}_0 \rightarrow \mGamma_0$ such that $\hat{\mW} \rightarrow_p \mW$.
\end{proof}

We now prove the results in Theorem \ref{thm:irf_distr_misspec}. To do so, we first show that \eqref{eq:phi_decomp} holds.
\begin{proof}[Proof of Eq. \eqref{eq:phi_decomp}]
    By definition of $\mR_0$, we can write
    \begin{align*}
        \vb = \vb^0 + \mR_0(\vbeta - \vbeta_0) + o(||\vbeta - \vbeta_0),
    \end{align*}
    for some local deviations from $\vb^0$ by $\vbeta, \vbeta^0\in\SR^r$. By a first-order expansion of $\vk(\vb)$, we get that
    \begin{align*}
        \vk(\vb) = \vk(\vb^0) + \mJ(\vbeta - \vbeta_0) + o(||\vbeta - \vbeta_0||).
    \end{align*}
    Under local identification and full column rank $\mJ$, we have that $\hat{\mB}_k = \mB_k + O_p(T^{-1/2})$ (Lemma \ref{lem:coeff_consist}), and thus, it holds that $o(||\hat{\vbeta} - \vbeta_0||) = o_p(T^{-1/2})$. This gives that
    \begin{align} \label{eq:mvexp_beta}
        \hat{\vphi}_K = \vk(\vb^0) + \mJ(\hat{\vbeta} - \vbeta^0) + o_p(T^{-1/2}).
    \end{align}
    Moreover, the first-order condition of \eqref{eq:md-problem} is that
    \begin{align}\label{eq:foc_mdprob}
        \mJ^\prime(\hat{\vbeta})\hat{\mW}\left(\hat{\vphi}_U - \vk(\hat{\vb})\right) = \vzeros.
    \end{align}
    We know $\mJ(\hat{\vbeta}^0) = \mJ + o_p(1)$. By plugging \eqref{eq:mvexp_beta} into \eqref{eq:foc_mdprob} yields
    \begin{align*}
        \mJ^\prime \hat{\mW}\left(\hat{\vphi}_U - \vphi_0 - \mJ(\hat{\vbeta}-\vbeta_0)\right) = \vzeros.
    \end{align*}
    If we multiply by $\sqrt{T}$ and use that $\hat{\mW} = \mW + o_p(1)$, we get that
    \begin{align*}
        \mJ^\prime\mW\mJ\sqrt{T}(\hat{\vbeta} - \vbeta_0) = \mJ^\prime \mW \sqrt{T}(\hat{\vphi}_U - \vphi_0) + o_p(1).
    \end{align*}
    Since we know that $\sqrt{T}(\hat{\vphi}_U - \vphi_0) \ind \mZ + \vd$, we have that
    \begin{align*}
        \sqrt{T}(\hat{\vbeta}-\vbeta_0) \ind ( \mJ^\prime\mW\mJ)^{-1}\mJ^\prime \mW(\mZ + \vd).
    \end{align*}
    Finally, by expanding $\vk(\hat{\vbeta})$, we have that $\sqrt{T}(\hat{\vphi}_K - \vphi_0) = \mJ\sqrt{T}(\hat{\vbeta}-\vbeta_0) + o_p(1)$, and hence,
    \begin{align*}
        \sqrt{T}(\hat{\vphi}_K - \vphi_0) \ind \mP_W(\mZ+\vd).
    \end{align*}
    Noting that $\sqrt{T}(\vphi_T - \vphi_0) = \vd$ gives us the desired result in \eqref{eq:phi_decomp}.
\end{proof}

\begin{lemma}\label{lem:eta_dist}
    Suppose Assumptions \ref{a1:iid}-\ref{a4:robust_identification} hold. Let $\veta \equiv (\vsigma^\prime, \vtheta^\prime)^\prime$ and $\hat{\veta} = (\hat{\vsigma}^\prime, \hat{\vtheta}^\prime)$. Then, as $T\rightarrow\infty$, it holds that
    \begin{align*}
        \sqrt{T}(\hat{\veta} - \veta_0)\ind \vzeta_{\eta}, \quad \E(\vzeta_{\eta})=\vzeros,
    \end{align*}
where $\vzeta_{\eta}$ does not depend on $\vd$.
\end{lemma}
\begin{proof}[Proof of Lemma \ref{lem:eta_dist}]
    Let $\hat{\vu}_t = \vy_t - \hat{\mPhi}_K\vy_{t-1}$. Then we get that
    \begin{align*}
        \hat{\mSigma} &= \frac{1}{T}\sum^T_{t=2}\vu_t\vu_t^\prime\\
        &= \frac{1}{T-1}\sum^T_{t=2}\vu_t\vu_t^\prime - \left(\hat{\mPhi}_K - \mPhi_T\right)\left(\frac{1}{T-1}\sum^T_{t=2}\vy_{t-1}\vu_t^\prime\right) - \left(\frac{1}{T-1}\sum^T_{t=2}\vu_t\vy_{t-1}^\prime\right)\left(\hat{\mPhi}_K - \mPhi_T\right)^\prime \\
        &\quad + \left(\hat{\mPhi}_K - \mPhi_T\right)\left(\frac{1}{T-1}\sum^T_{t=2}\vy_{t-1}\vy_{t-1}^\prime\right)\left(\hat{\mPhi}_K - \mPhi_T\right)^\prime.
    \end{align*}
    By \eqref{eq:phi_decomp} we know that $\hat{\mPhi}_K - \mPhi_T = O_p(T^{-1/2})$. Furthermore, we know that $(T-1)^{-1}\sum^T_{t=2}\vy_{t-1}\vu_t^\prime = O_p(T^{-1/2})$ and $(T-1)^{-1}\sum^T_{t=2}\vy_{t-1}\vy_{t-1}^\prime = O_p(1)$ by LLN. Therefore, we have that
    \begin{align*}
        \hat{\mSigma} = \frac{1}{T}\sum^T_{t=2}\vu_t\vu_t^\prime + O_p(T^{-1}).
    \end{align*}
    Hence,
    \begin{align*}
        \sqrt{T}(\hat{\mSigma}-\mSigma) = \frac{1}{\sqrt{T}}\sum^T_{t=2}(\vu_t\vu_t^\prime - \mSigma) + o_p(1),
    \end{align*}
    which does not contain $\vd$ on the right-hand side. Under Assumption \ref{a4:robust_identification}, the asymptotic distribution of $\hat{\vtheta}$ loads only on $\sqrt{T}(\hat{\vsigma}-\vsigma^0)$ and is therefore also free from $\vd$. By the delta method, 
    \begin{align*}
        \sqrt{T}(\hat{\veta}-\veta_0) = \mA_{\eta}\sqrt{T}(\vec(\hat{\mSigma} - \mSigma)) + o_p(1),
    \end{align*}
    for a suitable derivative $\mA_{\eta}$. The asymptotic distribution of $\hat{\veta}$ does not depend on $\vd$. In other words, the misspecification in propagation does not bias the covariance estimator at first order.
\end{proof}

\begin{proof}[Proof of Theorem \ref{thm:irf_distr_misspec}]
    We have defined $g_{\vphi,h} \equiv \partial \psi_h/\partial \vphi\Big |_{(\vphi_0, \veta_0)}$ and $g_{\veta,h} \equiv \partial \psi_h/\partial \veta\Big|_{(\vphi_0, \veta_0)}$. Since $\vphi_T - \vphi_0 = O(T^{-1/2})$, it holds that the derivatives evaluated at $(\vphi_T, \veta_0)$ differs from the derivatives at $(\vphi_0, \veta_0)$ by $o(1)$. Therefore, a first-order Taylor expansion of $\psi_{h}$ around the true parameters yields
    \begin{align*}
        \sqrt{T}(\hat{\psi}_h - \psi_{h,T}) = g_{\phi,h}^\prime \sqrt{T}(\hat{\vphi_K} - \vphi_T) + g_{\eta,h}^\prime\sqrt{T}(\hat{\veta}-\veta_0) + o_p(1).
    \end{align*}
    By \eqref{eq:phi_decomp} and Lemma \ref{lem:eta_dist}, we obtain that
    \begin{align*}
        \sqrt{T}(\hat{\psi}_h - \psi_{h,T}) = g_{\phi,h}^\prime\mP_W\mZ - g_{\phi,h}^\prime(\mI-\mP_W)\vd + g_{\eta,h}^\prime\vzeta_{\eta} + o_p(1).
    \end{align*}
    The asymptotic distribution can therefore be characterized as
    \begin{align*}
        \sqrt{T}(\hat{\psi}_h - \psi_{h,T}) \ind \calN(b_h(\vd), \sigma_h^2),
    \end{align*}
    with $b_h(\vd) = -\vg_{\phi,h}^\prime(\mI -\mP_W)\vd$ and asymptotic variance 
    \begin{align*}
        \sigma^2_h = \vg_{\phi,h}^\prime\mP_W\mV_{\phi}\mP_W^\prime \vg_{\phi,h} + \vg^\prime_{\eta,h}\mV_{\eta}\vg_{\eta,h} + 2\vg_{\phi,h}^\prime\mP_W\mV_{\phi \eta}\vg_{\eta,h}.
    \end{align*}
    At $\mW = \mV_{\phi}^{-1}$, we have $\mP_W = \mJ(\mJ^\prime\mV_{\phi}^{-1}\mJ)^{-1}\mJ^\prime\mV^{-1}_{\phi}$, and thus,
    \begin{align*}
        \mP_W\mV_{\phi}\mP_W^\prime = \mJ(\mJ^\prime\mV^{-1}_{\phi}\mJ)^{-1}\mJ^\prime.
    \end{align*}
    Hence, we obtain that
    \begin{align*}
        \sigma^2_h = \vg_{\phi,h}^\prime \mJ(\mJ^\prime\mV^{-1}_{\phi}\mJ)^{-1}\mJ^\prime \vg_{\phi,h} + \vg^\prime_{\eta,h}\mV_{\eta}\vg_{\eta,h} + 2\vg_{\phi,h}^\prime\mP_W\mV_{\phi \eta}\vg_{\eta,h}.
    \end{align*}
\end{proof}
The result in Theorem \ref{thm:irf_distr_misspec} is a pointwise result in $\vd$, whereas the result in Theorem \ref{thm:robust_baci_cov} is uniform. Therefore, in order to prove Theorem \ref{thm:robust_baci_cov}, we establish a uniform extension of Theorem \ref{thm:irf_distr_misspec} in the lemma below.
\begin{lemma}\label{lem:unif_irf_distr_miss}
    Suppose the assumptions in Theorem \ref{thm:irf_distr_misspec} hold. Fix $c>0$ and a horizon $h$. Moreover, suppose $\sigma^2_h>0$. For each $\vd\in\mathcal{D}(c)$, let $P_{T,d}$ denote the DGP corresponding to $\vphi_{T,d} = \vphi_0 + T^{-1/2}\vd$, and let $\psi_{h,T,d}\equiv \psi_h(\phi_{T,d},\eta_0)$. Then,
    \begin{align*}
        \lim_{T\rightarrow\infty}\sup_{\vd\in\mathcal{D}(c)}\sup_{x\in\SR}\Bigg|P_{T,d}\left(\frac{\sqrt{T}(\hat{\psi}_h - \psi_{h,T,d}) - b_h(\vd)}{\sigma_h} \leq x\right) - \mPhi(x)\Bigg| = 0,
    \end{align*}
    where $b_h(\vd)$ and $\sigma^2_h$ are defined in Theorem \ref{thm:irf_distr_misspec}. Moreover, for every $\epsilon>0$,
    \begin{align}
        &\sup_{\vd\in\mathcal{D}(c)}P_{T,d}(|\hat{\sigma}_h - \sigma_h| > \epsilon)\rightarrow 0;\\
        &\sup_{\vd\in\mathcal{D}(c)}P_{T,d}(|\hat{\bar{b}}_h(c) - \bar{b}_h(c)| > \epsilon)\rightarrow 0.
    \end{align}
\end{lemma}
\begin{proof}[Proof of Lemma \ref{lem:unif_irf_distr_miss}]
    To prove Lemma \ref{lem:unif_irf_distr_miss} it is sufficient to establish the results for an arbitrary sequence $\vd_T \in \mathcal{D}(c)$, with $\vphi_{T,d_T} = \vphi_0 + \vd_T/\sqrt{T}$. Throughout the proof, the probability statements are under $P_{T,d_T}$. Since $\mH$ is positive definite and $\vd_T \in \mathcal{D}(c)$, we know that $\{\vd_T\}$ is bounded, and hence
    \begin{align*}
        \vphi_{T,d_T}=\vphi_0+\frac{\vd_T}{\sqrt{T}}\rightarrow \vphi_0.
    \end{align*}
    Since $\mPhi^0$ is strictly stable, we know that all $\mPhi_{T,d}$ with $\vd\in\mathcal{D}(c)$ lie in a stable neighborhood for sufficiently large $T$. Together with Assumption \ref{a1:iid}, this gives that
    \begin{align}
        \hat{\mGamma}_0 - \mGamma_0 = \left(\hat{\mGamma}_0 - \E_{T,d_T}(\vy_t\vy_t^\prime)\right) + \left(\E_{T,d_T}(\vy_t\vy_t^\prime) - \mGamma_0\right) = o_p(1) + o(1). \label{eq:unif-gamma}
    \end{align}
    Moreover, $\{(\vy_{t-1}\otimes\mI_n)\vu_t\}$ is a martingale difference triangular
    array. Its conditional variance converges by \eqref{eq:unif-gamma}, and Assumption 1
    implies we can use the Lindeberg central limit theorem such that
    \begin{align}
        \frac{1}{\sqrt{T}}\sum_{t=2}^T
        (\vy_{t-1}\otimes\mI_n)\vu_t
        \ind
        \calN(\vzeros,\mGamma_0\otimes\mSigma). \label{eq:unif-mclt}
    \end{align}
    Thus, by the same OLS representation as in Lemma \ref{lem:aux_unrestr_gaus},
    \begin{align}
        \sqrt{T}(\hat{\vphi}_U-\vphi_{T,d_T})
        \ind \calN(\vzeros,\mV_\phi). \label{eq:unif-ols}
    \end{align}
    In the remainder, we apply the arguments in the proofs of \eqref{eq:phi_decomp}, Lemma
    \ref{lem:eta_dist}, and Theorem \ref{thm:irf_distr_misspec} along the sequence $\vd_T$. To see this, note that
    \begin{align*}
        \hat{\vphi}_U-\vphi_0
        =
        (\hat{\vphi}_U-\vphi_{T,d_T})
        + T^{-1/2}\vd_T
        =
        O_p(T^{-1/2}),
     \end{align*}
    so the consistency and local root-$T$ rate of the restricted estimator used in
    \eqref{eq:phi_decomp} remain valid. Furthermore, all Taylor remainders in these
    arguments are evaluated in an $O(T^{-1/2})$ neighborhood of
    $(\vphi_0,\veta_0)$ and are therefore $o_p(1)$ by the same differentiability
    assumptions as before. Finally, Lemma \ref{lem:eta_dist} shows that the first-order
    behavior of $\hat{\veta}$ depends only on the innovation terms and is unaffected by
    the local propagation misspecification. Hence, we can write
    \begin{align}
        \sqrt{T}(\hat{\vphi}_K-\vphi_{T,d_T})
        &=
        \mP_W\sqrt{T}(\hat{\vphi}_U-\vphi_{T,d_T})
        -(\mI-\mP_W)\vd_T+o_p(1),
        \label{eq:unif-projection}\\
        \sqrt{T}(\hat{\psi}_h-\psi_{h,T,d_T})
        &=
        b_h(\vd_T)
        +g_{\phi,h}^\prime\mP_W
        \sqrt{T}(\hat{\vphi}_U-\vphi_{T,d_T})
        +g_{\eta,h}^\prime\sqrt{T}(\hat{\veta}-\veta_0)
        +o_p(1).
        \label{eq:unif-irf-expansion}
    \end{align}
    The joint limiting distribution of the two stochastic terms in
    \eqref{eq:unif-irf-expansion} is the same as in Theorem
    \ref{thm:irf_distr_misspec} and does not depend on $\vd_T$. Therefore,
    \begin{align}
        \frac{
        \sqrt{T}(\hat{\psi}_h-\psi_{h,T,d_T})-b_h(\vd_T)}
        {\sigma_h}
        \ind \calN(0,1)
        \label{eq:seq-irf}
    \end{align}
    for every deterministic sequence $\vd_T\in\mathcal{D}(c)$. 
    
    We now obtain uniformity. We prove this by contradiction. Suppose the first conclusion of Lemma \ref{lem:unif_irf_distr_miss} is false. There would exist $\epsilon>0$, a subsequence $T_j$, and
    $\vd_j\in\mathcal{D}(c)$ such that
    \begin{align*}
        \sup_{x\in\SR}
        \left|
        P_{T_j,d_j}\left(
        \frac{
        \sqrt{T_j}(\hat{\psi}_h-\psi_{h,T_j,d_j})-b_h(\vd_j)}
        {\sigma_h}\leq x
        \right)-\mPhi(x)
        \right|>\epsilon.
    \end{align*}
    However, since $\vd_j \in \mathcal{D}(c)$, the sequence $\{\vd_j\}$ must be such that \eqref{eq:seq-irf} holds. Since the limiting distribution function $\Phi(\cdot)$ is continuous, Pólya's theorem implies convergence uniformly in $x$, which is a contradiction.
    Hence, we can conclude that
    \begin{align*}
        \sup_{\vd\in\mathcal{D}(c)}\sup_{x\in\SR}
        \left|
        P_{T,d}\left(
        \frac{
        \sqrt{T}(\hat{\psi}_h-\psi_{h,T,d})-b_h(\vd)}
        {\sigma_h}\leq x
        \right)-\mPhi(x)
        \right|\to0.
    \end{align*}
    It remains us to establish uniform consistency of the plug-in quantities $\hat{\sigma}_h$ and $\hat{\bar{b}}_h(c)$. First, we know that along every deterministic sequence $\vd_T\in\mathcal{D}(c)$, \eqref{eq:unif-gamma}, the consistency of $\hat{\mSigma}$ (Lemma \ref{lem:eta_dist}), and the consistency of the restricted parameter estimators
    imply, by continuous mapping,
    \begin{align*}
        \hat{\sigma}_h\inp\sigma_h.
    \end{align*}
    Since this holds along every sequence $\vd_T\in\mathcal{D}(c)$, we can conclude that
    \begin{align*}
        \sup_{\vd\in\mathcal{D}(c)}
        P_{T,d}(|\hat{\sigma}_h-\sigma_h|>\epsilon)\to0.
    \end{align*}

    Second, similarly, for $\bar{b}_h(c)$ it holds that
   \begin{align*}
        \bar b_h(c)
        =
        \frac{c\sqrt{\vell_h^\prime\mH^{-1}\vell_h}}{\sigma_h},
    \end{align*}
    where $\vell_h=(\mI-\mP_W)^\prime g_{\phi,h}$. Since $\hat{\bar b}_h(c)$ is obtained by replacing the population quantities by their consistent estimators, the continuous mapping theorem implies that $\hat{\bar b}_h(c)\inp\bar b_h(c)$ along every sequence
    $\vd_T\in\mathcal{D}(c)$. Therefore, we can conclude that
    \begin{align*}
        \sup_{\vd\in\mathcal{D}(c)}
        P_{T,d}(|\hat{\bar b}_h(c)-\bar b_h(c)|>\epsilon)\to0.
    \end{align*}
    This concludes the proof.
\end{proof}

\begin{proof}[Proof of Theorem \ref{thm:robust_baci_cov}]
    Suppose that the conclusion fails. Then there exists some $\epsilon>0$, a subsequence $T_j$, and $d_j \in \mathcal{D}(c)$ such that
    \begin{align*}
        P_{T_j, d_j}\Big\{\psi_{h,T_j} \in \mathrm{CI}^{\mathrm{BA}}_{h}(c)\Big\} \leq 1-\alpha -\epsilon.
    \end{align*}
    We have assumed that $\mH$ is positive definite, so we have that $\mathcal{D}(c)$ is compact. Therefore, along the subsequence, $\vd_j \rightarrow \vd^*\in \mathcal{D}$. Applying Lemma \ref{lem:unif_irf_distr_miss} to $\vd_{T_j}=\vd_j$, we get that
\begin{align*}
    \frac{\sqrt{T}(\hat{\psi}_h - \psi_{h,T_j,d_j}) - b_h(\vd_j)}{\sigma_h} \ind \mathcal{N}(0,1).
\end{align*}
    Since $b_h(\vd_j) \rightarrow -\vell_h^\prime \vd^*$, Slutsky's theorem gives
    \begin{align*}
        \frac{\sqrt{T_j}(\hat{\psi}_h - \psi_{h,T_j, d_j})}{\sigma_h} \ind \calN(-\frac{\ell^\prime_h \vd^*}{\sigma_h}, 1).
    \end{align*}
    Since $d^* \in \mathcal{D}(c)$, we know that $|-\ell^\prime_h d^*/\sigma_h| \leq \bar{b}$. From the second part of Lemma \ref{lem:unif_irf_distr_miss}, we know that the plug-in estimators are consistent: $\hat{\sigma}_h \inp \sigma_h$ and $\hat{\bar{b}}\inp \bar{b}$. So, continuity of $cv_{\alpha}(\cdot)$ implies that
    \begin{align*}
        cv_{\alpha}(\hat{\bar{b}}) \inp cv_{\alpha}(\bar{b}).
    \end{align*}
    Therefore, the limiting coverage probability for this subsequence is given by
    \begin{align*}
        P\left(\Bigg|Z+\frac{\ell^\prime_h d^*}{\sigma_h}\Bigg| \leq cv_{\alpha}(\bar{b})\right).
    \end{align*}
    However, since $|-\ell^\prime_h d^*/\sigma_h| \leq \bar{b}$, it must hold by definition of $cv_{\alpha}(\bar{b})$ that
    \begin{align*}
        P\left(\Bigg|Z+\frac{\ell^\prime_h d^*}{\sigma_h}\Bigg| \leq cv_{\alpha}(\bar{b})\right) \geq 1-\alpha,
    \end{align*}
    which is a contradiction. Therefore, the theorem holds.
\end{proof}

\subsection{Computational details} \label{app:computation}
\subsubsection{Maximum likelihood estimation}
The log-likelihood function is defined in \eqref{eq:ll}. To compute the maximum-likelihood estimator (MLE), we can use the gradient conditions at the MLE, and iterate through their solutions. First, let us define the following quantities:
\begin{align}
    &\mB_{-k} = \bigotimes_{j\neq k}\mB_j \in \SR^{n/n_k \times n/n_k}, \quad \mOmega_{-k} = \bigotimes_{j\neq k}\mOmega_{j} \in \SR^{n/n_k \times n/n_k}\\
    & \mM_k = \mB_{-k}^\prime \mOmega_{-k}\mB_{-k} \in \SR^{n/n_k \times n/n_k}.
\end{align}
Second, we define the residual tensor and the mode-$k$ unfoldings. The residual tensor is
\begin{align}
\mcalU_t = \mcalY_t - \mcalY_{t-1}\times_1 \mB_1 \times_2 \cdots \times_d \mB_d,
\end{align}
so that, in vectorized form,
\begin{align}
\mathrm{vec}(\mcalU_t) = \mathrm{vec}(\mcalY_t) - \left(\bigotimes_{j=1}^d \mB_j\right)\mathrm{vec}(\mcalY_{t-1}) \in \mathbb{R}^{n}.
\end{align}
For any tensor $\mcalX_t\in\mathbb{R}^{n_1\times\cdots\times n_N}$, we denote by $\mX_{t(k)}\in\mathbb{R}^{n_k\times (n/n_k)}$ its mode-$k$ unfolding (matricization). Mode-$k$ works as follows: for a mode $k$ with dimension $n_k$, fix every index of of the tensor except the $k-th$, that is, fix index $(i_1, \ldots, i_{k-1}, i_{k+1}, \ldots, i_N)$. For that index, we stack the data in mode $k$ into a column. Hence, we obtain $n/n_k$ columns of $n_k$ entries. In other words, we rearrange data in tensor $\mcalX_t$ into a matrix along mode $k$.

Applying mode-$k$ unfolding to our model reduces the model a matrix regression, where $\mB_k$ multiplies from the left, and the combined effect of all other matrices, collected in $\mB_{-k}$, multiplies from the right:
\begin{align}
\mY_{t(k)} = \mB_k \mY_{t-1(k)} \mB_{-k}^\prime + \mU_{t(k)}.
\end{align}
The matrix regression form allows us to estimate each $\mB_k$ separately, rather than estimating a single $n\times n$ matrix. We can specify the gradient conditions at the MLE for the matrix parameters $(\mB_k, \mSigma_k)$ as follows:
\begin{align} \label{eq:rf_mle_gradient}
&\mB_k\left(\sum^T_{t=2}\mY_{t-1(k)}\mM_k\mY_{t-1(k)}^\prime\right) -  \sum^T_{t=2}\mY_{t(k)}\mOmega_{-k}\mB_{-k}\mY_{t-1(k)}^\prime = \mathbf{0}{n_k\times n_k},\\
&\frac{n}{n_k}(T-1)\mSigma_k
= \sum^T_{t=2}\mU_{t(k)}\mOmega_{-k}\mU^\prime_{t(k)},
\end{align}
for each $k=1,\dots,d$, where $\mM_k\in\mathbb{R}^{(n/n_k)\times(n/n_k)}$, $\mY_{t(k)}\in\mathbb{R}^{n_k\times (n/n_k)}$, and $\mSigma_k\in\mathbb{R}^{n_k\times n_k}$, ensuring all matrix products are dimensionally conformable.

Each of the gradient solution has closed-form solutions for the MLE. Each solution depends on the other parameters, so we need to iterate through the solutions until convergence. The iterations are given by
\begin{align}
    \mB_k &= \left(\sum^T_{t=2}\mY_{t(k)}\mOmega_{-k}\mB_{-k}\mY_{t-1(k)}^\prime\right)\left(\sum^T_{t=2}\mY_{t-1(k)}\mM_k\mY_{t-1(k)}^\prime\right)^{-1},\\
    \mSigma_k &= \frac{n_k}{n(T-1)}\sum^T_{t=2}\mU_{t(k)}\mOmega_{-k}\mU^\prime_{t(k)}.
\end{align}

\newpage
\section{Additional simulation results} \label{app:sims}
\begin{table}[!h]
\centering
\resizebox{\textwidth}{!}{%
\begin{tabular}{llllclllclll}
\toprule
                                  & \multicolumn{11}{c}{Average RMSE}                                               \\
                                  \midrule
$T$\textbackslash{}($n_1$, $n_2$) & \multicolumn{3}{c}{(2,3)} & & \multicolumn{3}{c}{(5,5)} & & \multicolumn{3}{c}{(6,9)} \\
\cmidrule(l){2-4} \cmidrule(l){6-8} \cmidrule(l){10-12} 
                                  & $\phi=1$          & $\phi=0.5$         & $\phi=0.05$   &       & $\phi=1$          & $\phi=0.5$         & $\phi=0.05$       &  &  $\phi=1$        & $\phi=0.5$         & $\phi=0.05$ \\
                                  \cmidrule(l){2-4} \cmidrule(l){6-8} \cmidrule(l){10-12}
100                               & 21.884 & 21.392 & 20.930 & & 7.527 & 6.156 & 5.600 & & 18.304 & 10.663 & 9.060\\
250                               & 8.094 & 7.620 & 7.623 & & 3.614 & 2.265 & 2.034 & &  14.075 & 5.863 & 4.464\\
500                               & 3.897 & 3.809 & 3.670 & & 2.484 & 1.187 & 0.982 & & 13.836 & 4.609 & 3.801\\
1000                              & 1.915 & 1.819 & 1.858 & & 1.963 & 0.651 & 0.488 & & 10.002 & 3.846 & 3.244\\
\midrule
   & \multicolumn{11}{c}{St. dev. RMSE}                                               \\
                                  \midrule
$T$\textbackslash{}($n_1$, $n_2$) & \multicolumn{3}{c}{(2,3)} & & \multicolumn{3}{c}{(5,5)} & &\multicolumn{3}{c}{(6,9)} \\
\cmidrule(l){2-4} \cmidrule(l){6-8} \cmidrule(l){10-12} 
                                  & $\phi=1$          & $\phi=0.5$         & $\phi=0.05$   &       & $\phi=1$          & $\phi=0.5$         & $\phi=0.05$      &   &  $\phi=1$          & $\phi=0.5$         & $\phi=0.05$ \\
                                  \cmidrule(l){2-4} \cmidrule(l){6-8} \cmidrule(l){10-12}
100                               & 1.344 & 1.287 & 1.265 &  & 0.441 & 0.368 & 0.341 & & 0.447 & 0.384 & 0.338\\
250                               & 0.784 & 0.750 &  0.725 & & 0.375 & 0.221 & 0.208 & & 0.350 & 0.149 & 0.161\\
500                               & 0.545 & 0.532 & 0.516 & & 0.398 & 0.173 & 0.135 & & 0.389 & 0.163 & 0.155\\
1000                              & 0.376 & 0.371 & 0.352 & & 0.420 & 0.139 & 0.095 & & 0.371 & 0.089 & 0.078\\
\bottomrule
\end{tabular}
}
\caption{This table reports the average root mean square error (RMSE) and its standard deviation for the structural MD estimator for stationarity level $\rho=0.3$, across different values of the sample size $T$ and dimension sizes $(n_1, n_2)$.  All reported values have been multiplied by $10^{2}$.}
\label{tab:sim_rmse_03}
\end{table}

\begin{table}[!h]
\centering
\resizebox{\textwidth}{!}{%
\begin{tabular}{llllclllclll}
\toprule
                                  & \multicolumn{11}{c}{Average RMSE}                                               \\
                                  \midrule
$T$\textbackslash{}($n_1$, $n_2$) & \multicolumn{3}{c}{(2,3)} & & \multicolumn{3}{c}{(5,5)} & & \multicolumn{3}{c}{(6,9)} \\
\cmidrule(l){2-4} \cmidrule(l){6-8} \cmidrule(l){10-12} 
                                  & $\phi=1$          & $\phi=0.5$         & $\phi=0.05$   &       & $\phi=1$          & $\phi=0.5$         & $\phi=0.05$       &  &  $\phi=1$        & $\phi=0.5$         & $\phi=0.05$ \\
                                  \cmidrule(l){2-4} \cmidrule(l){6-8} \cmidrule(l){10-12}
100                               & 21.023 & 20.874 &  20.811 & & 7.535 & 5.900 & 5.662 & & 16.347 & 9.975 & 7.610\\
250                               & 8.041 & 7.834 & 7.674 & & 3.714 & 2.259 & 2.034 & & 11.839 & 5.924 & 4.809\\
500                               & 3.843 & 3.645 & 3.799 & & 2.452 & 1.176 & 0.977 & & 12.315 & 4.851 & 3.850\\
1000                              &  1.866 & 1.876 & 1.874 & & 1.917 & 0.656 & 6.097 & & 10.820 & 4.338 & 3.540\\
\midrule
   & \multicolumn{11}{c}{St. dev. RMSE}                                               \\
                                  \midrule
$T$\textbackslash{}($n_1$, $n_2$) & \multicolumn{3}{c}{(2,3)} & & \multicolumn{3}{c}{(5,5)} & &\multicolumn{3}{c}{(6,9)} \\
\cmidrule(l){2-4} \cmidrule(l){6-8} \cmidrule(l){10-12} 
                                  & $\phi=1$          & $\phi=0.5$         & $\phi=0.05$   &       & $\phi=1$          & $\phi=0.5$         & $\phi=0.05$      &   &  $\phi=1$          & $\phi=0.5$         & $\phi=0.05$ \\
                                  \cmidrule(l){2-4} \cmidrule(l){6-8} \cmidrule(l){10-12}
100                               & 1.280 & 1.326 & 1.284 & & 0.432 & 0.344 & 0.350 & & 0.449 & 0.315 & 0.345\\
250                               & 0.786 & 0.738 & 0.769 & & 0.394 & 0.216 & 0.204 & & 0.494 & 0.240 & 0.234\\
500                               & 0.525 & 0.514 & 0.520 & & 0.390 & 0.168 & 0.137 & & 0.457 & 0.184 & 0.132\\
1000                              & 0.368 & 0.361 & 0.359 & & 0.420 & 0.152 & 0.156 & & 0.320 & 0.163 & 0.090 \\
\bottomrule
\end{tabular}
}
\caption{This table reports the average root mean square error (RMSE) and its standard deviation for the structural MD estimator for stationarity level $\rho=0.8$, across different values of the sample size $T$ and dimension sizes $(n_1, n_2)$.  All reported values have been multiplied by $10^{2}$.}
\label{tab:sim_rmse_08}
\end{table}

\begin{figure}
    \centering
    \includegraphics[width=\linewidth]{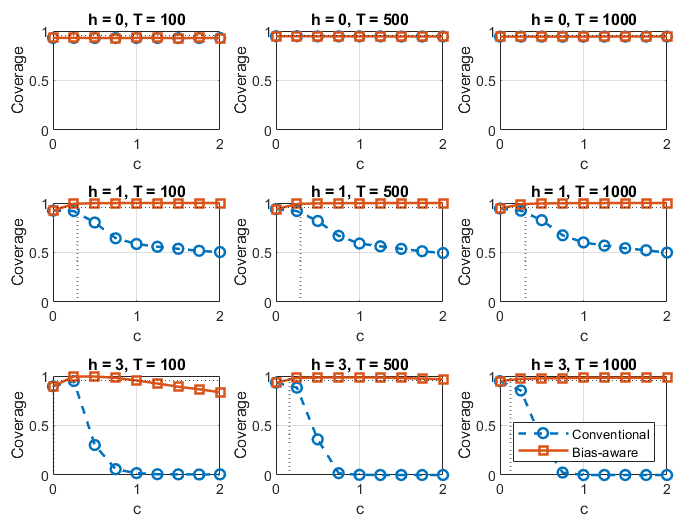}
    \caption{This figure reports the coverage results for the robust bias-aware confidence intervals (Section \ref{sec:robust}) and the conventional confidence intervals (Section \ref{sec:estim and props}) at horizons $h=0,1,3$ (rows) for different values of $T$ (columns).}
    \label{fig:sim_robust_robust_conv_app}
\end{figure}

\newpage
\bibliographystyleapndx{chicago}
\bibliographyapndx{bib.bib}

\end{document}